\documentclass[twocolumn]{aastex62}
\def\gsim{\ \raise 3pt \hbox{$\rangle$} \kern -8.5pt \raise -2pt \hbox{$\sim$}\ }
\usepackage{graphicx,natbib,color,amsmath,subfigure}
\usepackage{longtable}
\usepackage{ulem}
\newcommand{\blank}[1]{}
\usepackage{color}

\def\sdo{{\textit{SDO}}}
\def\mw{{microwave}}

\def\SDO{{\textit{SDO}}}

\journalinfo{Accepted for publication in The Astrophysical Journal December 02, 2022}

\begin{document}
\title{Strongest coronal magnetic fields in solar cycles 23-24: probing, statistics, and implications}

\author[0000-0003-2990-1390]{Viktor V. Fedenev}
\affil{Institute of Solar-Terrestrial Physics SB RAS, Lermontov st., 126a, Irkutsk, 664033 Russia}
\author[0000-0002-1107-7420]{Sergey A. Anfinogentov}
\affil{Institute of Solar-Terrestrial Physics SB RAS, Lermontov st., 126a, Irkutsk, 664033 Russia}
\author[0000-0001-5557-2100]{Gregory D. Fleishman}
\affil{Center For Solar-Terrestrial Research, New Jersey Institute of Technology, Newark, NJ 07102}

\begin{abstract}
Strong coronal magnetic field, when present, manifests itself as bright microwave sources at high frequencies produced by gyroresonant (GR) emission mechanism in thermal coronal plasma. The highest frequency at which this emission is observed is proportional to the absolute value of the strongest coronal magnetic field on the line of sight. Although no coronal magnetic field larger than roughly 2,000\,G was expected, recently the field at least twice larger has been reported. Here, we report a search for and statistical study of such strong coronal magnetic fields using high-frequency GR emission.
A historic record of spatially resolved \mw\ observations at high frequencies, 17 and 34\,GHz, is available from Nobeyama RadioHeliograph for more than 20 years (1995--2018). Here we employ this data set to identify sources of bright GR emission at 34\,GHz and perform a statistical analysis of the identified GR cases to quantify the strongest coronal magnetic fields during two solar cycles. We found that although active regions with the strong magnetic field are relatively rare (less than 1\% of all active regions), they appear regularly on the Sun. These active regions are associated with prominent manifestations of solar activity. 
\end{abstract}

\keywords{Sun: magnetic fields---Sun: radio radiation---sunspots---Sun: general---Sun: corona---methods: data analysis}

\section{Introduction}

Strongest large-scale solar magnetic fields are concentrated in active regions (ARs) that host most of the solar activity such as flares and eruptions.
The energy supporting this activity is stored in the coronal magnetic field, which is a dominant form of energy in a  low-$\beta$ coronal environment.
Therefore, extreme events, such as exceptionally powerful solar flares or eruptions, require proportionally high level of the free magnetic energy in the host AR. 
Thus, identification of active regions with exceptionally strong magnetic field in the corona is of crucial importance for studying and predicting extraordinary solar events.

At the photospheric level,
a typical AR can host the magnetic field with the absolute values up to around $\sim$3,000\,G.
However, in some rare cases,  it can be much stronger, reaching 5,000\,G or even more \citep{2006SoPh..239...41L, 2018ApJ...852L..16O, 2018AGUFMSH41C3646O,Wang_2018}.

Coronal magnetic field is created by the corresponding photospheric field (that can be viewed as a boundary condition) and electric currents flowing in the coronal volume.
Thus, high photospheric magnetic field  will also manifest itself at higher heights of the solar atmosphere---the chromosphere and corona. In the chromosphere, measurements of the full vector, or the absolute value, of the magnetic field are available from only a few case studies due to both observational difficulties and sophisticated time consuming inversion process \citep[see, e.g., the review by][]{2018GMS...235...43F}. The longest set of regular data are available from the Synoptic Optical Long-term Investigation of the Sun (SOLIS) facility \citep{2003SPIE.4853..194K, 2011SPIE.8148E..09B}, but provides only the line-of-sight (LOS) component of the chromospheric magnetic field in the weak-field approximation. 

\begin{figure*}
    \centering
    \includegraphics[width=\linewidth]{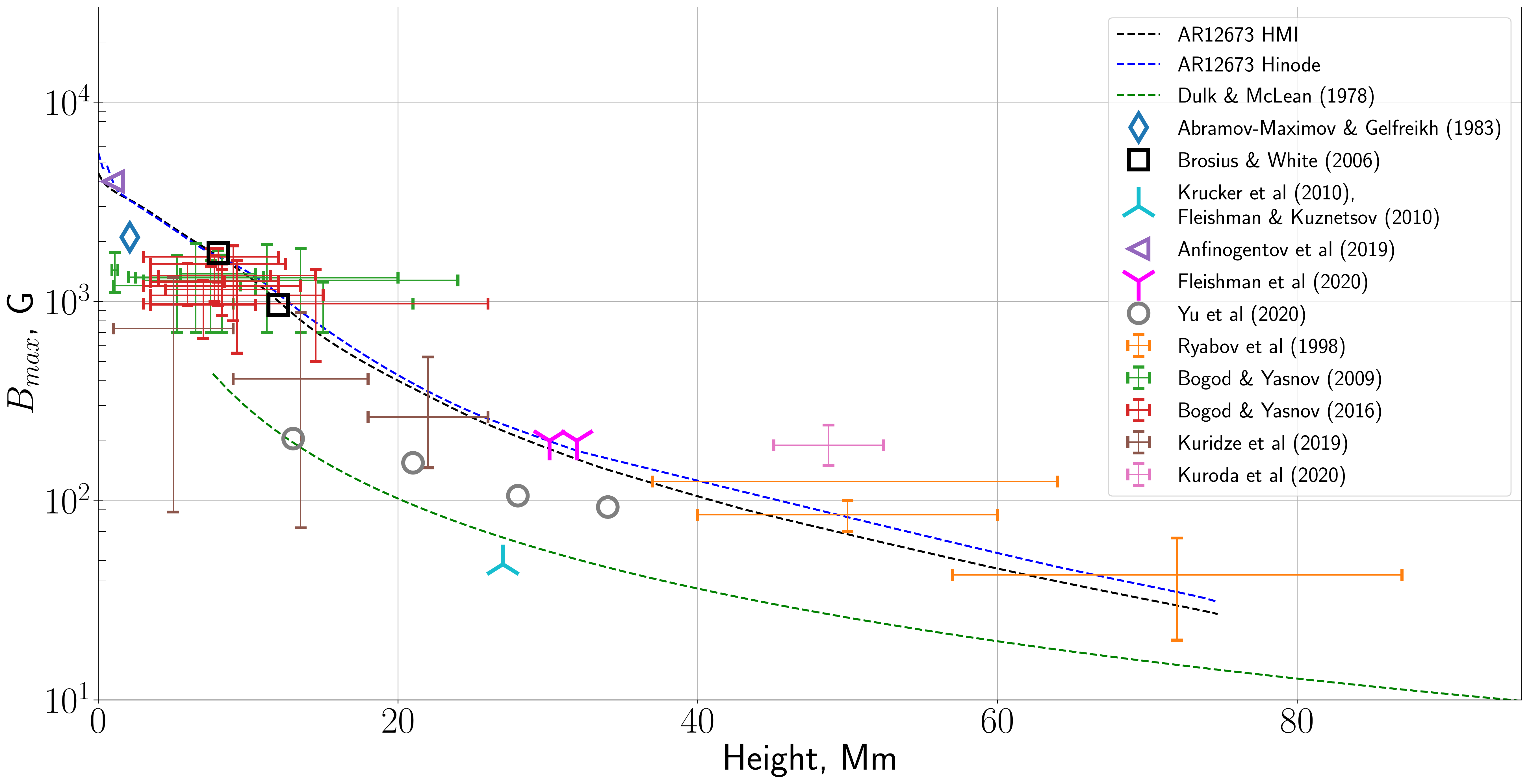}
    \caption{Selected set of coronal magnetic field measurements using radio methods
    \citep{2006ApJ...641L..69B, 2010ApJ...714.1108K,Fl_Kuzn_2010, 2019ApJ...880L..29A, 2020Sci...367..278F, 2020FrASS...7...22K, 2020ApJ...900...17Y, 1983SvAL....9..132A, 1999SoPh..185..157R, 2009AstBu..64..372B, 2016SoPh..291.3317B} and optical Zeeman method        \citep{2019ApJ...874..126K}. The meaning of the symbols are shown in the panel. The figure also displays a phenomenological height dependence of a typical AR magnetic field proposed by \citet{1978SoPh...57..279D} and the dependence of the strongest coronal magnetic field on the height derived by \citet{2019ApJ...880L..29A} for AR\,12673.
    }
    \label{fig:field_measurement_history}
    \label{f_all_coronal_B}
\end{figure*}

Historically, the coronal magnetic field at various heights was evaluated using indirect methods. This includes extrapolations from the photospheric data and the use of coherent metric or decimetric emissions. More direct probing was occasionally available from Zeeman splitting in the cool plasma, e.g., of the solar prominences, or based on the microwave data. Early measurements and estimates were compiled together by \citet{1978SoPh...57..279D}, who proposed a general relation $B=0.5 [(R/R_\sun) -1)]^{-1.5}$\,G in the range $1.02<(R/R_\sun)<10$, which was consistent with all individual inputs within a factor of 3; see the dashed green curve in Figure\,\ref{f_all_coronal_B}.

More recently, some of the traditional techniques for coronal magnetic field probing were refined, while new ones have been proposed. 
Direct  coronal magnetic field measurements in the optical range using Zeeman effect are possible
if suitable spectral lines are formed at coronal heights.
For instance, \citet{2019ApJ...874..126K} measured magnetic field in  coronal loops filled with a cold chromosheric-temperature  material formed due to the coronal rain phenomenon (plasma cooling due to thermal instability).
This permitted conventional CaII Zeeman chromospheric magnetography in the weak field approximation, but at coronal heights. Another approach is to utilize spectropolarimetric observations of \lq \lq forbidden\rq\rq\  coronal infrared lines which have much better ratio between Zeeman splitting and thermal broadening than spectral lines in visible range \citep{2004ApJ...613L.177L}.

The microwave emission from solar active regions was firstly observed in 1959 by  \citet{1959AnAp...22....1K}. A few years later,  \citet{1962ApJ...136..975K,1962SvA.....6....3Z} demonstrated,  that high-intensity microwave radiation from solar active regions is related to gyroresonance (GR) opacity. GR is produced by thermal electrons gyrating in the ambient magnetic field. GR radiation at a given frequency $f$
is formed in several thin gyro layers, where the absolute value of the magnetic field corresponds to one of  low harmonics of the local gyro frequency $f_{Be} = \frac{eB}{2\pi m_e c}$.

Often, bright microwave sources are observed in the vicinity of a neutral line delimiting positive and negative polarities of an active region \citep{1977ApJ...213..278K,1986ApJ...301..460A,1993A&A...270..509A}. Such sources  are usually referred to as peculiar or neutral line sources (NLS). Based on unusually high brightness temperature of the NLSs, an idea of a steady nonthermal electron population was proposed \citep[e.g.,][]{1993A&A...270..509A}. However, the
neutral line is a preferable location of magnetic flux-ropes, which can be characterized by the increased absolute value of the magnetic field due to its twist.
The strong magnetic field value and the associated electric current may result in higher plasma temperatures.
These support gyroresonant nature of NLS.
Indeed, numerical simulations of the microwave emission above neutral line \citep{2014SoPh..289.1215Y,2016SoPh..291..823K} demonstrate that gyroresonance emission is a likely  mechanism of NLSs in active regions and can be efficient at the third, fourth, and and even fifth gyro harmonic.
Similar conclusion has been obtained in several observational studies  of neutral line sources \citep{1982ApJ...253L..49A, 2006PASJ...58...21U, 2008SoPh..249..315U}  where the authors attributed some of the investigated sources to the gyroresonance emission. 

In the case of non-monotonic variation of the physical
parameters along the line of sight (e.g. when the line of sight crosses a hot coronal loop), spectral lines (also called cyclotron lines) can be formed as an enhanced microwave brightness in one or several narrow frequency bands \citep{1975SoPh...44..461Z, 1980IAUS...86...87Z}.
Observations at a single frequency do not allow us, however, to discriminate the cyclotron lines from NLS and usual sunspot-associated GR sources.

The coronal magnetic field in flaring loops can be obtained from analysis of  microwave imaging spectroscopy data on gyrosynchrotron emission associated with solar flares.
The model spectral fitting of these data yields maps of the absolute value of the magnetic field and its direction relative to the line of sight (LOS) \citep{Gary_etal_2013,2020Sci...367..278F, 2020FrASS...7...22K}.

Besides analysis of gyroresonance and gyrosynchrotron emission, magnetic field can be estimated by the analysis of thermal free-free emission \citep{1980SoPh...67...29B,2000A&AS..144..169G,2017A&A...601A..43L} and by utilising circular polarization inversion  
phenomenon \citep[e.g.,][]{1999SoPh..185..157R}. For the detailed description of advanced methods in radio magnetography, see the review by \cite{2021alissandrakis}.

In addition to these direct measurements, there are several  indirect approaches. 
One of them employs coronal MHD-seismology. It allows measuring the coronal magnetic field by  analysis of kink oscillations of coronal loops seen in EUV \citep{2001A&A...372L..53N}. 
The obtained value is interpreted as a weighted average along the coronal loop with a dominant contribution from a segment near the loop top.
Yet, at the present time there is no reliable routine method for direct measuring the magnetic field vector in solar corona at an arbitrary location. 

Figure\,\ref{f_all_coronal_B} compiles several measurements of the coronal magnetic field using the mentioned techniques. For the reference, it also shows the \citet{1978SoPh...57..279D} relationship along with the largest magnetic field values vs the heights. The latter are derived from the  non-linear force-free field (NLFFF) extrapolation of the coronal magnetic field reconstructed from the photospheric magnetograms \citep{2019ApJ...880L..29A}, which represents the state-of-the-art of the coronal magnetic field modeling \citep[e.g.,][]{2000ApJ...540.1150W}.

In this study we focus on the magnetic field measurements in the low corona using thermal gyroresonant emission utilizing a two-solar-cycle long database of the Nobeyama RadioHeliograph \citep[NoRH,][]{1994IEEEP..82..705N}. Typically, gyro emission of solar active regions is  observed in the range of 1--20\,GHz. The major contribution to this emission comes from gyroresonant layers corresponding to the second (ordinary mode) and third (extraordinary mode) gyro harmonics.
Microwave radiation at these harmonics is typically optically thick and, thus, the electron temperature in the outermost optically thick gyro layer is equivalent to the brightness temperature of the corresponding microwave source. GR emission from higher harmonics is typically optically thin, so it  contributes less to the  microwave brightness.

Thus, observations of a bright ($\gtrsim1$\,MK), optically thick GR source at a certain frequency $f$ is a direct evidence that the magnetic field in the corona is at least as high as $B=\frac{ 2 \pi m_e c}{n e}f$, where $n$ is the harmonic number; most likely $n=3$. Therefore, bright gyroresonant emission at unusually high microwave frequencies implies the existence of proportionally high magnetic field values in the corona.

One such case was discovered by \citet{2019ApJ...880L..29A}. They detected an optically thick GR emission associated with AR 12673 which produced the most powerful solar flare in 24$^{th}$ solar cycle.
As far as we are aware, the GR source in AR\,12673  is the only yet reported case of the observed optically thick GR emission at 34 GHz (GR-34 for short), which, if corresponds to the third gyro harmonics, implies the coronal magnetic field in excess of 4,000\,G.
For this AR, \citet{2019ApJ...880L..29A} built two 3D models of the coronal magnetic field based on the Helioseismic and Magnetic Imager of Solar Dynamics Observatory (SDO/HMI) and Spectropolarimeter of Solar Optical Telescope on-board Hinode spacecraft (Hinode SOT/SP) photospheric magnetograms.
The maximal magnetic field upon height derived from the models and plotted in Fig.~\ref{fig:field_measurement_history} are consistent with the NoRH measurement.
Several  coronal magnetic field measurements shown in Fig.~\ref{fig:field_measurement_history}, which utilized Zeeman, gyroresonant, and gyrosynchrotron diagnostics, are rather close to the highest values of the magnetic field predicted by the NLFFF models.
This implies that there might be other ARs with extremely high coronal magnetic fields manifesting themselves in GR-34 emission.

Using the well-studied case of AR~12673 \citep{2019ApJ...880L..29A} as a guide, we search for other bright GR-34 sources with similar properties in other active regions.
To find such active regions, we study full-disk microwave imaging observations available from the Nobeyama Radioheliograph (NoRH) at 17 \& 34\,GHz for about two  solar cycles. Such a long duration of the database permits us to find even rather rare events and investigate their statistical properties.

The paper is organized as follows. In Section~\ref{sec:AR12673}, we overview the reported GR-34 case in AR~12673. The observational data used in this work is described in Section~\ref{sec:obs_data}. Our workflow for analysing the entire set of NoRH data is presented in Section~\ref{sec:data_analysis}. A statistical study of the ARs potentially exhibiting GR emission at 34 GHz is given in Section~\ref{sec:statistical_analysis}. Finally, we discuss the obtained results  in Section~\ref{sec:discussion_and_conclusion}.

\section{NOAA AR 12673 on 2017-Sep-06}
\label{sec:AR12673}
Since AR\,12673 is the only reported case of the GR emission at 34\,GHz, we use it as a reference while searching for GR sources at this frequency in other ARs.
The reported GR-34 source was observed on 2017-Sep-06 directly revealing the magnetic field within 3,000--4,000\,G at the transition region \citep{2019ApJ...880L..29A}, while the photospheric magnetic field exceeded 5,000\,G  \citep{Wang_2018}. This GR-34 source displayed the brightness temperature of more than 150,000\,K almost all day and, thus, stood out cleanly against the solar disk with the brightness temperature of (1--2)$\times10^4$\,K.

\begin{figure}[htbp] \centering
    \includegraphics[width=\linewidth]{"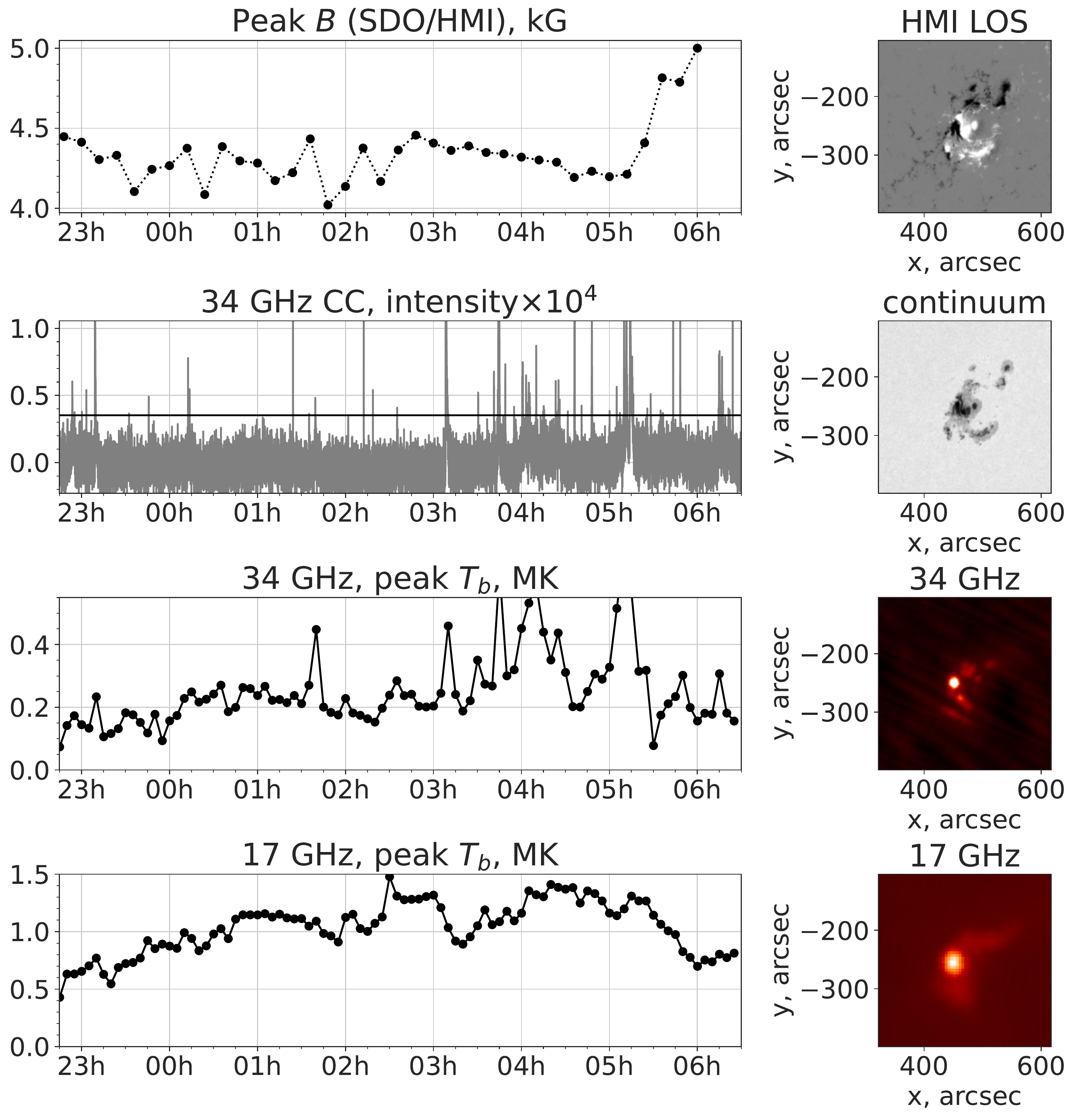"}
    \caption{A set of data for the GR-34 case observed on 2017-Sep-06. Left panels demonstrate time series of the peak magnetic field, the correlation curve for 34\,GHz, and peak brightness temperatures at 34 and 17\,GHz. Horizontal line in the CC panel shows the radio burst filtering threshold described in Section \ref{sec:data_analysis}. Right panels show the line of sight SDO/HMI magnetogram, SDO/HMI continuum image, and NoRH images at 34 and 17\,GHz. SDO/HMI images were taken at 02:00, radio images were taken at 02:04. The maximum brightness temperatures at the radio images are 163\,877\,K and 1\,151\,161\,K for 34 and 17\,GHz, respectively.}
    \label{fig2017}
\end{figure}

We put together relevant observational properties of AR 12673 in Figure~\ref{fig2017}, which displays the maximum absolute values of the photospheric magnetic field reported by SDO/HMI, the correlation curve (see below in Section\,\ref{sec:mw_data}), and the peak brightness temperatures of each image at 34\,GHz and 17\,GHz. The photospheric magnetic field in the AR is higher than 4,000\,G and, by the end of the NoRH observations,  increases toward 5,000\,G, which is the upper bound that HMI pipeline can report \citep{2014SoPh..289.3483H}. Correlation curve values display numerous outliers (short peaks), which implies numerous radio bursts during the day.

The observed radio brightness fluctuates indicative of variations of thermal properties of the gyroresonance plasma layers, although some of enhancements are certainly due to flares.
The radio images at both 17 and 34\,GHz are dominated by a compact source with a nearly circular shape and an apparent area of approximately two NoRH beams. \citet{2019ApJ...880L..29A} devised a 3D magneto-thermal model that reproduced the observed properties of GR emission at both 17 and 34\,GHz. Microwave emission synthesised from the model at 34\,GHz displays several very compact locations of optically thick GR emission with the brightness temperature of several MK. Being convolved with the NoRH beam, whose area is noticeably larger than any of the GR-34 source, the peak brightness temperature falls well below 1\,MK. The shape and size of the synthesized convolved source match well the observed ones remaining rather compact, comparable with the beam size. We conclude that even  sources with moderate observed brightness temperature ($\lesssim0.5$\,MK) can be optically thick if they are compact. 

The observational properties of the GR-34 source described above, namely, long lifetime, brightness temperature above $10^5$\,K, and the small area are interpreted as manifestation of a bright optically thick unresolved gyroresonant source at 34\,GHz.

This AR demonstrated extraordinary activity. It produced several X-class flares, many M-class flares, and eruptions. It was associated with nine sunquakes \citep{2020ApJ...895...76S} and two solar energetic particle (SEP) events \citep{sepevents}.

\section{Observational data}
\label{sec:obs_data}

\subsection{Microwave data}
\label{sec:mw_data}

The main source of data used in this study is the NoRH full-disk microwave imaging observations of the Sun  at 17\,GHz and 34\,GHz \citep{1994IEEEP..82..705N}. Here, we focus on the 34\,GHz data while use the 17\,GHz data for the context as needed.

NoRH data products include the radio images, the correlation curves, as well as the raw visibilities.
The data is available for every observational day from 22:45\,UT to 06:30\,UT regardless the time of year.
The cadence is 1 second for the correlation curves  and 10 seconds for the raw  data \citep{1994IEEEP..82..705N}.
    Pre-synthesized images are available  as 10-minutes images for 17\,GHz, while the cadence of 34\,GHz images is one per day.
 
Although the NoRH archive contains  only one  34\,GHz image per observational day, a user can synthesize images  at both frequencies with a  high cadence down to 10\,s.
The radio images used in this work were synthesized from raw data using the corresponding NoRH package routines from SolarSoftWare \citep[SSW, ][]{1998SoPh..182..497F} library of programming language IDL (Interactive Data Language).
Raw datasets were automatically downloaded from NoRH database by the package routines.

The NoRH correlation curves (CC) are the sums of correlation coefficients of every pair of radioheliograph antennas excluding short-base pairs \citep{1994IEEEP..82..705N}. For our study they were also downloaded from the NoRH database. This quantity is directly related to the radio flux from compact radio sources as it involves only high spacial frequencies.
In the absence of flares and radio bursts, the correlation curve fluctuates around a constant value.
Fast changes in emission intensity, such as bursts, lead to changes in the correlation curve, causing peaks of various heights and duration.
In our study we used radio images to identify bright sources -- the GR-34 candidates, while employed the correlation curves to flag and eliminate flaring episodes.

\subsection{Photospheric Magnetograms}

The presence of optically thick GR radiation at $n=3$ at 34\,GHz implies the magnetic field of at least 4,000\,G at the transition region. Photospheric field values need to be higher than that. To check if an unusually strong photospheric magnetic field is present in the selected ARs, we employ photospheric magnetograms.

 SDO/HMI \citep{2012SoPh..275....3P} produces full-disk maps of the magnetic field vector with high time (12 minutes) and spatial ($\approx 1$ arcsecond, spacing $\approx 0.5$ arcseconds per pixel) resolution starting 2010.

Hinode Solar Optical telescope Spectro-Polarimeter (Hinode SOT/SP) data is available since 2006 and provides limited field of view (FOV) magnetograms with a spatial resolution down to 0.3 arcseconds per pixel.
Compared with SDO/HMI data, magnetograms from Hinode SOT/SP offer lower cadence, but the sensitivity and spatial resolution of Hinode SOT/SP data are much better \citep{2008SoPh..249..167T} than those of \SDO/HMI data because of higher spatial and spectral resolution of raw spectrapolarimetric data.
The absolute values of the magnetic field reported by Hinode SOT/SP for the same ARs are often higher than  \SDO/HMI ones, since the former resolves finer spatial structures with potentially higher values of the magnetic field.
Data from both instruments may contain artifacts in the areas of very strong magnetic field; see an example in \citet{2019ApJ...880L..29A}.
Both Hinode SOT/SP and \SDO/HMI measure strong magnetic fields with the upper limit of around 5,000\,G. Here we use the routinely available \SDO/HMI data as a primary tool for inspecting the photospheric magnetic fields, while employ the Hinode SOT/SP data to cross-check and confirm the \SDO/HMI results as needed. 

For earlier cases that occurred  before Hinode and \sdo, we used the line of sight (LOS) observations made with Michelson Doppler Imager instrument on-board Solar and Heliospheric Observatory (SOHO/MDI) \citep{1995SoPh..162..129S}. The limitations of this instrument do not allow for reliable measurements of very strong magnetic fields, the upper limit is around 3,000\,G. However it provides us information about magnetic morphology of the active regions and allows us to estimate the lower bound of the magnetic field value. 
Occasionally, vector magnetic measurements could be available from ground-based observations  \citep[e.g.,][]{2006SoPh..239...41L}.

\subsection{Other data sources}
Soft X-ray light curves from GOES were used to identify flaring activity.
Active regions as well as the times of solar flares, needed to filter them out, were obtained from the Heliophysics Events Knowledgebase  \citep[HEK,][]{2012SoPh..275...67H}.

\section{Data analysis}
\label{sec:data_analysis}

\begin{figure}[htbp] \centering
    \includegraphics[width=\linewidth]{"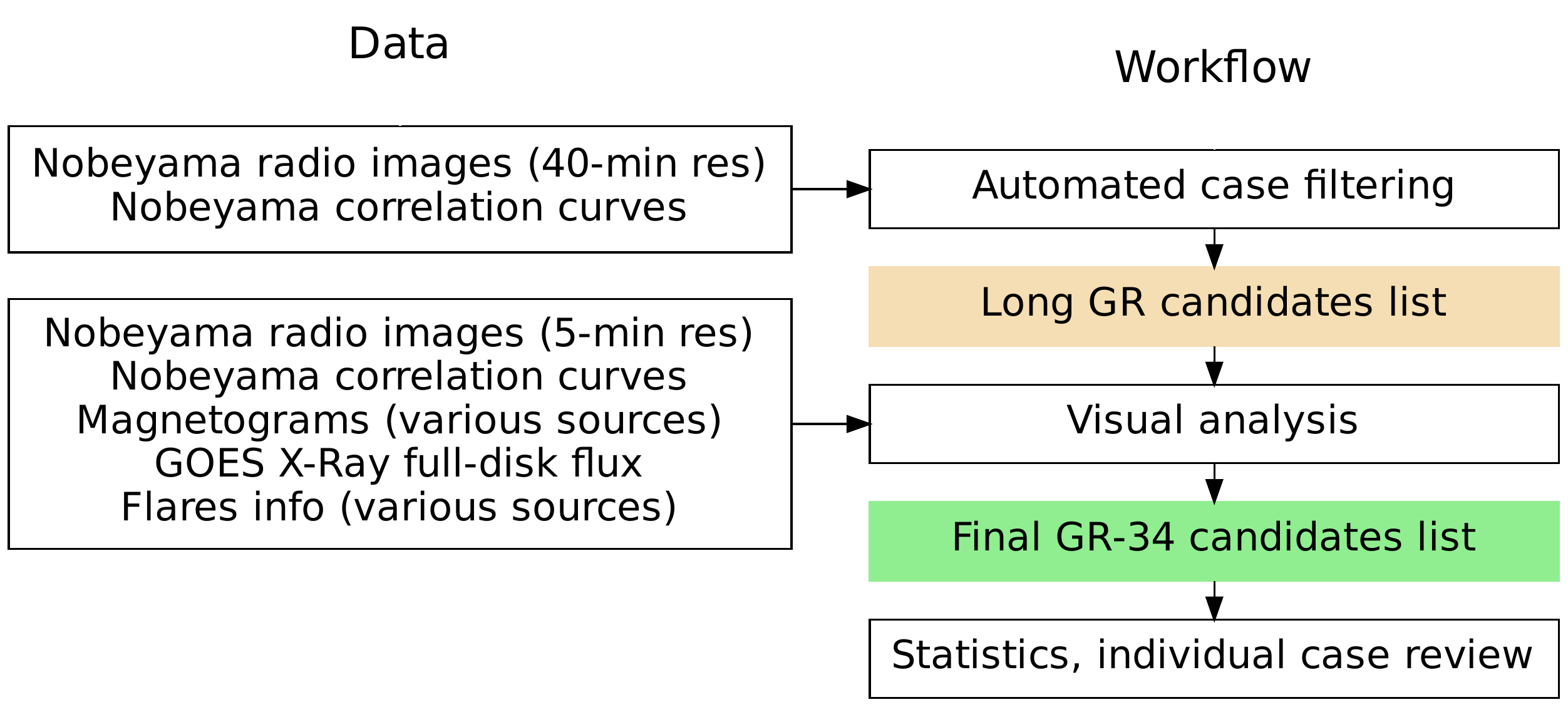"}
    \caption{Flow chart illustrating the process of selecting potential GR-34 candidates and data usage in this study.}
    \label{workflow}
\end{figure}

Guided by the reference  case of NOAA 12673, we developed an algorithm searching for similar bright GR-34 sources in the entire NoRH dataset from 1995 to 2017. Year 2018 and later is available in archives, but we excluded it from the analysis because of poor data quality. The workflow (see Fig.~\ref{workflow}) includes automated identification of long-living radio sources on images in the NoRH database with the peak brightness above a certain threshold and manual filtration of false alarm cases such as flares, glitches etc. Confirmed GR-34 candidates then receive a more detailed analysis. In addition, we check several ``famous'' ARs that attracted extraordinary attention in the literature.

\subsection{Automated selection and  filtration of the event}
Quiet sun brightness temperature at 34 GHz is about 10,000\,K. ARs and flare radio sources can have higher values  up to several millions Kelvin. In the reference event (6th September 2017) the values between 100,000\,K and 500,000\,K were observed while mostly remaining above 150,000\,K. In this study the threshold 100,000\,K was chosen for the target gyroresonant events to get rid of most of free-free emission sources, while maintaining likely gyroresonance cases for subsequent manual analysis. At this stage we check only the global maximum values of intensity for each radio image.
During this step a long list of GR-34 emission candidates is formed.

\subsubsection{Selection algorithm}

 To capture all potential GR-34 sources, we synthesized full-disk images with 40 minute cadence from raw visibilities.
 We compromise between higher cadences and lower data download size, and the 40 minute cadence has been proven sufficient to catch the GR-34 days-candidates for further analysis.
 We select all days when the brightness temperature exceeds the threshold of 100,000\,K in at least two (not necessarily consecutive) images generated with the 40 min cadence.

To filter out bright sources related to flares, we check that CC value remains within the noise threshold at the corresponding time. This ensures that our algorithm detects non-flaring cases with the brightness temperature above the predefined threshold.

To make sure that the brightness temperature remains  high for most of the  observation day, we also require that the peak brightness temperature of 80\,\% of selected 40 minute cadence images is greater than 60,000\,K. This is an empirical technique to reduce the number of false-positive cases before manual analysis.

The automated selection/filtration algorithm is given below:

\begin{enumerate}
\item Download 34\,GHz NoRH correlation curves for each day between 1995 and 2017
\item {Download 34\,GHz raw NoRH visibilities data and synthesize 40-minute cadence radio images for the same time range}
\item For every observational day:
\begin{enumerate}
    \item Calculate noise threshold for correlation curve.
    \item Compute maximum value of the brightness temperature for each image
    \item If the peak brightness temperature exceeds  100,000\,K for at least two images during the day and correlation coefficient remains below the noise threshold at the corresponding times, continue to the next step
    \item If 80\,\% of all peak brightness temperatures obtained on step (3b) exceed 60,000\,K, the corresponding radio source is selected as a GR-34 candidate
\end{enumerate}
\end{enumerate}

\subsubsection{Filtering out bursty episodes}

From our reference
case AR12673 (see Fig. \ref{fig2017}) we know that there could be a  bursty activity on top of thermal GR emission, which manifests itself in the form of large short spikes on the correlation curve. Thus, we need to automatically identify those spikes that may indicate radio burst activity. This task requires that the noise level of the correlation curves is robustly determined.

We checked that a majority of the values of correlation curve during the day are approximately normally distributed around their mean value. This type of distribution takes place in quiet days (without flares), but in the presence of any significant bursts its shape has a long tail towards higher values (see Fig. \ref{quietsun}b). In order to correctly determine the noise level threshold under these conditions, we use a robust method as follows:

\begin{figure}[htbp] \centering
    \includegraphics[width=\linewidth]{"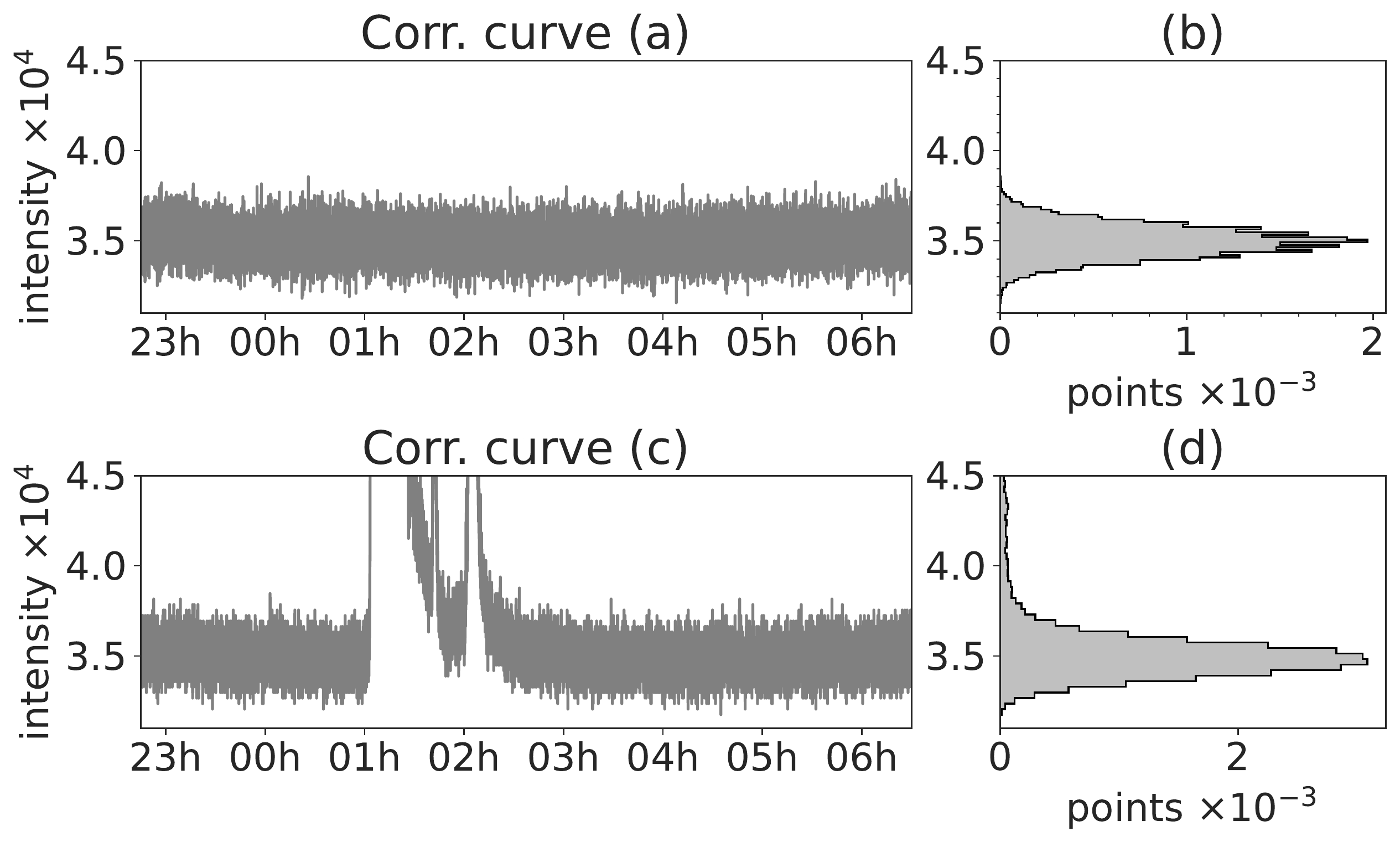"}
    \caption{Examples of 34\,GHz correlation curves and the histograms of their values for a quiet (a, b) and a bursty (c, d) day. The histogram at the right represents the occurrence distribution of the correlation curve values at the left. Panel (b) displays a symmetric bell-shaped distribution of noise, while panel (d) features a tail around higher values because of a burst seen in panel (c).}
    \label{quietsun}
\end{figure}

\begin{enumerate}

\item Compute median value ($med$) of the correlation curve during the day
\item Pick up all CC values below median and compute their standard deviation ($std$)

\item Set the threshold as $thr = med + 6 \, std$ to minimize the false-positive detection of flares
\end{enumerate}

This algorithm works well on the correlation curves with burst spikes or without them, because the bottom half of the distribution was found to be represented by the noise.

We tried different approaches to filter out the bursts and found that the adopted approach with the correlation curves is more reliable than the use of already available event lists (like HEK or Solar Flare Database \citep{2017ApJS..231....6S}), because some of the radio bursts present in the correlation curve are not listed in the databases. Also, the cadence of correlation curve (1\,s) exceeds the cadence of synthesized radio images, which permits us to identify even rather short bursts.

 After automated processing of several thousands of observational days, the algorithm found less than 100 candidates which were further analysed manually.

\subsection{Visual analysis}
The automated algorithm described above was designed to identify all potential gyroresonant sources, but it is unable to  filter out all false alarm cases such as corrupted data and some of the flare radio burst events.

Thus, it is essential to perform manual analysis in order to exclude such cases to form a clean final list of GR-34 source candidates. For this purpose we obtained a more detailed information on the ARs in the initial list by synthesising radio images at 34\,GHz and 17\,GHz with the 5 minute cadence from the raw data. To reduce noise, the 5 minute cadence images were averaged over ten 10-second intervals with norh\_synth options nfrint=10, nfrcal=10.

\subsubsection{Corrupted data removal}
\begin{figure}[htbp] \centering
    \includegraphics[width=\linewidth]{"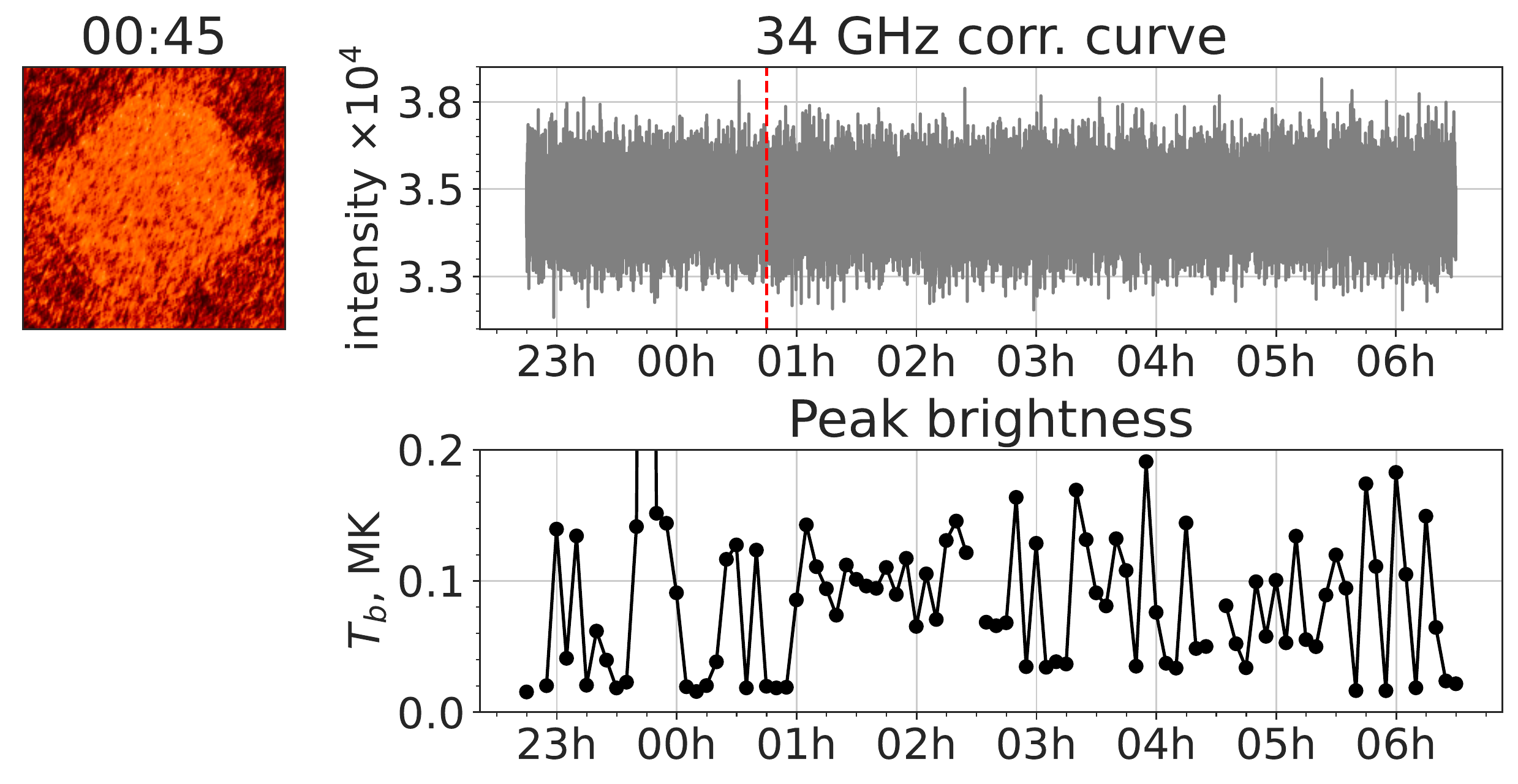"}
    \caption{Example of corrupted data from Nobeyama Radioheliograph. Image is  distorted, values of the maximum brightness temperature are meaningless. The actual values are concentrated around 20,000\,K.}
    \label{baddata}
    \end{figure}

In the initial list of the GR-34 event candidates there were days with failed imaging data, where neither the solar disk nor radio sources can be recognized in the images. The automated algorithm picked up such events because the correlation data for such days was normally distributed and showed a ``quiet day'' signature; while the maximum brightness plots had many outliers (see example in fig.~\ref{baddata}). The ARs automatically selected based on such   corrupted data were manually excluded from further analysis.
    
\subsubsection{Flares  and bursts}

\begin{figure}[htbp] \centering
    \includegraphics[width=\linewidth]{"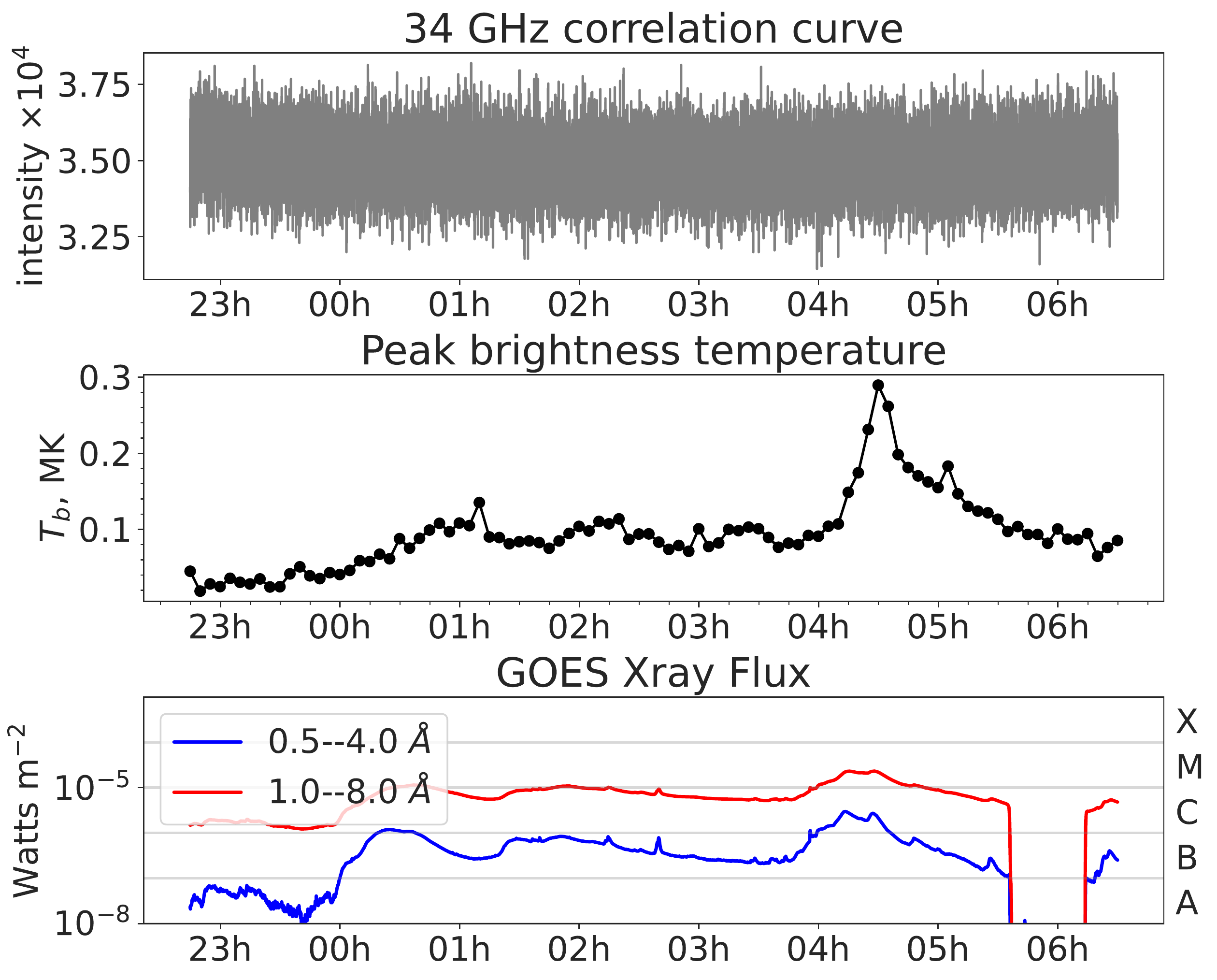"}
    \caption{Example of a solar flare (M1.6 event observed 5 of September, 2011) not visible in NoRH 34\,GHz correlation data (upper panel), but clearly seen in the peak brightness temperature at 34\,GHz (middle panel) and GOES X-Ray flux (bottom panel) plots around 4:30 UT as a sharp spike followed by a quasi-exponential decay.}
    \label{silentcc}
    \end{figure}

The radio images alone do not allow for unambiguous distinguishing of GR sources from long duration flares.
In order to filter out flaring episodes remaining after initial automated filtration, we visually investigated evolution of the correlation curves and peak brightness temperature during the day for every GR-34 candidate. However, spatially extended solar flares may not show up in the correlation curves. Example of such an event is presented in Fig. \ref{silentcc}.

\begin{figure*}[htbp] \centering
    \includegraphics[width=\linewidth]{"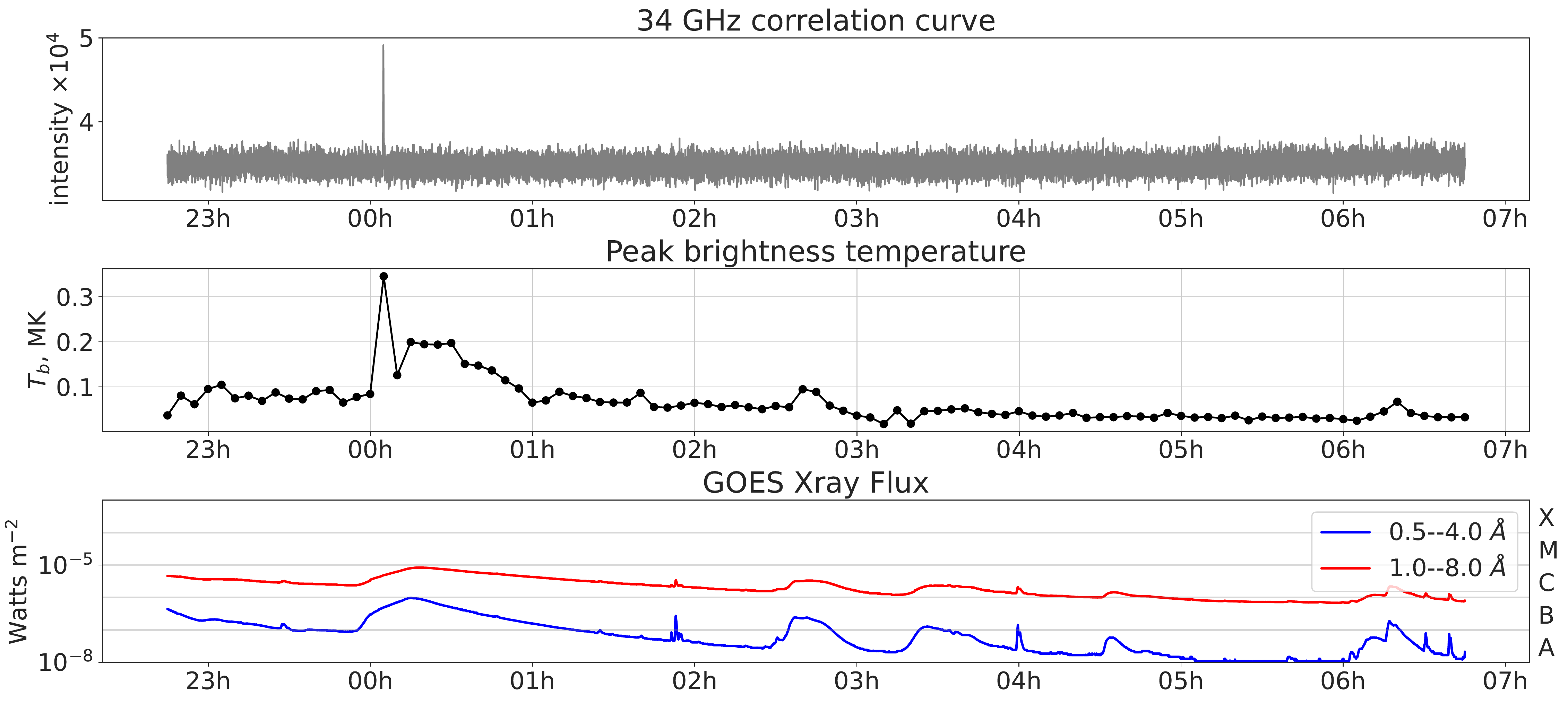"}
    \caption{Flaring activity in different datasets (observation date: 2003-Aug-15). Top panel is NoRH correlation curve; middle panel is the per-image brightness temperature of a radio source --- a flare-related increase is seen; bottom panel is GOES X-Ray flux. The distinctive spike can be noticed in the correlation curve, while the peak brightness temperature plot shows a smooth fall-down (for about 1 hour) after the short spike at around 00:00 UT.}
    
    \label{flareexample}
    \end{figure*}

In such cases, we additionally employ GOES X-Ray light curves. Any flare-caused rise of the radio brightness temperature is also accompanied by a rise of the soft X-ray flux associated with this long duration flare event (see Fig. \ref{flareexample}). If there were no significant flares in the selected period of time, the soft X-Ray flux does not show an enhancement.

Having all radio images with 5 minute cadence available, we analyze evolution of the peak brightness temperature for all cases in the automatically obtained  list of GR-34 candidates. If the brightness temperature falls below 60,000\,K after a solar flare and raises up above this value only if another solar flare occurred, then the GR-34 candidate is removed from the list. This empirical visual analysis threshold of 60,000\,K was chosen to be consistent with the step (3d) of the general automated filtration scheme.
 
\subsection{Microwave source sizes}

Expecting very small (near or below the NoRH spatial resolution) radio source size for GR-34 emission, we consider  evolution of the sizes and shapes of the sources over time as well as their relationships with the measured values of the brightness temperature.
 
 We estimate the source area by counting pixels above the threshold 0.8 of the peak brightness temperature and  compare it to the area of the Nobeyama Radioheliograph beam computed using the same threshold (see Fig. \ref{spatialanalysis}). To avoid negatively impacting sampling effects, both observed images and beams were interpolated to higher resolution grid. Note, that selected threshold is larger than typically used half-maximum level. The rationale of selecting this higher threshold is the following:

\begin{figure*}[htbp] \centering
    \includegraphics[width=\linewidth]{"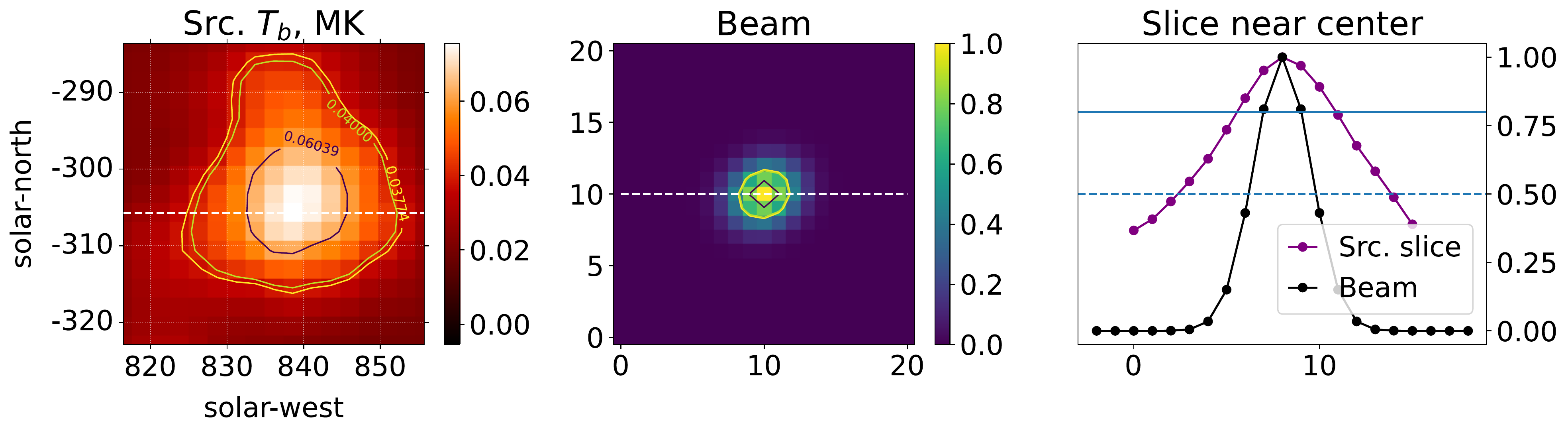"}
    \caption{An example of NOAA AR\,11515 (2012-Jul-07, 23:59 UT) radio source compared to the NoRH beam at 34\,GHz. The microwave image  is shown in the left panel. The NoRH beam computed for the corresponding observation time is presented in the center panel. Right panel shows brightness temperature profiles taken along the horizontal white dashed lines (in pixels) for the source and the beam. Brightness temperature values (vertical axis) in this comparison are normalized to unity to show the difference in the sizes between the microwave source and the beam. Dashed blue line stands for 0.5 of the maximum value, solid blue line stands for 0.8 of  the maximum value.}
    \label{spatialanalysis}
\end{figure*}

 \begin{itemize}
     \item CLEAN artifacts and remaining side lobes may exceed half-maximum level in some cases
     \item The main compact radio source is often observed on top of a large extended brightness temperature enhancement (sometimes also exceeding half-maximum value).
 \end{itemize}

\subsection{Are there double (or multiple) spatial sources?}

In some days of GR-34 candidates list there is another microwave source besides the selected one. Most of such sources are related to flaring radio emission and are located in other active regions. Nevertheless, these sources have an impact on the plots of the peak brightness during the flares, but they are not critical to the filtration process. We can easily filter out parasitic sources due to availability of other instruments (not only Nobeyama) and due to short lifespan of their flares.

\subsection{``Famous'' active regions}

Some ARs receive disproportionally more attention in the literature than others. This is the case of ARs associated with strongest flares, or covering a large area at the solar surface, or something else. Some of those ARs are in the Table (e.g., 10486, 10488, 12192, and 12673), while others are not. 

To make sure that our search algorithm has not overlooked such ARs, we manually analyzed several ``famous'' ARs. We selected the following ARs. AR 10652 is the one from which VLA measurements of the magnetic field have been performed by \citet{2006ApJ...641L..69B} that match well the curve of the peak magnetic field values vs height; see Figure\,\ref{f_all_coronal_B}. AR 10720 hosted one of the well-known powerful X-class flare on 2005-Jan-20. AR 10930 is known as a host of a series of strong flares in December 2006 including a record-breaking spike burst exceeding 1 million sfu level \citep{2021GMS...262..141G, 2019arXiv190109262G}. AR 11158 is one of the first ARs in solar cycle 24 observed with a new fleet of instruments including \SDO. Finally, AR 12209 is known by its unusually large area on the Sun, for which multi-frequency VLA observations are available.

\begin{figure}[htbp] \centering
    \includegraphics[width=\linewidth]{"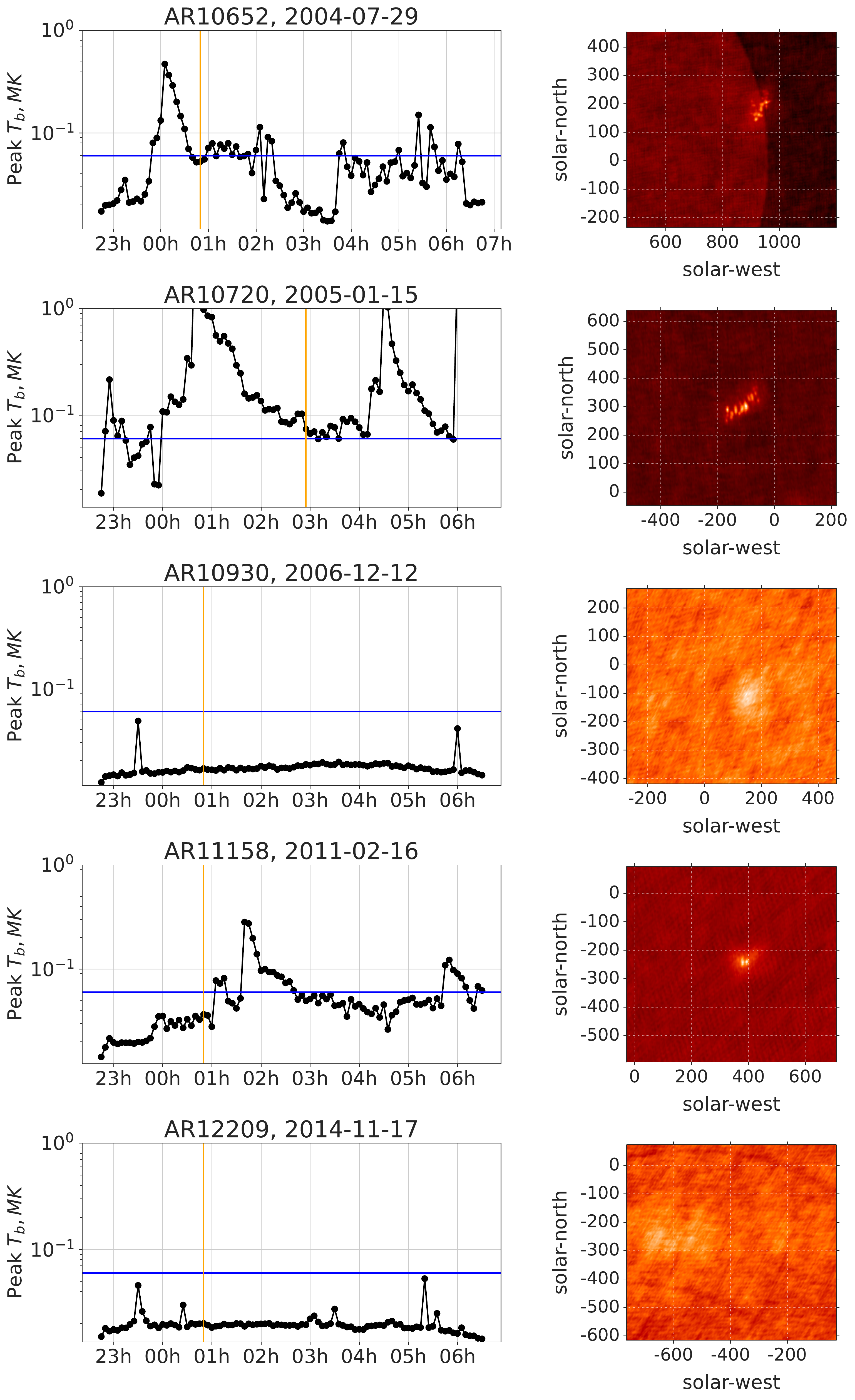"}
    \caption{Brightness temperatures of microwave sources of well-known active regions. Left panels represent evolution of the peak brightness temperature. Right panels are images of these sources selected at the times highlighted by a vertical orange line in corresponding left panels. Blue line shows 60,000\,K threshold used in the automated filtration algorithm.}
    \label{famousars}
    \end{figure}

Figure\,\ref{famousars} shows brightness temperature evolution and representative examples of the microwave images at  34\,GHz in these ARs. Although in some of those images we can see a compact bright source, it is typically below than the adopted minimum brightness threshold of 60,000\,K other than during flares seen as enhancements in the brightness temperature evolution. This implies that these ARs should not be added to the final list of GR-34 candidates.

\subsection{Results}

Table~\ref{results-table} presents the final  list of the selected GR-34 cases. It contains 57 days from 1995 up to 2017, which corresponds to 27 unique ARs.
It reports the date of the GR-34 emission detection, the AR number, and location rounded to the nearest integer.

Then, Table~\ref{results-table} reports daily minimum and daily median values of the per-image peak brightness temperatures of the images computed with 5-minute cadence. 
Daily minimal and median values of the peak brightness temperatures were calculated after excluding flare times according to Heliophysics Events Knowledgebase data. We do not show maximum brightness temperature for the whole day, because it  can still be affected by a flaring emission even though we took measures to exclude it to the extent possible. 
Also, we excluded from the analysis all individual images where the peak brightness temperature lies below 25,000\,K as potentially corrupted and incorrect.
Temperatures are rounded to 0.001 MK. Both 34\,GHz and 17\,GHz images for computing daily median and minimal brightness temperature statistics were taken in Stokes I polarization, because Stokes V is not available at 34 GHz. We checked that the degree of polarization in the brightness center of the 17 GHz source is typically low.

Table~\ref{results-table} contains information on the daily maximum photospheric magnetic field value. Magnetic fields are rounded by 1 terminating digit.  

Table~\ref{results-table} reports if the given case likely represents the optically thick or thin GR-34 emission and the association of the given AR with a sunquake, SEP event(s), and the strongest flare produced by the given AR. 

\section{Statistical analysis of the data set}
\label{sec:statistical_analysis}

Having the final list of the GR-34 microwave sources and their determined parameters we perform statistical study of the ARs in terms of the following characteristics:

\begin{enumerate}

\item Fraction of GR-34 ARs over all ARs;

\item Fraction of optically thick GR-34 ARs over all GR-34 ARs and over all ARs;

\item Distributions of the GR-34 ARs over visible solar disk;
\item Per-image peak brightness temperature of the microwave sources as well as its daily median and minimum values;
\item Lifetime of the GR-34 sources;
\item Relative area  of the microwave source from which the radiation comes  in comparison to instrumental beam area;
\item Association of the corresponding ARs with flaring activity, sunquakes, and solar energetic particle (SEP) events;
\item Association with the solar cycle phase.
\end{enumerate}

\subsection{General statistical data. Optically thick GR emission at 34\,GHz}

We found that less than 1\,\% of all ARs produce GR-34 emission ($27/4774 \times 100\% \approx 0.57\%$). Even though the total number of such ARs is modest, they play an important role as many of them are associated with powerful flares or other forms of enhanced solar activity (see below).

An important question is if the observed GR-34 source is optically thick or thin: the optically thick source is most likely produced at the third gyroharmonic, which has an immediate implication to estimate the magnetic field at the transition region. However, bright optically thin GR-source might be produced at the fourth or fifth one (see e.g. \citet{2019CosRe..57....1K}). Our estimates for the fourth harmonic using Eq. (20) from \citet{2014ApJ...781...77F} reveal that the optical depth $\tau \sim 1$ at 34 GHz is achieved, for instance, at the temperature of $T \sim 5-6$ MK, electron density $n_e \sim 10^{10} cm^{-3}$, and magnetic field inhomogeneity scale $L_B \sim 10^{9}$\,cm, which are plausible numbers in the low corona.

The brightness temperature of the optically thick source is equal to the thermal electron temperature in the gyro layer that is above 1\,MK in the corona. Most of our sources display brightness temperature below the coronal values. From our reference case we know that this could happen because the instrumental beam is larger than the size of the bright GR-34 source. In such a case, the bright flux of the compact  radio source is distributed over a large area, which reduces the measured brightness temperature proportionally. From this consideration we conclude (see Section\,\ref{sec:AR12673}) that the optically thick cases must either show the brightness temperature above 1\,MK or be compact (barely resolved), or both.  

\begin{figure} \centering 
    \includegraphics[width=\linewidth]{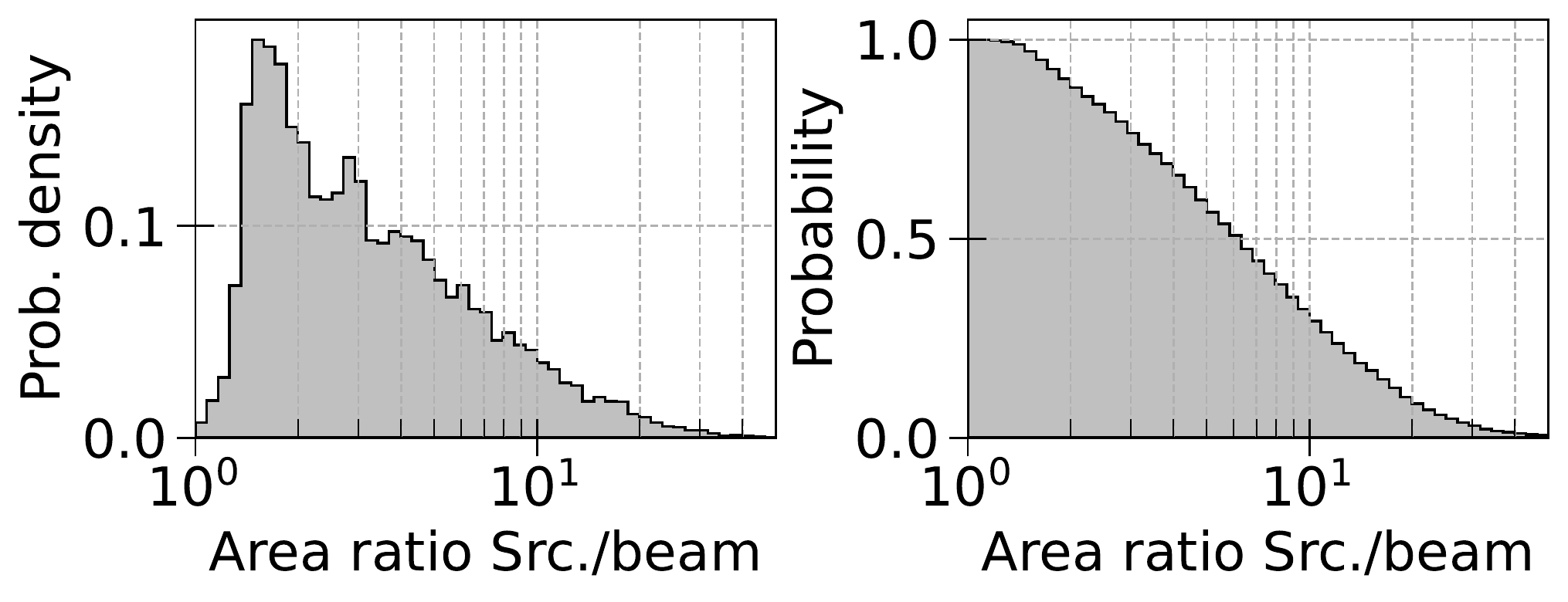}
    \caption{Distribution of the source-to-beam area ratios across all images, both with and without flares,  for all days listed in Table \ref{results-table}.
    Left panel shows the histogram of the source area divided by the beam area computed at the level of 0.8. 
    The histogram is normalized to the probability density. A cumulative histogram corresponding to the same quantity and normalised to probability distribution is shown in the right panel.}
    \label{size_distribution}
\end{figure}

To reveal optically thick GR-34 cases among the selected events, we perform statistical analysis of the source area along with the beam size and measured brightness temperature.
  For most ($\sim 75$\,\%) of the images the ratio of radio source area to the beam area exceeds 3.
 Compact sources having area below 3 beam areas may be associated with either flaring or optically thick GR-34 emission.
The histogram of the distribution of the source-to-beam area ratios for all 57 GR-34 candidates is given in Figure~\ref{size_distribution}.

\begin{figure*}[htbp] \centering
    \includegraphics[width=\linewidth]{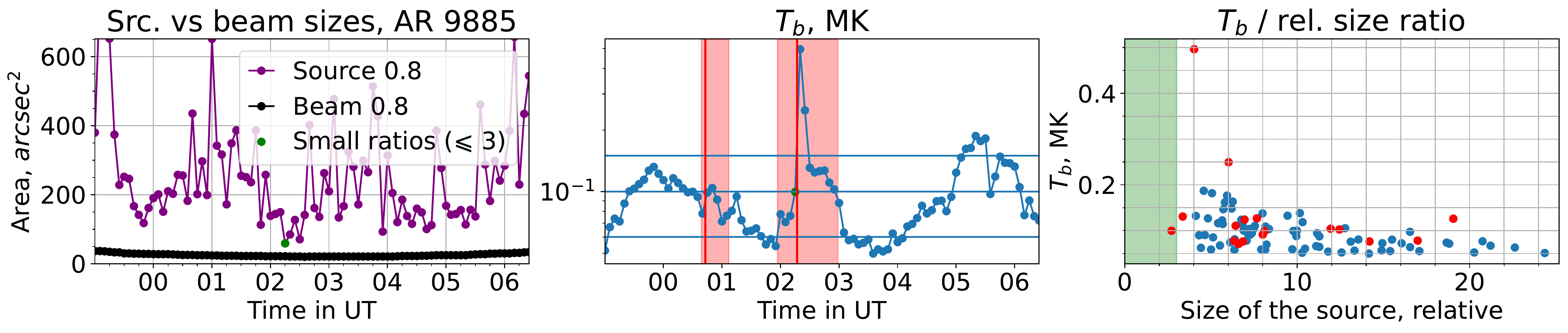}
    \includegraphics[width=\linewidth]{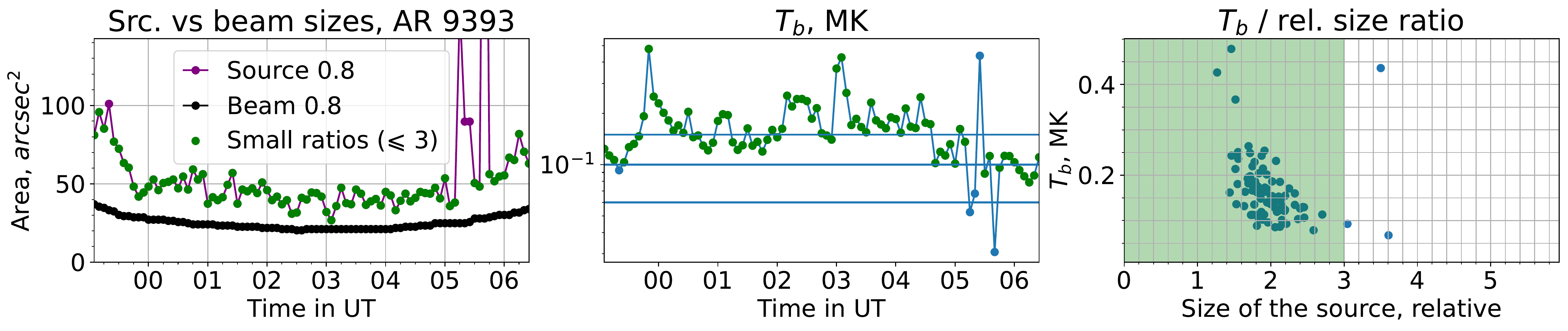}
    \includegraphics[width=\linewidth]{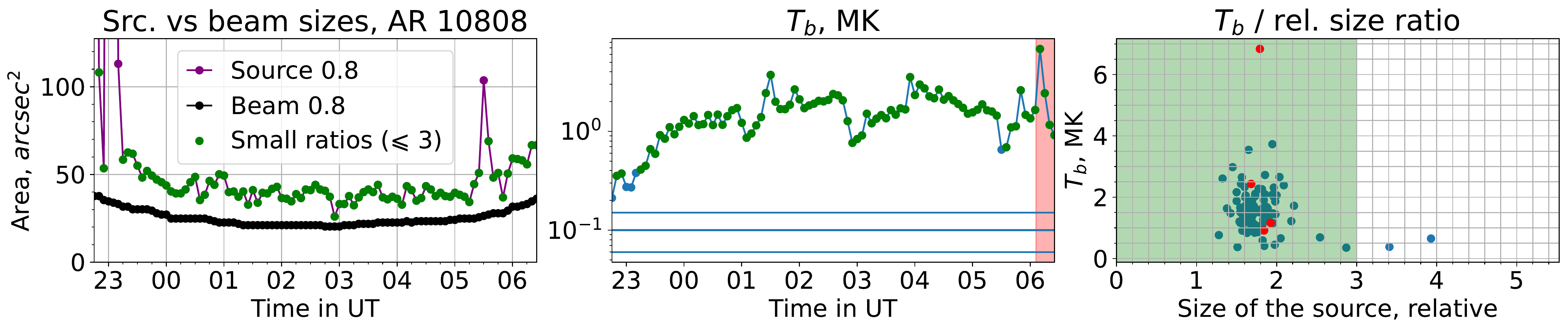}
    \includegraphics[width=\linewidth]{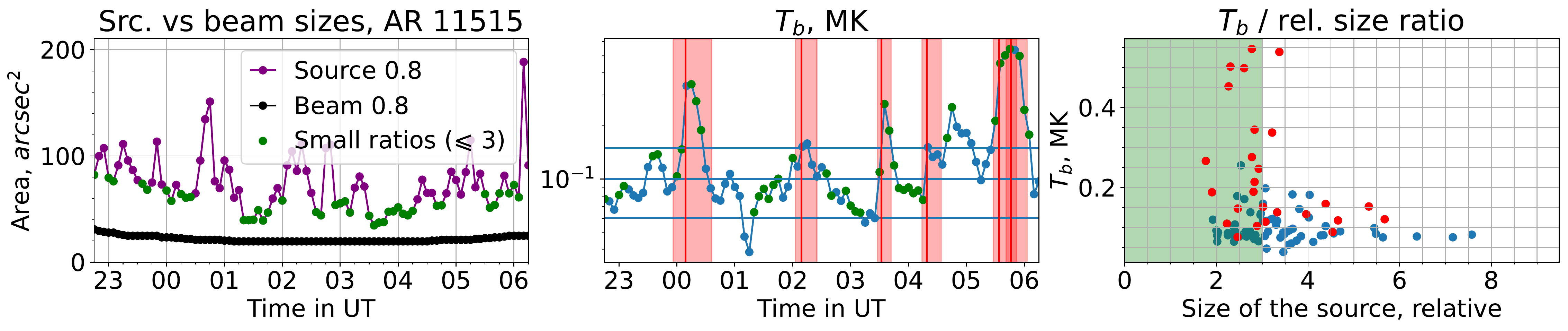}
    \includegraphics[width=\linewidth]{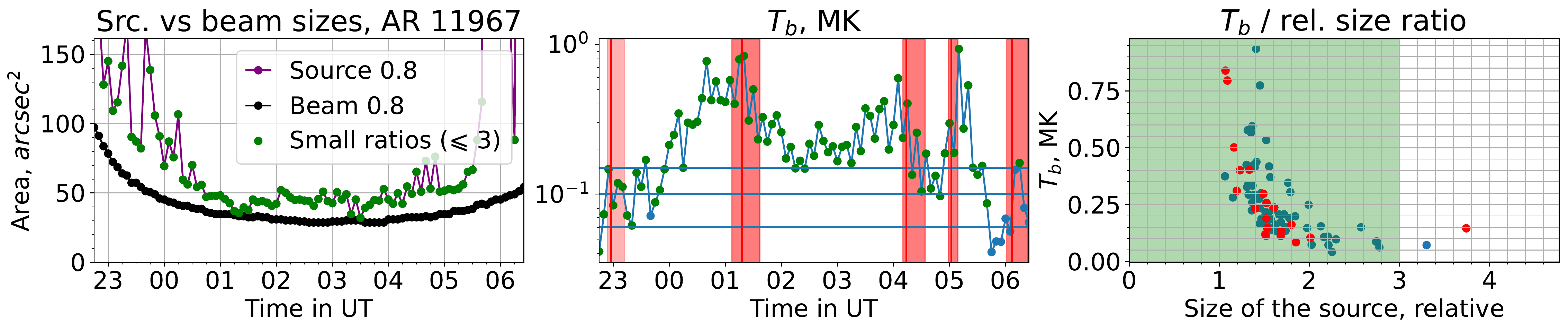}
    \includegraphics[width=\linewidth]{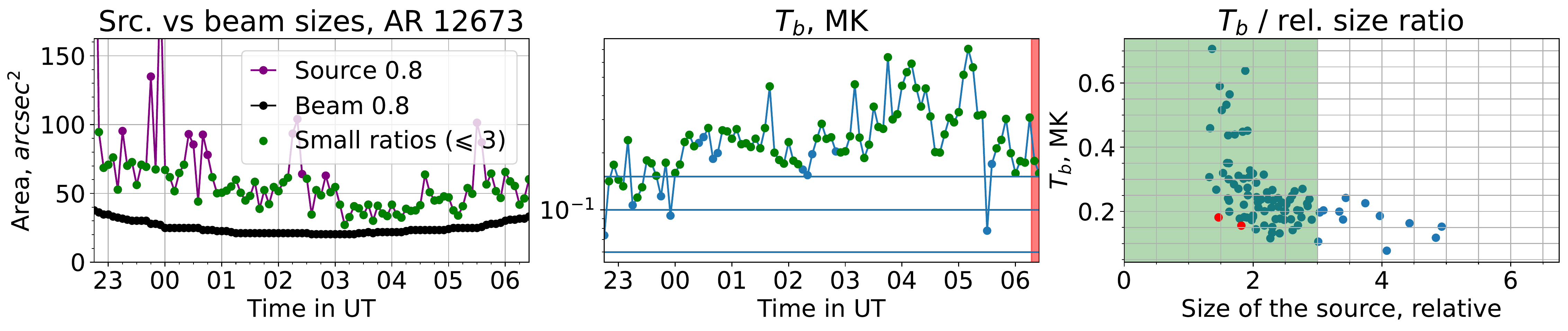}
    \caption{NoRH radio source area, NoRH beam size area, and source peak brightness temperature relationship at 34 GHz. The beam area and the radio source area above 0.8 of the maximum values are used. Green colors indicate source areas less than 3 beam areas. Red colors feature solar flare episodes reported by HEK. Blue horizontal lines correspond to 0.06 MK, 0.1 MK, and 0.15 MK respectively. The top row (AR 9885) shows an example of a large, likely an optically thin source. Other cases (NOAA ARs 9393, 10808, 11515, 11967 and 12673) correspond to optically thick GR sources. See all 57 active regions in the online figure set.}
    \label{spatialanalysis-timelines}
\end{figure*}

A typical example of the evolution of the source and beam  areas, brightness temperature, and their relationships is given in top panels of Fig. \ref{spatialanalysis-timelines}: the source size varies broadly, while the brightness temperature does not correlate with the source size. Only a minor fraction of images displays a truly compact source.

\figsetstart
\figsetnum{11}
\figsettitle{Microwave sources, their $T_b$ and size-to-beam ratios}
\figsetgrpstart
\figsetgrpnum{11.1}
\figsetgrptitle{AR8644}
\figsetplot{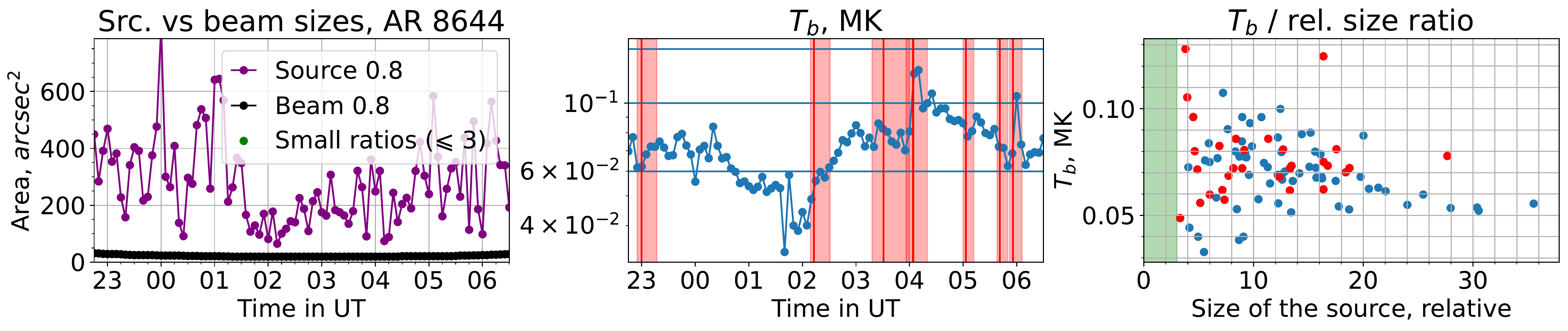}
\figsetgrpnote{NoRH radio source area, NoRH beam size area, and source peak brightness temperature relationship at 34 GHz. The beam area and the radio source area above 0.8 of the maximum values are used. Green colors indicate source areas less than 3 beam areas. Red colors feature solar flare episodes reported by HEK. Blue horizontal lines correspond to 0.06 MK, 0.1 MK, and 0.15 MK respectively.}
\figsetgrpend

\figsetgrpstart
\figsetgrpnum{11.2}
\figsetgrptitle{AR9017}
\figsetplot{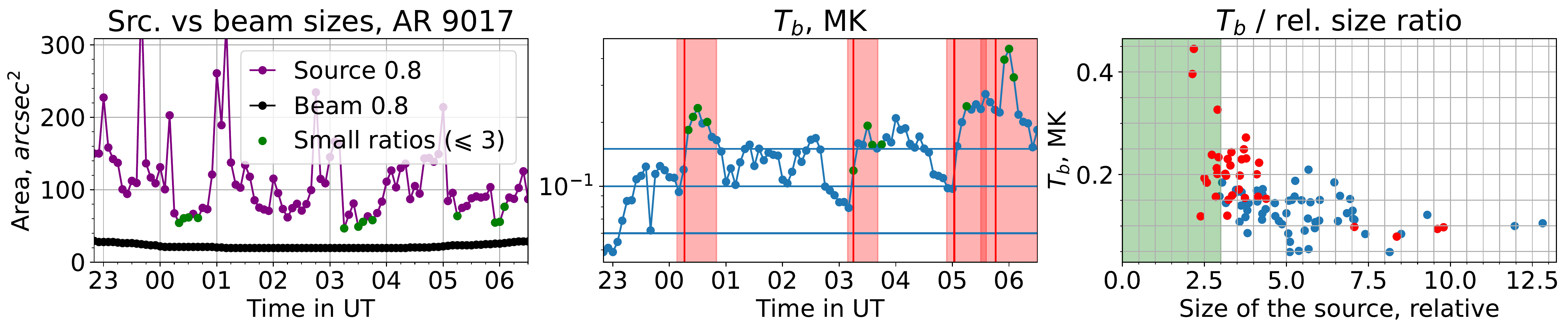}
\figsetgrpnote{NoRH radio source area, NoRH beam size area, and source peak brightness temperature relationship at 34 GHz. The beam area and the radio source area above 0.8 of the maximum values are used. Green colors indicate source areas less than 3 beam areas. Red colors feature solar flare episodes reported by HEK. Blue horizontal lines correspond to 0.06 MK, 0.1 MK, and 0.15 MK respectively.}
\figsetgrpend

\figsetgrpstart
\figsetgrpnum{11.3}
\figsetgrptitle{AR9077}
\figsetplot{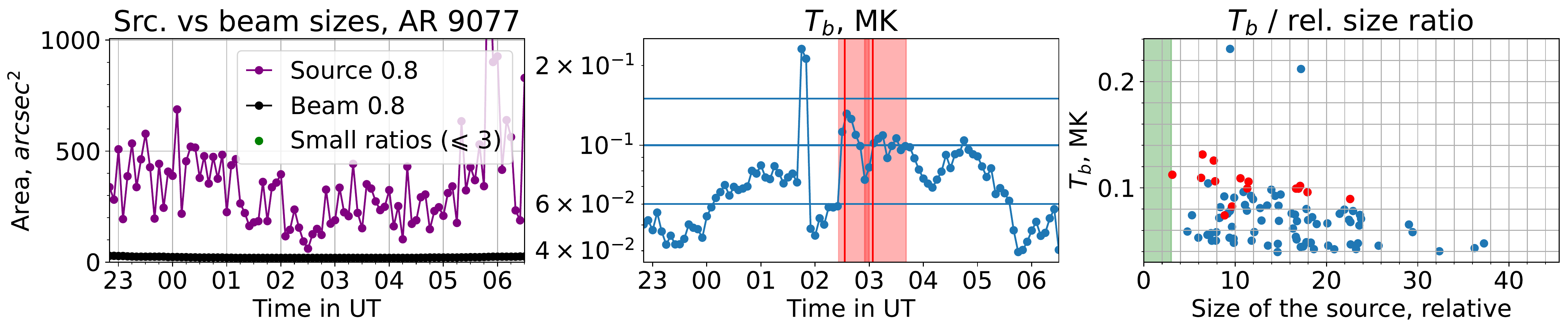}
\figsetgrpnote{NoRH radio source area, NoRH beam size area, and source peak brightness temperature relationship at 34 GHz. The beam area and the radio source area above 0.8 of the maximum values are used. Green colors indicate source areas less than 3 beam areas. Red colors feature solar flare episodes reported by HEK. Blue horizontal lines correspond to 0.06 MK, 0.1 MK, and 0.15 MK respectively.}
\figsetgrpend

\figsetgrpstart
\figsetgrpnum{11.4}
\figsetgrptitle{AR9077}
\figsetplot{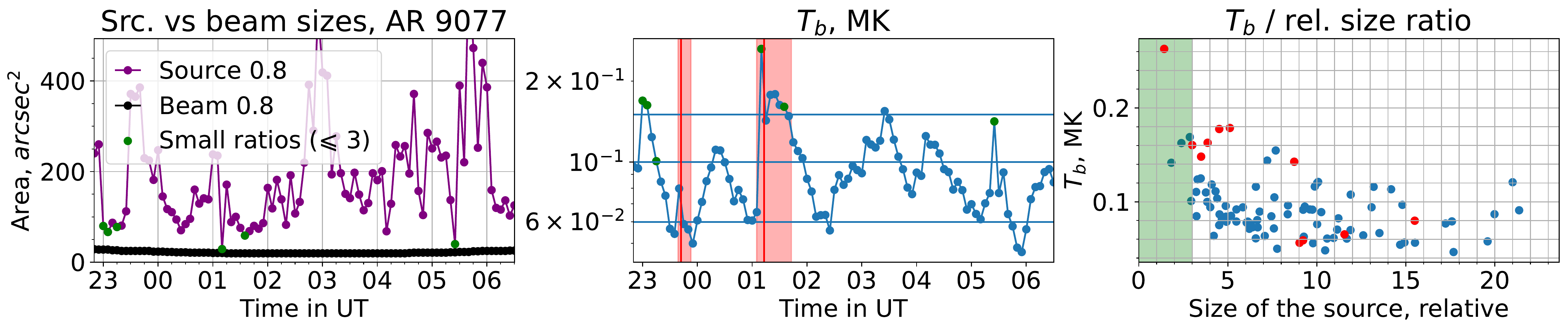}
\figsetgrpnote{NoRH radio source area, NoRH beam size area, and source peak brightness temperature relationship at 34 GHz. The beam area and the radio source area above 0.8 of the maximum values are used. Green colors indicate source areas less than 3 beam areas. Red colors feature solar flare episodes reported by HEK. Blue horizontal lines correspond to 0.06 MK, 0.1 MK, and 0.15 MK respectively.}
\figsetgrpend

\figsetgrpstart
\figsetgrpnum{11.5}
\figsetgrptitle{AR9066}
\figsetplot{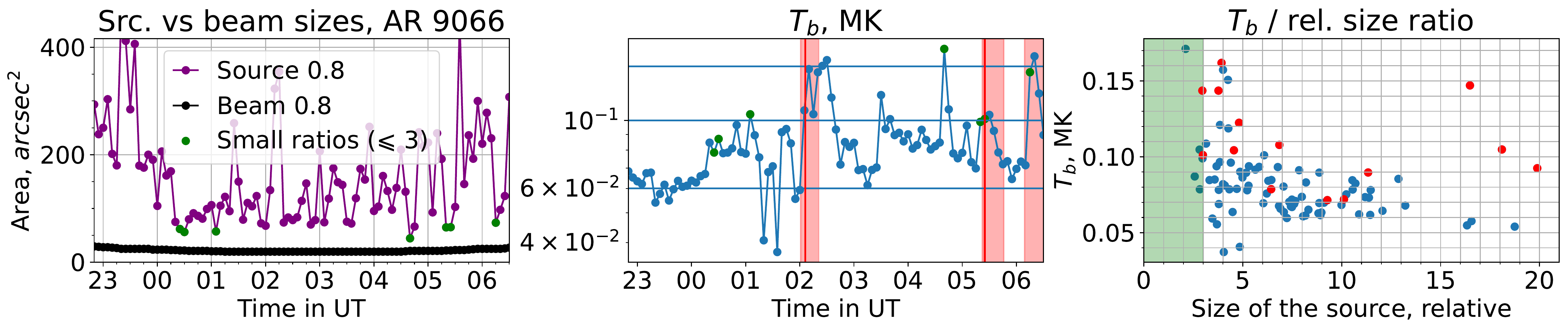}
\figsetgrpnote{NoRH radio source area, NoRH beam size area, and source peak brightness temperature relationship at 34 GHz. The beam area and the radio source area above 0.8 of the maximum values are used. Green colors indicate source areas less than 3 beam areas. Red colors feature solar flare episodes reported by HEK. Blue horizontal lines correspond to 0.06 MK, 0.1 MK, and 0.15 MK respectively.}
\figsetgrpend

\figsetgrpstart
\figsetgrpnum{11.6}
\figsetgrptitle{AR9393}
\figsetplot{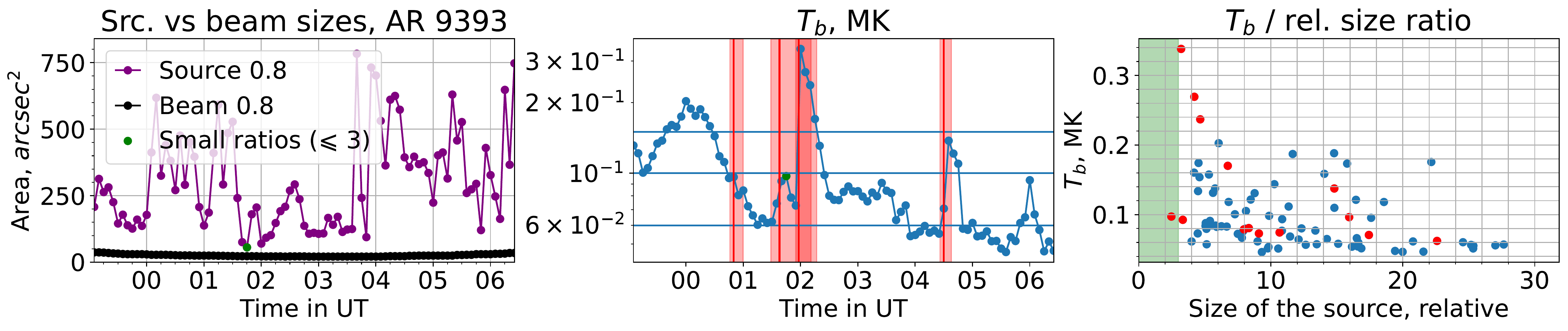}
\figsetgrpnote{NoRH radio source area, NoRH beam size area, and source peak brightness temperature relationship at 34 GHz. The beam area and the radio source area above 0.8 of the maximum values are used. Green colors indicate source areas less than 3 beam areas. Red colors feature solar flare episodes reported by HEK. Blue horizontal lines correspond to 0.06 MK, 0.1 MK, and 0.15 MK respectively.}
\figsetgrpend

\figsetgrpstart
\figsetgrpnum{11.7}
\figsetgrptitle{AR9393}
\figsetplot{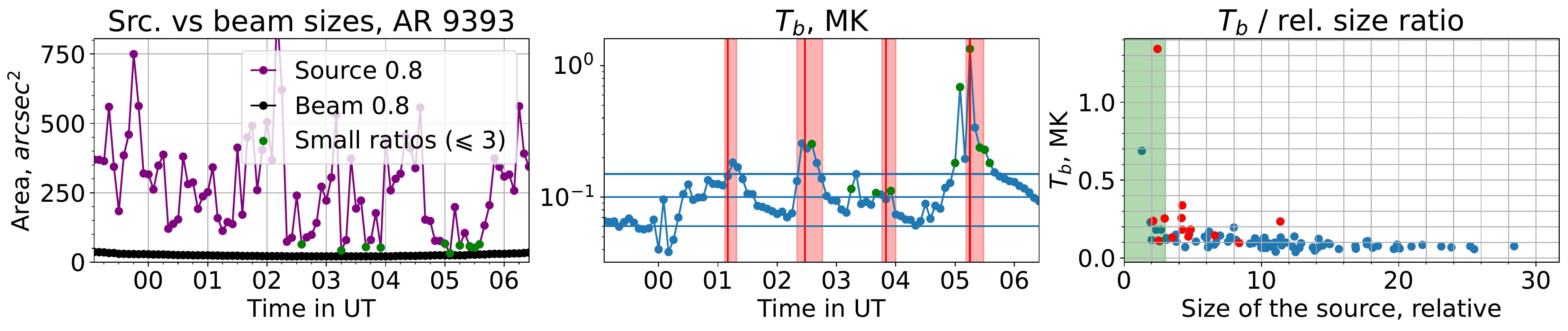}
\figsetgrpnote{NoRH radio source area, NoRH beam size area, and source peak brightness temperature relationship at 34 GHz. The beam area and the radio source area above 0.8 of the maximum values are used. Green colors indicate source areas less than 3 beam areas. Red colors feature solar flare episodes reported by HEK. Blue horizontal lines correspond to 0.06 MK, 0.1 MK, and 0.15 MK respectively.}
\figsetgrpend

\figsetgrpstart
\figsetgrpnum{11.8}
\figsetgrptitle{AR9393}
\figsetplot{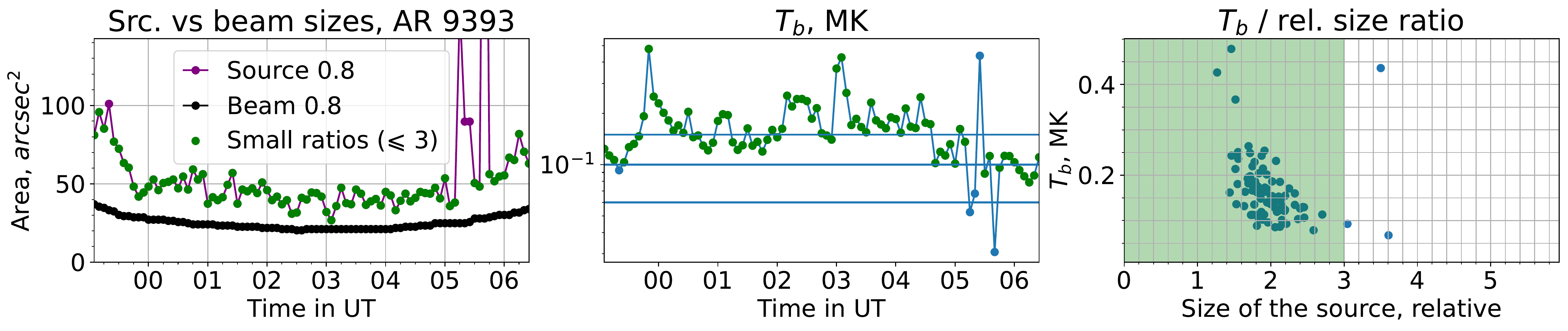}
\figsetgrpnote{NoRH radio source area, NoRH beam size area, and source peak brightness temperature relationship at 34 GHz. The beam area and the radio source area above 0.8 of the maximum values are used. Green colors indicate source areas less than 3 beam areas. Red colors feature solar flare episodes reported by HEK. Blue horizontal lines correspond to 0.06 MK, 0.1 MK, and 0.15 MK respectively.}
\figsetgrpend

\figsetgrpstart
\figsetgrpnum{11.9}
\figsetgrptitle{AR9393}
\figsetplot{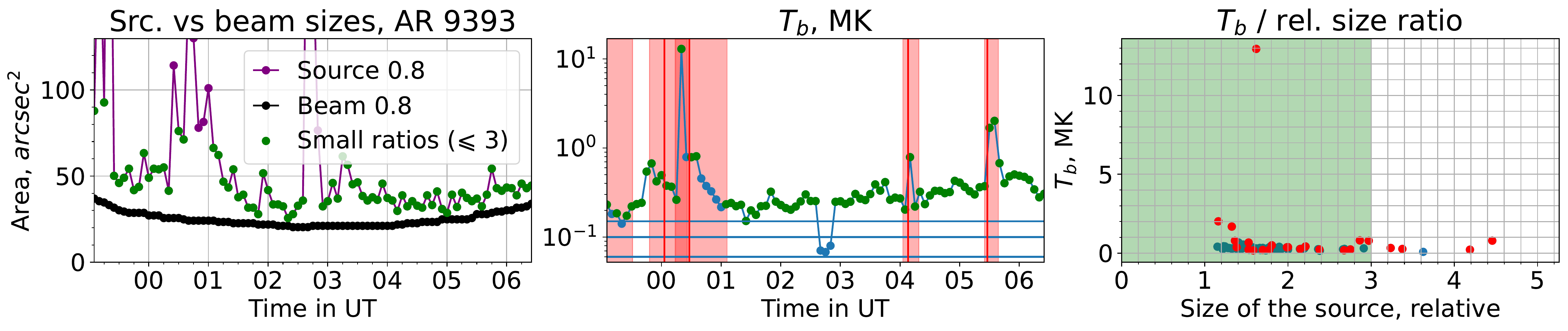}
\figsetgrpnote{NoRH radio source area, NoRH beam size area, and source peak brightness temperature relationship at 34 GHz. The beam area and the radio source area above 0.8 of the maximum values are used. Green colors indicate source areas less than 3 beam areas. Red colors feature solar flare episodes reported by HEK. Blue horizontal lines correspond to 0.06 MK, 0.1 MK, and 0.15 MK respectively.}
\figsetgrpend

\figsetgrpstart
\figsetgrpnum{11.10}
\figsetgrptitle{AR9393}
\figsetplot{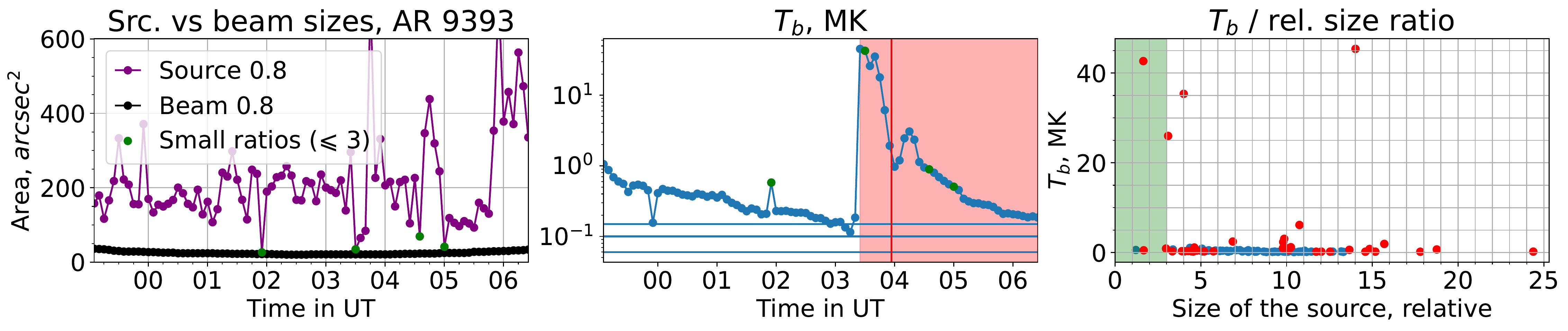}
\figsetgrpnote{NoRH radio source area, NoRH beam size area, and source peak brightness temperature relationship at 34 GHz. The beam area and the radio source area above 0.8 of the maximum values are used. Green colors indicate source areas less than 3 beam areas. Red colors feature solar flare episodes reported by HEK. Blue horizontal lines correspond to 0.06 MK, 0.1 MK, and 0.15 MK respectively.}
\figsetgrpend

\figsetgrpstart
\figsetgrpnum{11.11}
\figsetgrptitle{AR9393}
\figsetplot{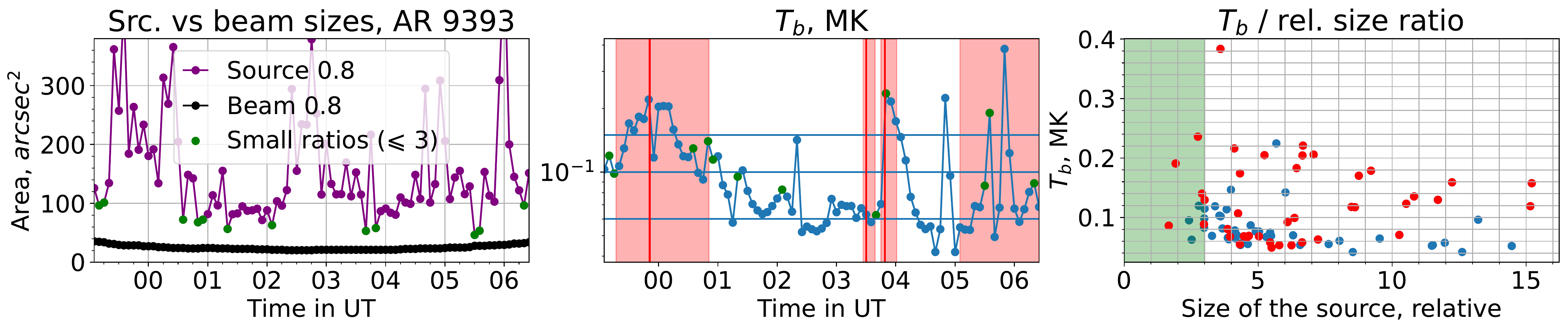}
\figsetgrpnote{NoRH radio source area, NoRH beam size area, and source peak brightness temperature relationship at 34 GHz. The beam area and the radio source area above 0.8 of the maximum values are used. Green colors indicate source areas less than 3 beam areas. Red colors feature solar flare episodes reported by HEK. Blue horizontal lines correspond to 0.06 MK, 0.1 MK, and 0.15 MK respectively.}
\figsetgrpend

\figsetgrpstart
\figsetgrpnum{11.12}
\figsetgrptitle{AR9591}
\figsetplot{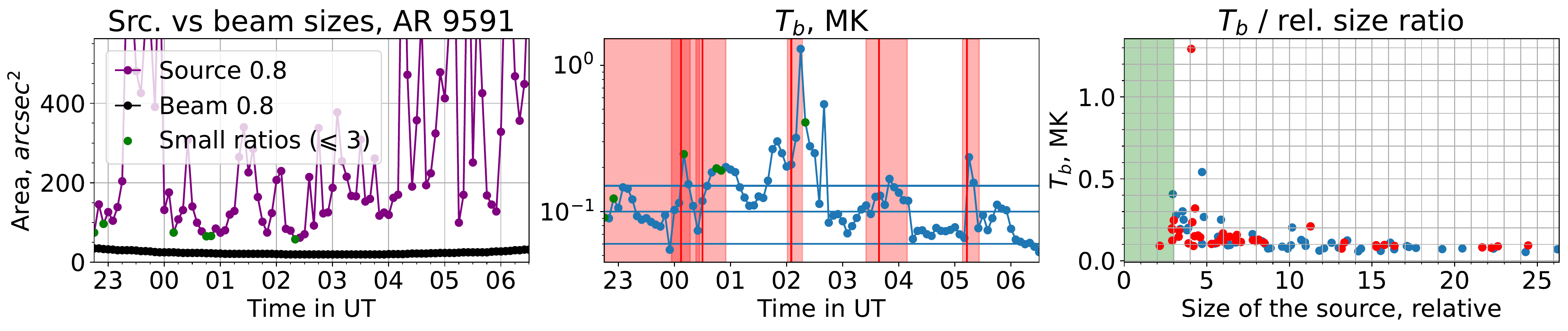}
\figsetgrpnote{NoRH radio source area, NoRH beam size area, and source peak brightness temperature relationship at 34 GHz. The beam area and the radio source area above 0.8 of the maximum values are used. Green colors indicate source areas less than 3 beam areas. Red colors feature solar flare episodes reported by HEK. Blue horizontal lines correspond to 0.06 MK, 0.1 MK, and 0.15 MK respectively.}
\figsetgrpend

\figsetgrpstart
\figsetgrpnum{11.13}
\figsetgrptitle{AR9608}
\figsetplot{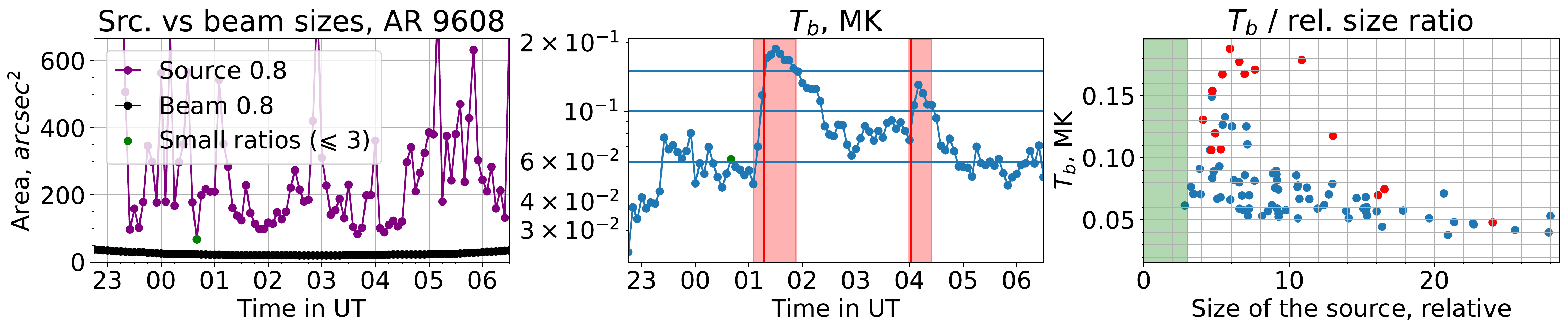}
\figsetgrpnote{NoRH radio source area, NoRH beam size area, and source peak brightness temperature relationship at 34 GHz. The beam area and the radio source area above 0.8 of the maximum values are used. Green colors indicate source areas less than 3 beam areas. Red colors feature solar flare episodes reported by HEK. Blue horizontal lines correspond to 0.06 MK, 0.1 MK, and 0.15 MK respectively.}
\figsetgrpend

\figsetgrpstart
\figsetgrpnum{11.14}
\figsetgrptitle{AR9690}
\figsetplot{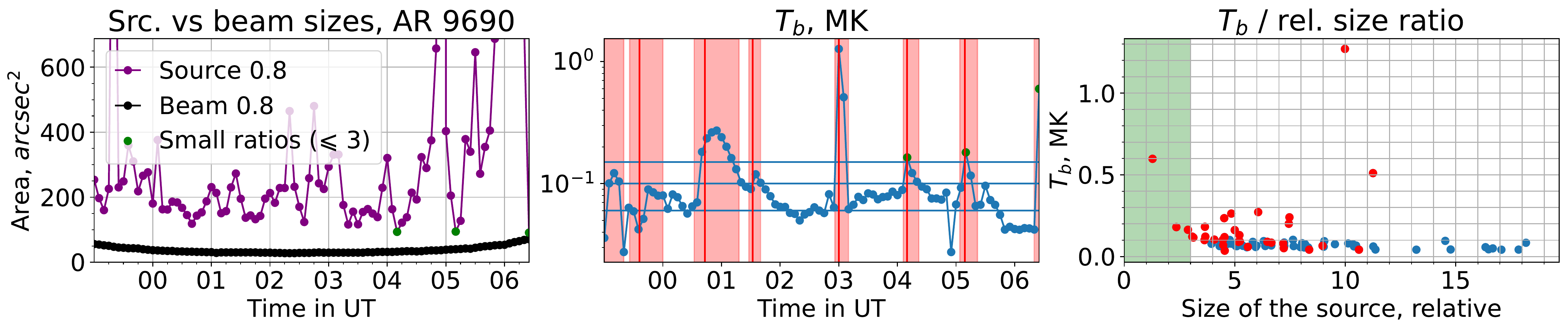}
\figsetgrpnote{NoRH radio source area, NoRH beam size area, and source peak brightness temperature relationship at 34 GHz. The beam area and the radio source area above 0.8 of the maximum values are used. Green colors indicate source areas less than 3 beam areas. Red colors feature solar flare episodes reported by HEK. Blue horizontal lines correspond to 0.06 MK, 0.1 MK, and 0.15 MK respectively.}
\figsetgrpend

\figsetgrpstart
\figsetgrpnum{11.15}
\figsetgrptitle{AR9885}
\figsetplot{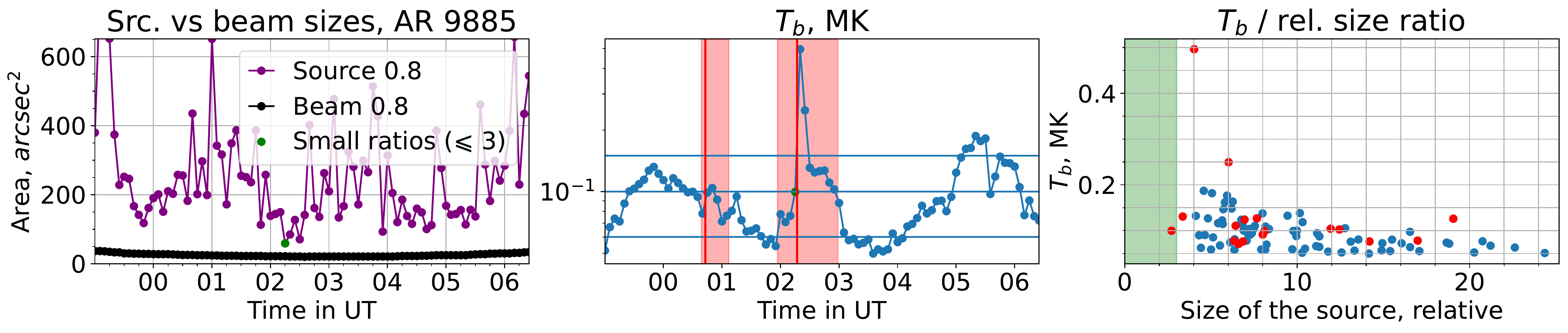}
\figsetgrpnote{NoRH radio source area, NoRH beam size area, and source peak brightness temperature relationship at 34 GHz. The beam area and the radio source area above 0.8 of the maximum values are used. Green colors indicate source areas less than 3 beam areas. Red colors feature solar flare episodes reported by HEK. Blue horizontal lines correspond to 0.06 MK, 0.1 MK, and 0.15 MK respectively.}
\figsetgrpend

\figsetgrpstart
\figsetgrpnum{11.16}
\figsetgrptitle{AR9893}
\figsetplot{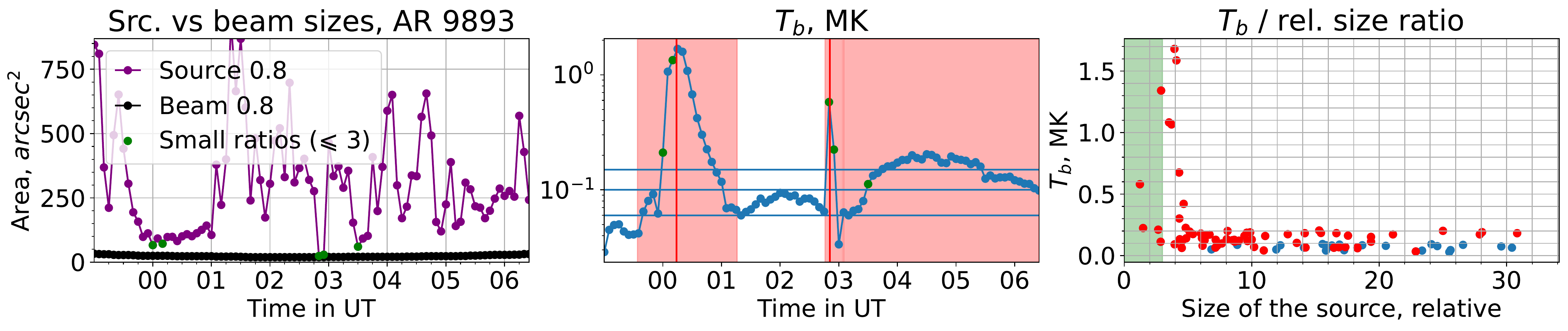}
\figsetgrpnote{NoRH radio source area, NoRH beam size area, and source peak brightness temperature relationship at 34 GHz. The beam area and the radio source area above 0.8 of the maximum values are used. Green colors indicate source areas less than 3 beam areas. Red colors feature solar flare episodes reported by HEK. Blue horizontal lines correspond to 0.06 MK, 0.1 MK, and 0.15 MK respectively.}
\figsetgrpend

\figsetgrpstart
\figsetgrpnum{11.17}
\figsetgrptitle{AR10095}
\figsetplot{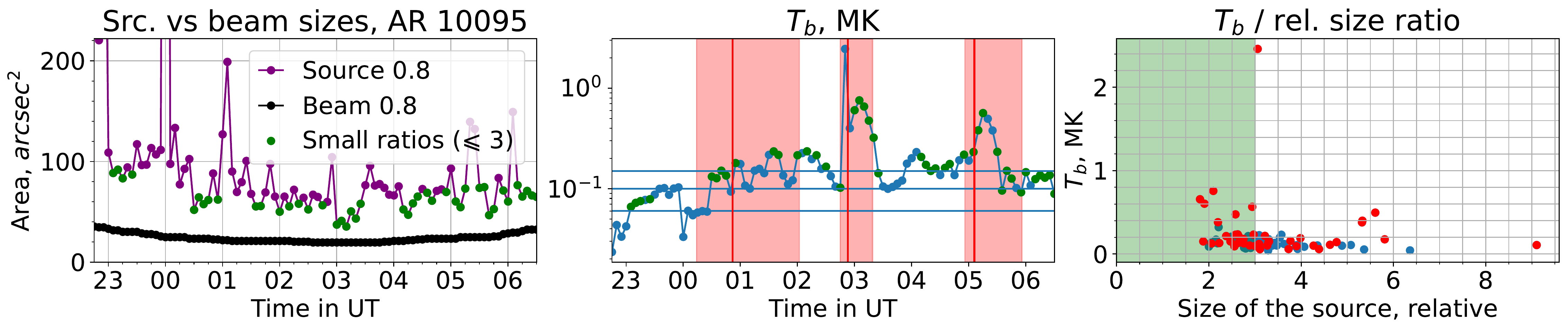}
\figsetgrpnote{NoRH radio source area, NoRH beam size area, and source peak brightness temperature relationship at 34 GHz. The beam area and the radio source area above 0.8 of the maximum values are used. Green colors indicate source areas less than 3 beam areas. Red colors feature solar flare episodes reported by HEK. Blue horizontal lines correspond to 0.06 MK, 0.1 MK, and 0.15 MK respectively.}
\figsetgrpend

\figsetgrpstart
\figsetgrpnum{11.18}
\figsetgrptitle{AR10095}
\figsetplot{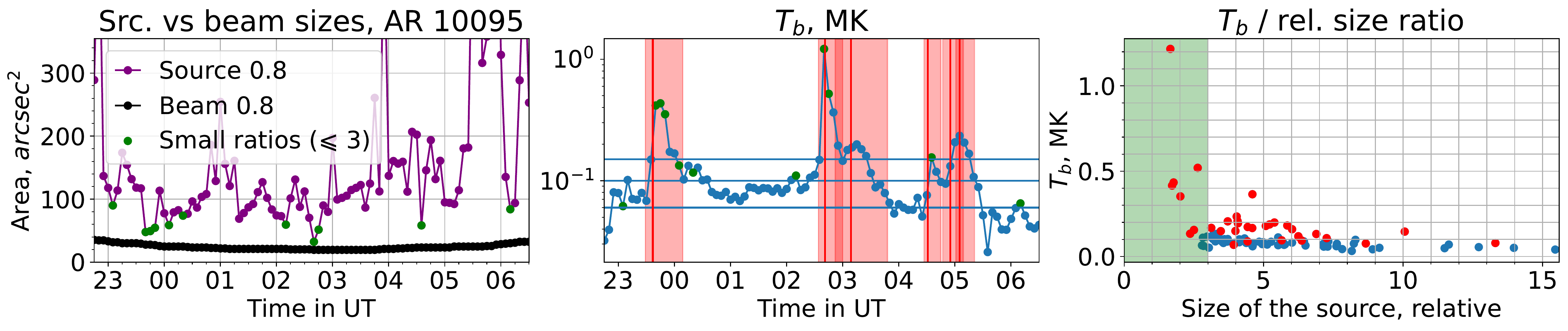}
\figsetgrpnote{NoRH radio source area, NoRH beam size area, and source peak brightness temperature relationship at 34 GHz. The beam area and the radio source area above 0.8 of the maximum values are used. Green colors indicate source areas less than 3 beam areas. Red colors feature solar flare episodes reported by HEK. Blue horizontal lines correspond to 0.06 MK, 0.1 MK, and 0.15 MK respectively.}
\figsetgrpend

\figsetgrpstart
\figsetgrpnum{11.19}
\figsetgrptitle{AR10095}
\figsetplot{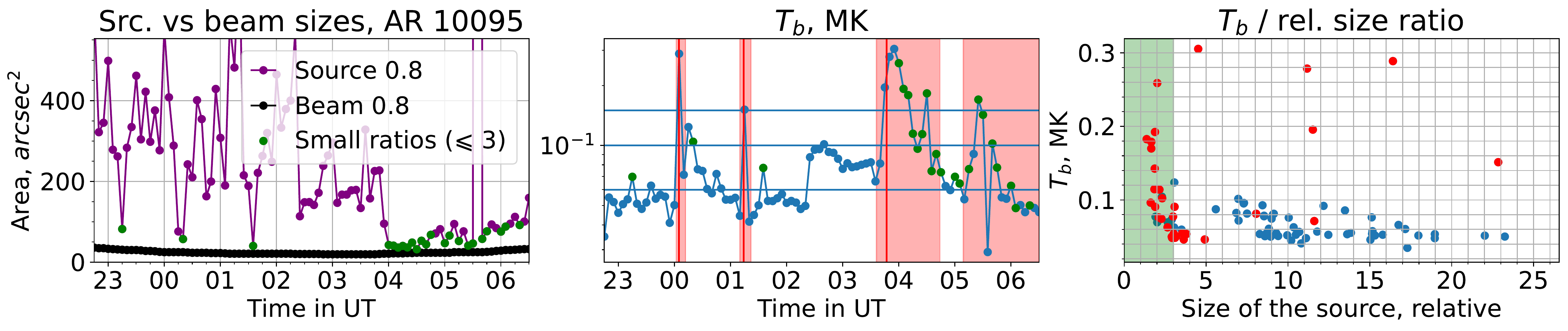}
\figsetgrpnote{NoRH radio source area, NoRH beam size area, and source peak brightness temperature relationship at 34 GHz. The beam area and the radio source area above 0.8 of the maximum values are used. Green colors indicate source areas less than 3 beam areas. Red colors feature solar flare episodes reported by HEK. Blue horizontal lines correspond to 0.06 MK, 0.1 MK, and 0.15 MK respectively.}
\figsetgrpend

\figsetgrpstart
\figsetgrpnum{11.20}
\figsetgrptitle{AR10375}
\figsetplot{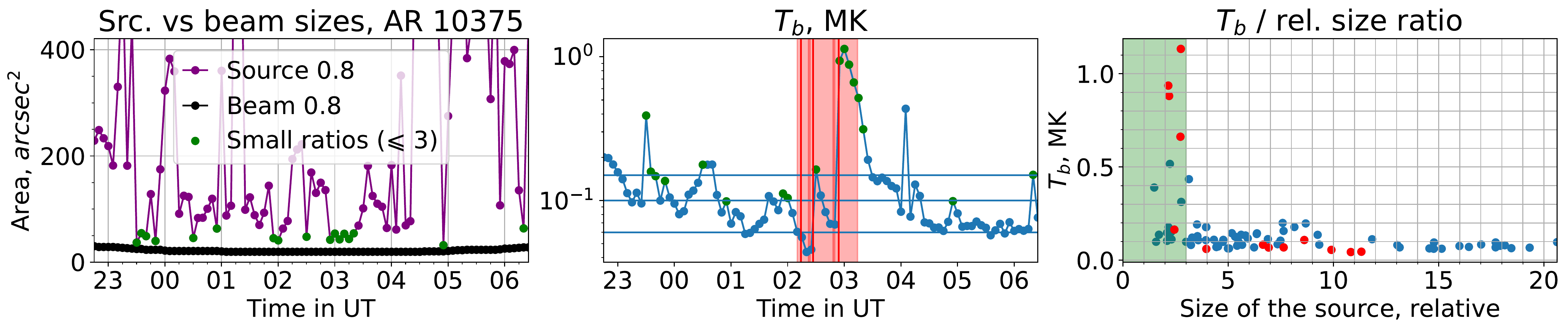}
\figsetgrpnote{NoRH radio source area, NoRH beam size area, and source peak brightness temperature relationship at 34 GHz. The beam area and the radio source area above 0.8 of the maximum values are used. Green colors indicate source areas less than 3 beam areas. Red colors feature solar flare episodes reported by HEK. Blue horizontal lines correspond to 0.06 MK, 0.1 MK, and 0.15 MK respectively.}
\figsetgrpend

\figsetgrpstart
\figsetgrpnum{11.21}
\figsetgrptitle{AR10375}
\figsetplot{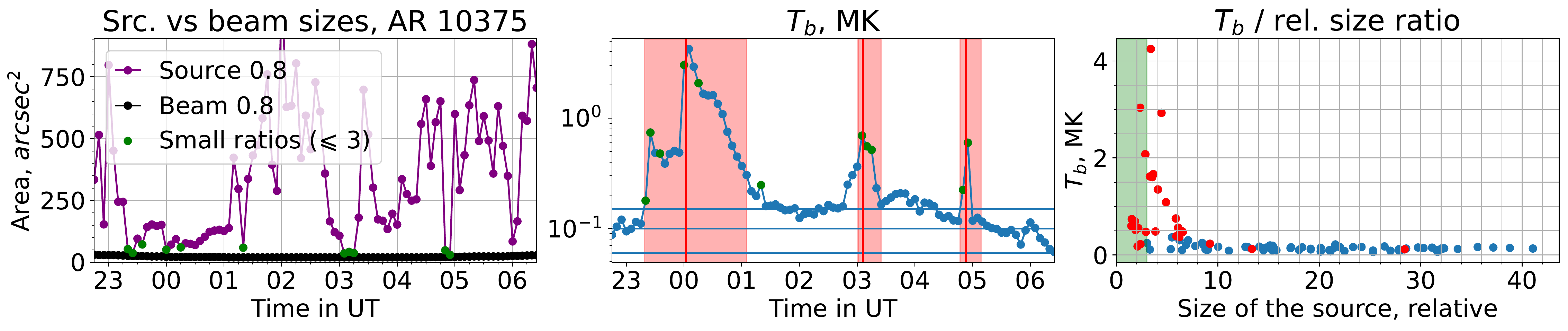}
\figsetgrpnote{NoRH radio source area, NoRH beam size area, and source peak brightness temperature relationship at 34 GHz. The beam area and the radio source area above 0.8 of the maximum values are used. Green colors indicate source areas less than 3 beam areas. Red colors feature solar flare episodes reported by HEK. Blue horizontal lines correspond to 0.06 MK, 0.1 MK, and 0.15 MK respectively.}
\figsetgrpend

\figsetgrpstart
\figsetgrpnum{11.22}
\figsetgrptitle{AR10375}
\figsetplot{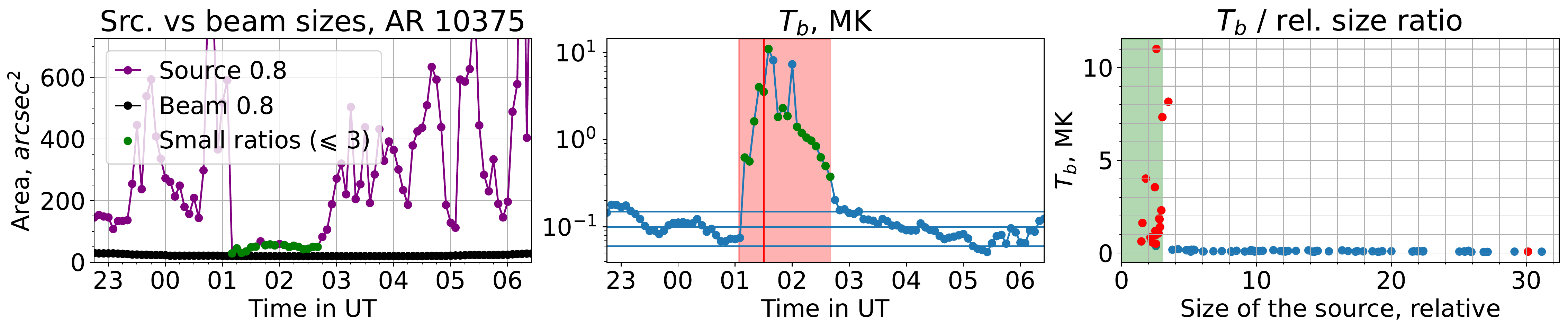}
\figsetgrpnote{NoRH radio source area, NoRH beam size area, and source peak brightness temperature relationship at 34 GHz. The beam area and the radio source area above 0.8 of the maximum values are used. Green colors indicate source areas less than 3 beam areas. Red colors feature solar flare episodes reported by HEK. Blue horizontal lines correspond to 0.06 MK, 0.1 MK, and 0.15 MK respectively.}
\figsetgrpend

\figsetgrpstart
\figsetgrpnum{11.23}
\figsetgrptitle{AR10375}
\figsetplot{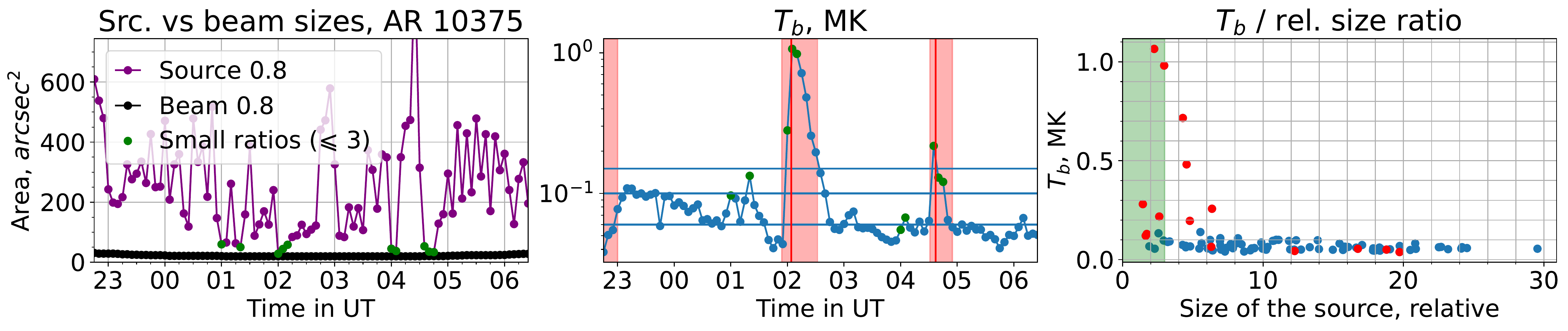}
\figsetgrpnote{NoRH radio source area, NoRH beam size area, and source peak brightness temperature relationship at 34 GHz. The beam area and the radio source area above 0.8 of the maximum values are used. Green colors indicate source areas less than 3 beam areas. Red colors feature solar flare episodes reported by HEK. Blue horizontal lines correspond to 0.06 MK, 0.1 MK, and 0.15 MK respectively.}
\figsetgrpend

\figsetgrpstart
\figsetgrpnum{11.24}
\figsetgrptitle{AR10486}
\figsetplot{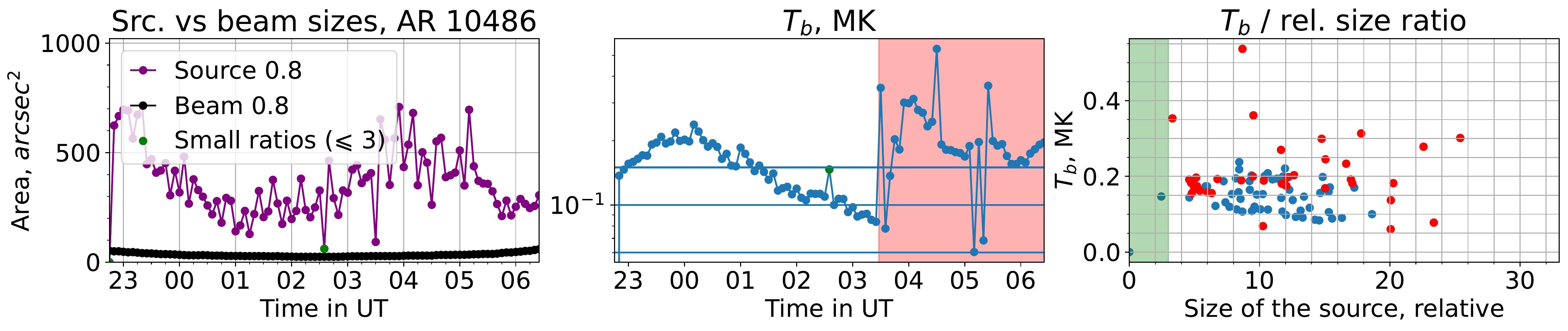}
\figsetgrpnote{NoRH radio source area, NoRH beam size area, and source peak brightness temperature relationship at 34 GHz. The beam area and the radio source area above 0.8 of the maximum values are used. Green colors indicate source areas less than 3 beam areas. Red colors feature solar flare episodes reported by HEK. Blue horizontal lines correspond to 0.06 MK, 0.1 MK, and 0.15 MK respectively.}
\figsetgrpend

\figsetgrpstart
\figsetgrpnum{11.25}
\figsetgrptitle{AR10486}
\figsetplot{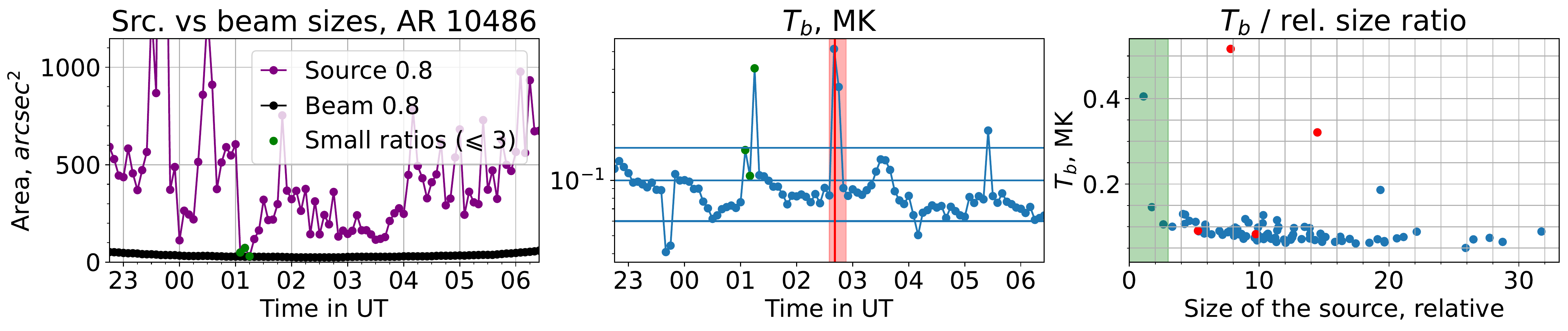}
\figsetgrpnote{NoRH radio source area, NoRH beam size area, and source peak brightness temperature relationship at 34 GHz. The beam area and the radio source area above 0.8 of the maximum values are used. Green colors indicate source areas less than 3 beam areas. Red colors feature solar flare episodes reported by HEK. Blue horizontal lines correspond to 0.06 MK, 0.1 MK, and 0.15 MK respectively.}
\figsetgrpend

\figsetgrpstart
\figsetgrpnum{11.26}
\figsetgrptitle{AR10486}
\figsetplot{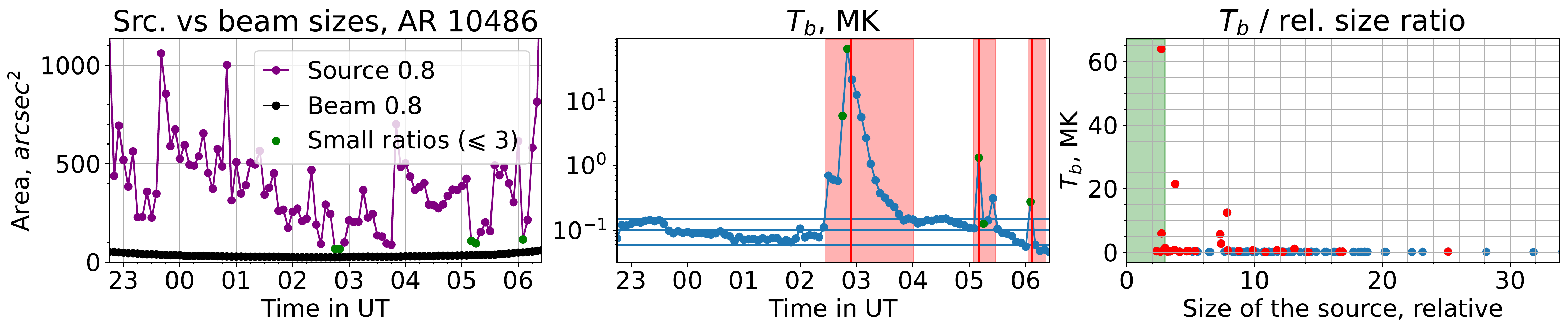}
\figsetgrpnote{NoRH radio source area, NoRH beam size area, and source peak brightness temperature relationship at 34 GHz. The beam area and the radio source area above 0.8 of the maximum values are used. Green colors indicate source areas less than 3 beam areas. Red colors feature solar flare episodes reported by HEK. Blue horizontal lines correspond to 0.06 MK, 0.1 MK, and 0.15 MK respectively.}
\figsetgrpend

\figsetgrpstart
\figsetgrpnum{11.27}
\figsetgrptitle{AR10488}
\figsetplot{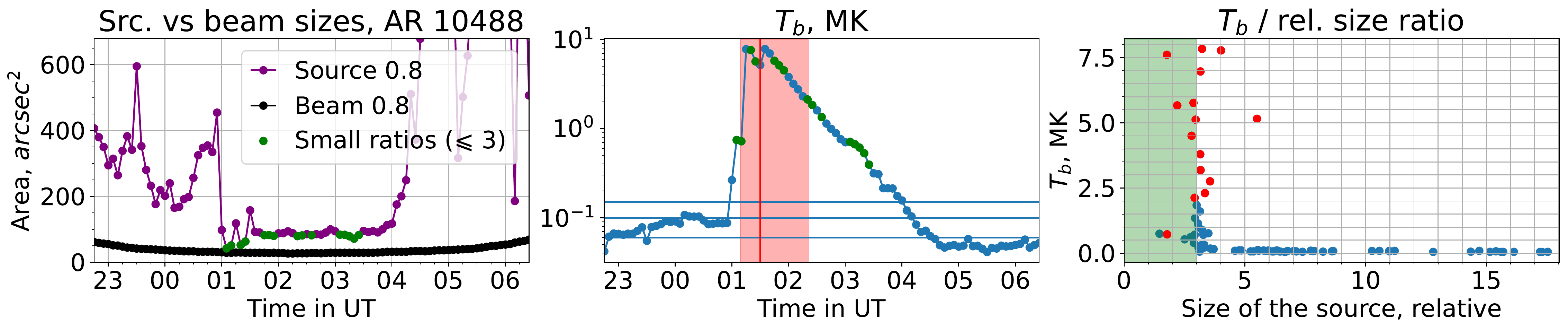}
\figsetgrpnote{NoRH radio source area, NoRH beam size area, and source peak brightness temperature relationship at 34 GHz. The beam area and the radio source area above 0.8 of the maximum values are used. Green colors indicate source areas less than 3 beam areas. Red colors feature solar flare episodes reported by HEK. Blue horizontal lines correspond to 0.06 MK, 0.1 MK, and 0.15 MK respectively.}
\figsetgrpend

\figsetgrpstart
\figsetgrpnum{11.28}
\figsetgrptitle{AR10488}
\figsetplot{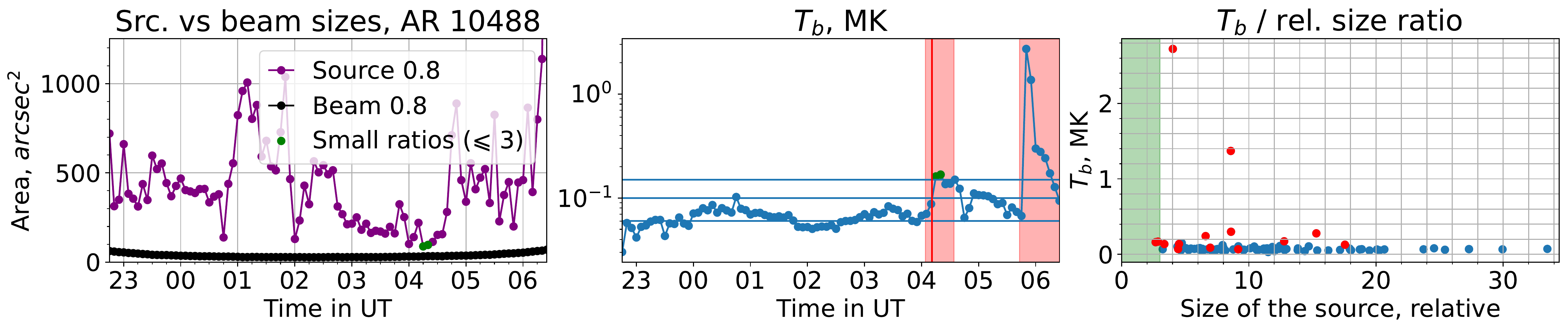}
\figsetgrpnote{NoRH radio source area, NoRH beam size area, and source peak brightness temperature relationship at 34 GHz. The beam area and the radio source area above 0.8 of the maximum values are used. Green colors indicate source areas less than 3 beam areas. Red colors feature solar flare episodes reported by HEK. Blue horizontal lines correspond to 0.06 MK, 0.1 MK, and 0.15 MK respectively.}
\figsetgrpend

\figsetgrpstart
\figsetgrpnum{11.29}
\figsetgrptitle{AR10656}
\figsetplot{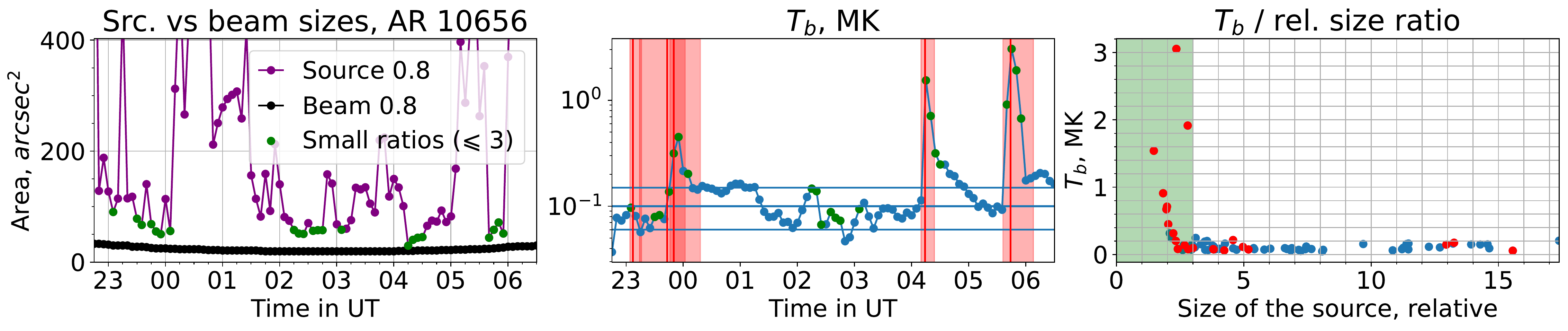}
\figsetgrpnote{NoRH radio source area, NoRH beam size area, and source peak brightness temperature relationship at 34 GHz. The beam area and the radio source area above 0.8 of the maximum values are used. Green colors indicate source areas less than 3 beam areas. Red colors feature solar flare episodes reported by HEK. Blue horizontal lines correspond to 0.06 MK, 0.1 MK, and 0.15 MK respectively.}
\figsetgrpend

\figsetgrpstart
\figsetgrpnum{11.30}
\figsetgrptitle{AR10656}
\figsetplot{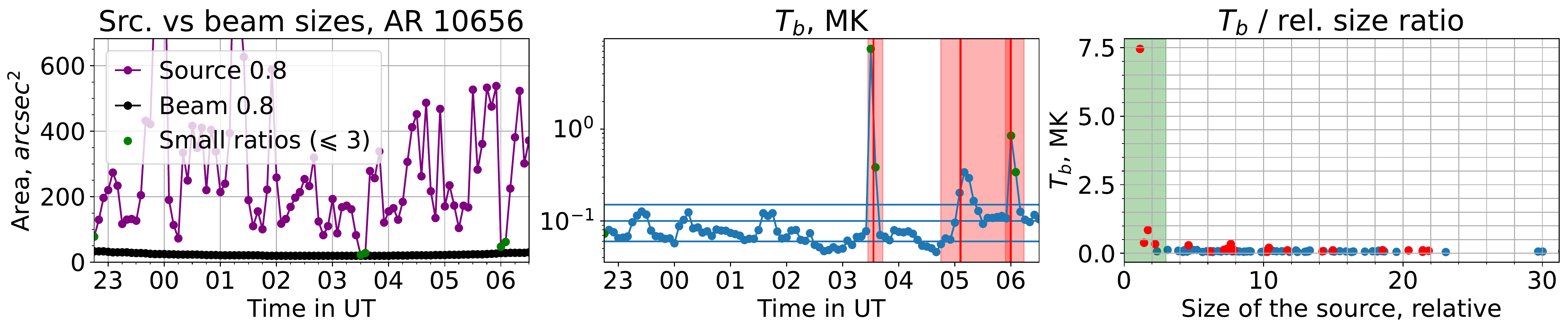}
\figsetgrpnote{NoRH radio source area, NoRH beam size area, and source peak brightness temperature relationship at 34 GHz. The beam area and the radio source area above 0.8 of the maximum values are used. Green colors indicate source areas less than 3 beam areas. Red colors feature solar flare episodes reported by HEK. Blue horizontal lines correspond to 0.06 MK, 0.1 MK, and 0.15 MK respectively.}
\figsetgrpend

\figsetgrpstart
\figsetgrpnum{11.31}
\figsetgrptitle{AR10656}
\figsetplot{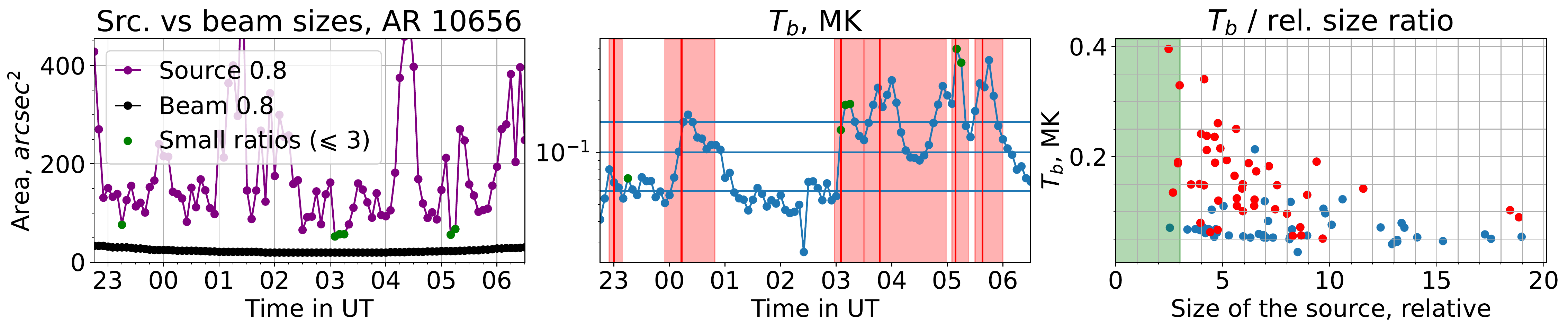}
\figsetgrpnote{NoRH radio source area, NoRH beam size area, and source peak brightness temperature relationship at 34 GHz. The beam area and the radio source area above 0.8 of the maximum values are used. Green colors indicate source areas less than 3 beam areas. Red colors feature solar flare episodes reported by HEK. Blue horizontal lines correspond to 0.06 MK, 0.1 MK, and 0.15 MK respectively.}
\figsetgrpend

\figsetgrpstart
\figsetgrpnum{11.32}
\figsetgrptitle{AR10656}
\figsetplot{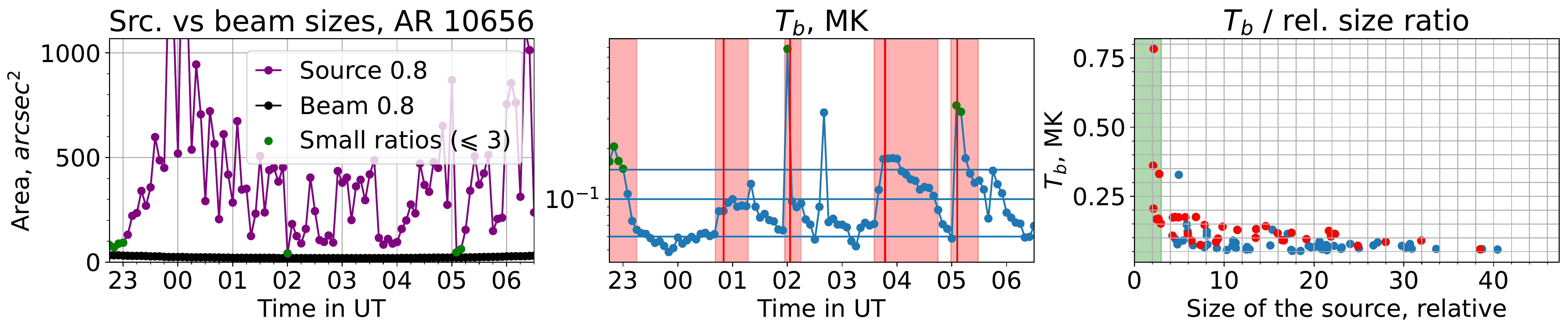}
\figsetgrpnote{NoRH radio source area, NoRH beam size area, and source peak brightness temperature relationship at 34 GHz. The beam area and the radio source area above 0.8 of the maximum values are used. Green colors indicate source areas less than 3 beam areas. Red colors feature solar flare episodes reported by HEK. Blue horizontal lines correspond to 0.06 MK, 0.1 MK, and 0.15 MK respectively.}
\figsetgrpend

\figsetgrpstart
\figsetgrpnum{11.33}
\figsetgrptitle{AR10656}
\figsetplot{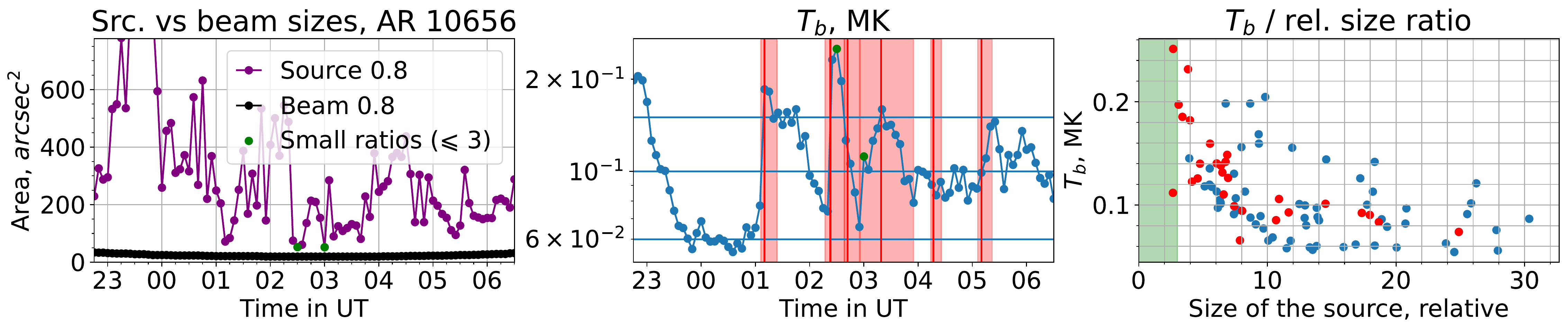}
\figsetgrpnote{NoRH radio source area, NoRH beam size area, and source peak brightness temperature relationship at 34 GHz. The beam area and the radio source area above 0.8 of the maximum values are used. Green colors indicate source areas less than 3 beam areas. Red colors feature solar flare episodes reported by HEK. Blue horizontal lines correspond to 0.06 MK, 0.1 MK, and 0.15 MK respectively.}
\figsetgrpend

\figsetgrpstart
\figsetgrpnum{11.34}
\figsetgrptitle{AR10656}
\figsetplot{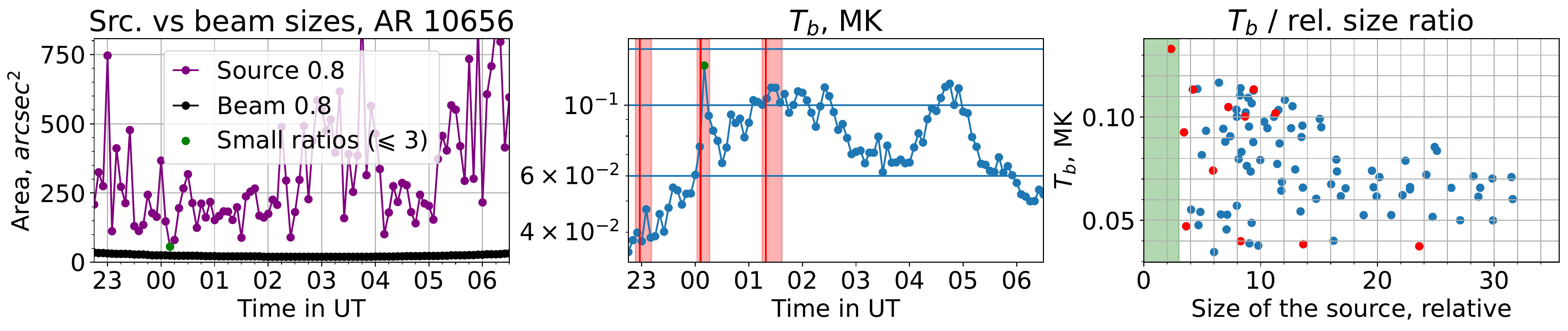}
\figsetgrpnote{NoRH radio source area, NoRH beam size area, and source peak brightness temperature relationship at 34 GHz. The beam area and the radio source area above 0.8 of the maximum values are used. Green colors indicate source areas less than 3 beam areas. Red colors feature solar flare episodes reported by HEK. Blue horizontal lines correspond to 0.06 MK, 0.1 MK, and 0.15 MK respectively.}
\figsetgrpend

\figsetgrpstart
\figsetgrpnum{11.35}
\figsetgrptitle{AR10786}
\figsetplot{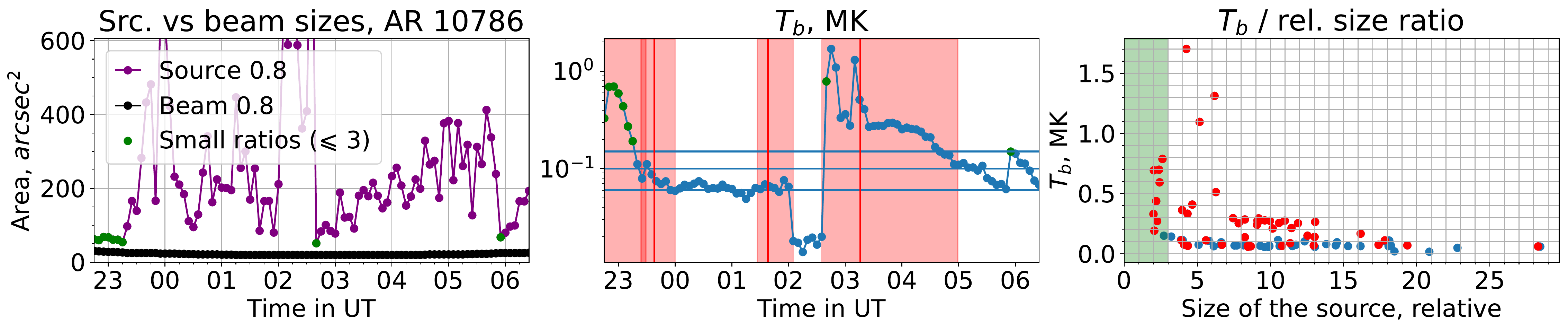}
\figsetgrpnote{NoRH radio source area, NoRH beam size area, and source peak brightness temperature relationship at 34 GHz. The beam area and the radio source area above 0.8 of the maximum values are used. Green colors indicate source areas less than 3 beam areas. Red colors feature solar flare episodes reported by HEK. Blue horizontal lines correspond to 0.06 MK, 0.1 MK, and 0.15 MK respectively.}
\figsetgrpend

\figsetgrpstart
\figsetgrpnum{11.36}
\figsetgrptitle{AR10808}
\figsetplot{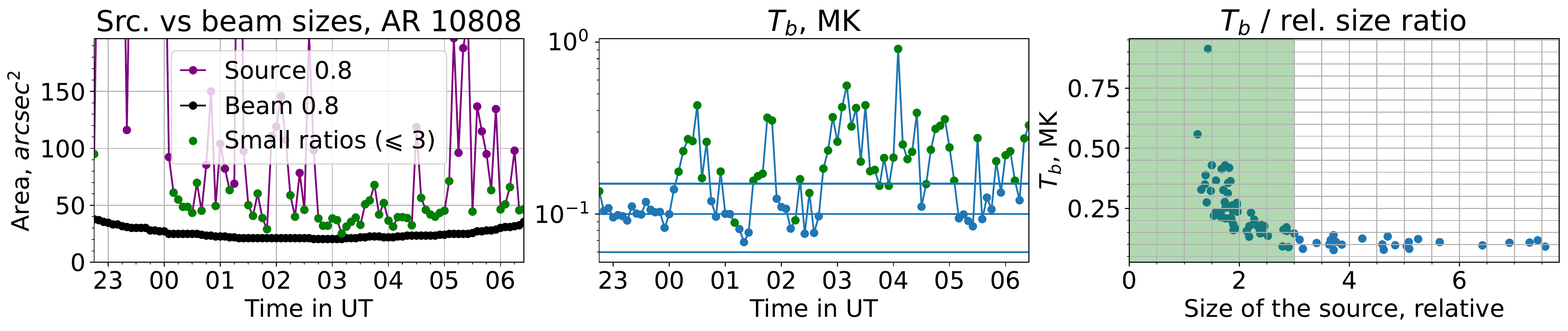}
\figsetgrpnote{NoRH radio source area, NoRH beam size area, and source peak brightness temperature relationship at 34 GHz. The beam area and the radio source area above 0.8 of the maximum values are used. Green colors indicate source areas less than 3 beam areas. Red colors feature solar flare episodes reported by HEK. Blue horizontal lines correspond to 0.06 MK, 0.1 MK, and 0.15 MK respectively.}
\figsetgrpend

\figsetgrpstart
\figsetgrpnum{11.37}
\figsetgrptitle{AR10808}
\figsetplot{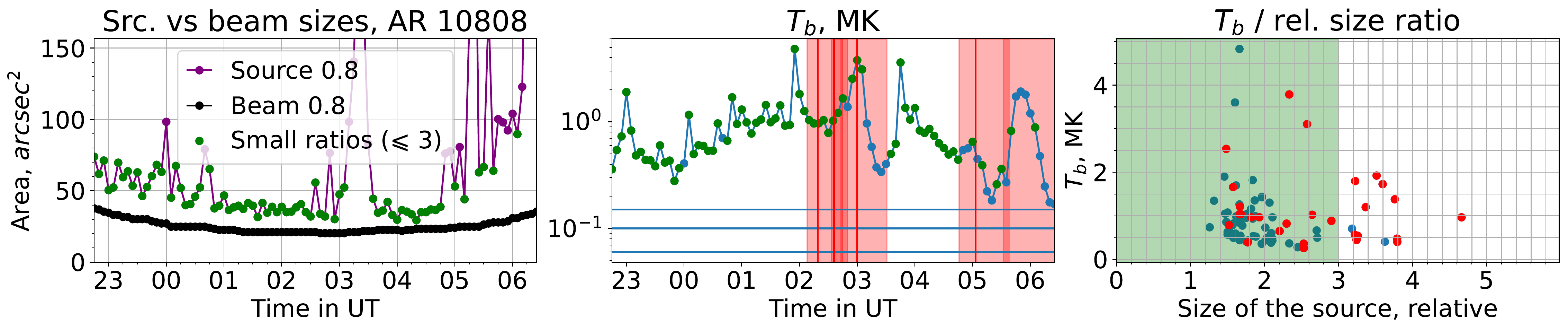}
\figsetgrpnote{NoRH radio source area, NoRH beam size area, and source peak brightness temperature relationship at 34 GHz. The beam area and the radio source area above 0.8 of the maximum values are used. Green colors indicate source areas less than 3 beam areas. Red colors feature solar flare episodes reported by HEK. Blue horizontal lines correspond to 0.06 MK, 0.1 MK, and 0.15 MK respectively.}
\figsetgrpend

\figsetgrpstart
\figsetgrpnum{11.38}
\figsetgrptitle{AR10808}
\figsetplot{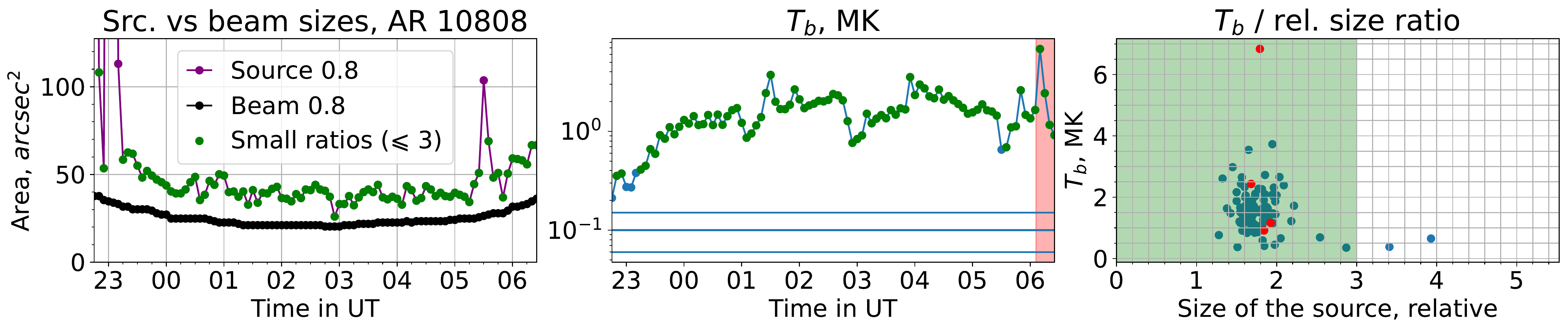}
\figsetgrpnote{NoRH radio source area, NoRH beam size area, and source peak brightness temperature relationship at 34 GHz. The beam area and the radio source area above 0.8 of the maximum values are used. Green colors indicate source areas less than 3 beam areas. Red colors feature solar flare episodes reported by HEK. Blue horizontal lines correspond to 0.06 MK, 0.1 MK, and 0.15 MK respectively.}
\figsetgrpend

\figsetgrpstart
\figsetgrpnum{11.39}
\figsetgrptitle{AR10808}
\figsetplot{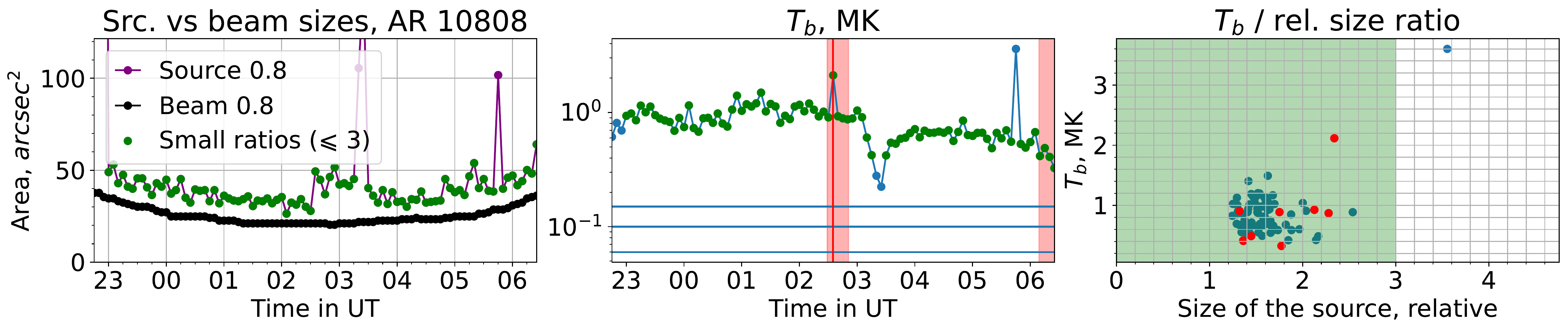}
\figsetgrpnote{NoRH radio source area, NoRH beam size area, and source peak brightness temperature relationship at 34 GHz. The beam area and the radio source area above 0.8 of the maximum values are used. Green colors indicate source areas less than 3 beam areas. Red colors feature solar flare episodes reported by HEK. Blue horizontal lines correspond to 0.06 MK, 0.1 MK, and 0.15 MK respectively.}
\figsetgrpend

\figsetgrpstart
\figsetgrpnum{11.40}
\figsetgrptitle{AR10808}
\figsetplot{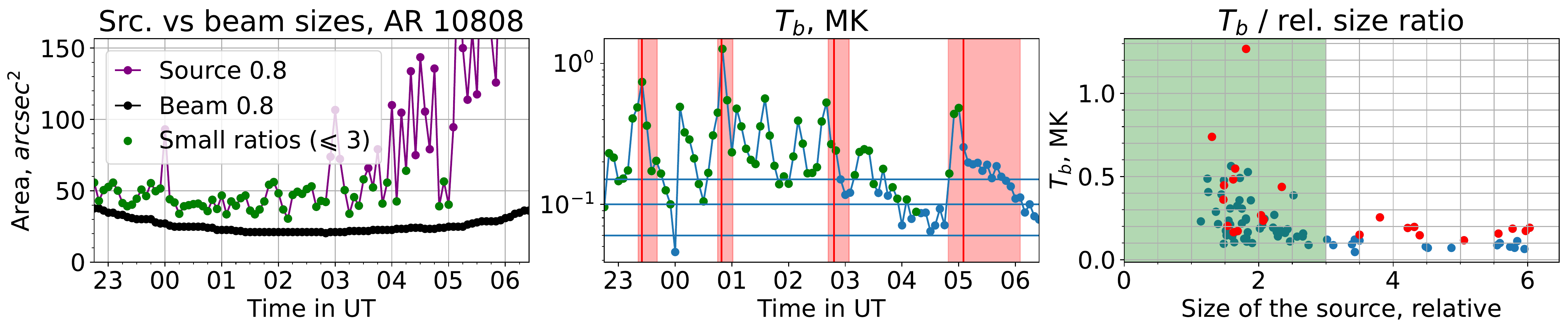}
\figsetgrpnote{NoRH radio source area, NoRH beam size area, and source peak brightness temperature relationship at 34 GHz. The beam area and the radio source area above 0.8 of the maximum values are used. Green colors indicate source areas less than 3 beam areas. Red colors feature solar flare episodes reported by HEK. Blue horizontal lines correspond to 0.06 MK, 0.1 MK, and 0.15 MK respectively.}
\figsetgrpend

\figsetgrpstart
\figsetgrpnum{11.41}
\figsetgrptitle{AR11302}
\figsetplot{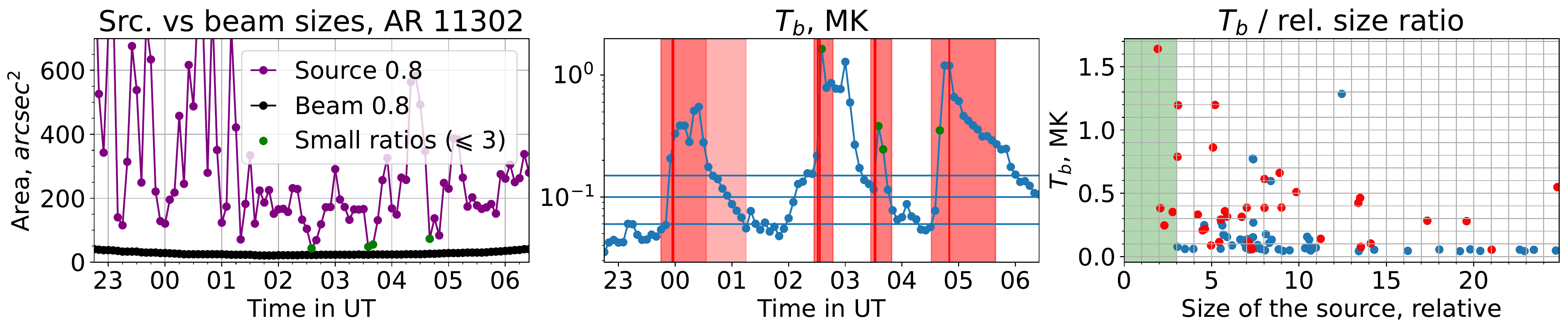}
\figsetgrpnote{NoRH radio source area, NoRH beam size area, and source peak brightness temperature relationship at 34 GHz. The beam area and the radio source area above 0.8 of the maximum values are used. Green colors indicate source areas less than 3 beam areas. Red colors feature solar flare episodes reported by HEK. Blue horizontal lines correspond to 0.06 MK, 0.1 MK, and 0.15 MK respectively.}
\figsetgrpend

\figsetgrpstart
\figsetgrpnum{11.42}
\figsetgrptitle{AR11302}
\figsetplot{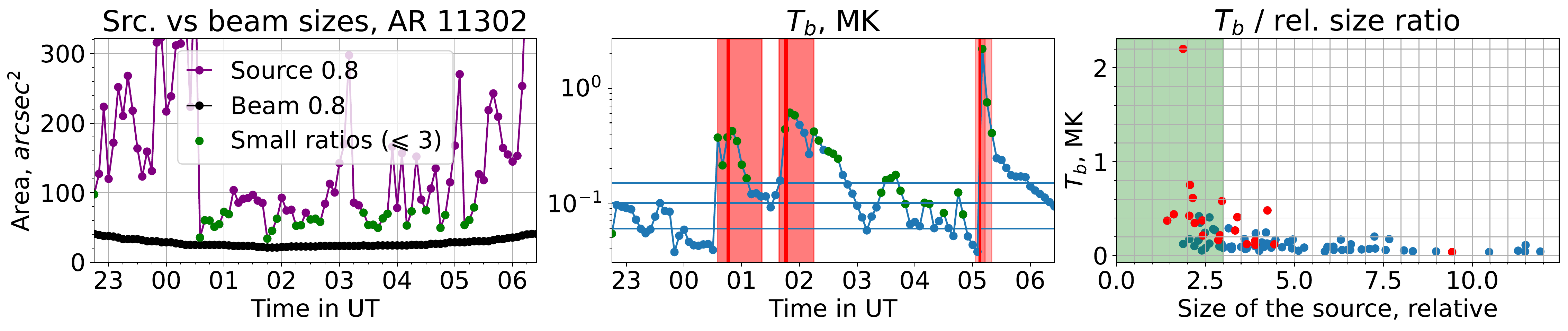}
\figsetgrpnote{NoRH radio source area, NoRH beam size area, and source peak brightness temperature relationship at 34 GHz. The beam area and the radio source area above 0.8 of the maximum values are used. Green colors indicate source areas less than 3 beam areas. Red colors feature solar flare episodes reported by HEK. Blue horizontal lines correspond to 0.06 MK, 0.1 MK, and 0.15 MK respectively.}
\figsetgrpend

\figsetgrpstart
\figsetgrpnum{11.43}
\figsetgrptitle{AR11429}
\figsetplot{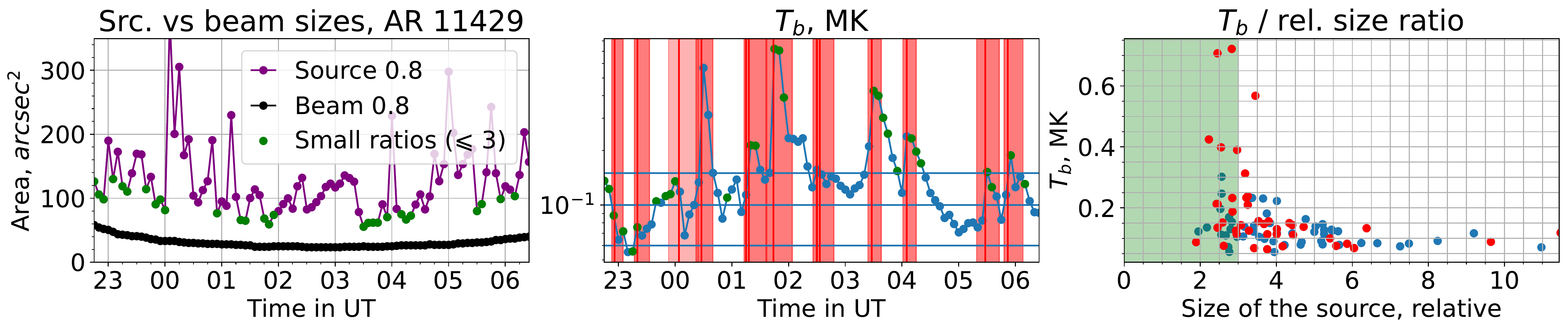}
\figsetgrpnote{NoRH radio source area, NoRH beam size area, and source peak brightness temperature relationship at 34 GHz. The beam area and the radio source area above 0.8 of the maximum values are used. Green colors indicate source areas less than 3 beam areas. Red colors feature solar flare episodes reported by HEK. Blue horizontal lines correspond to 0.06 MK, 0.1 MK, and 0.15 MK respectively.}
\figsetgrpend

\figsetgrpstart
\figsetgrpnum{11.44}
\figsetgrptitle{AR11429}
\figsetplot{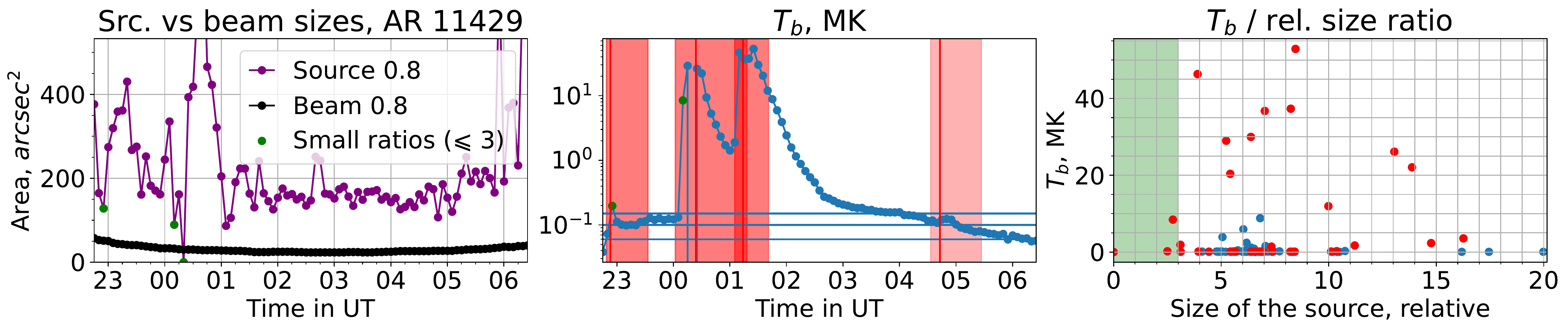}
\figsetgrpnote{NoRH radio source area, NoRH beam size area, and source peak brightness temperature relationship at 34 GHz. The beam area and the radio source area above 0.8 of the maximum values are used. Green colors indicate source areas less than 3 beam areas. Red colors feature solar flare episodes reported by HEK. Blue horizontal lines correspond to 0.06 MK, 0.1 MK, and 0.15 MK respectively.}
\figsetgrpend

\figsetgrpstart
\figsetgrpnum{11.45}
\figsetgrptitle{AR11515}
\figsetplot{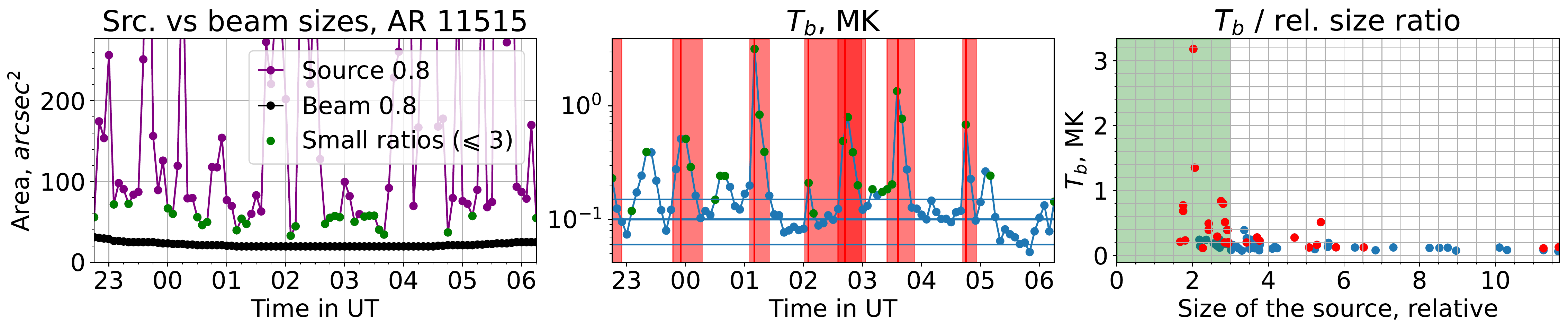}
\figsetgrpnote{NoRH radio source area, NoRH beam size area, and source peak brightness temperature relationship at 34 GHz. The beam area and the radio source area above 0.8 of the maximum values are used. Green colors indicate source areas less than 3 beam areas. Red colors feature solar flare episodes reported by HEK. Blue horizontal lines correspond to 0.06 MK, 0.1 MK, and 0.15 MK respectively.}
\figsetgrpend

\figsetgrpstart
\figsetgrpnum{11.46}
\figsetgrptitle{AR11515}
\figsetplot{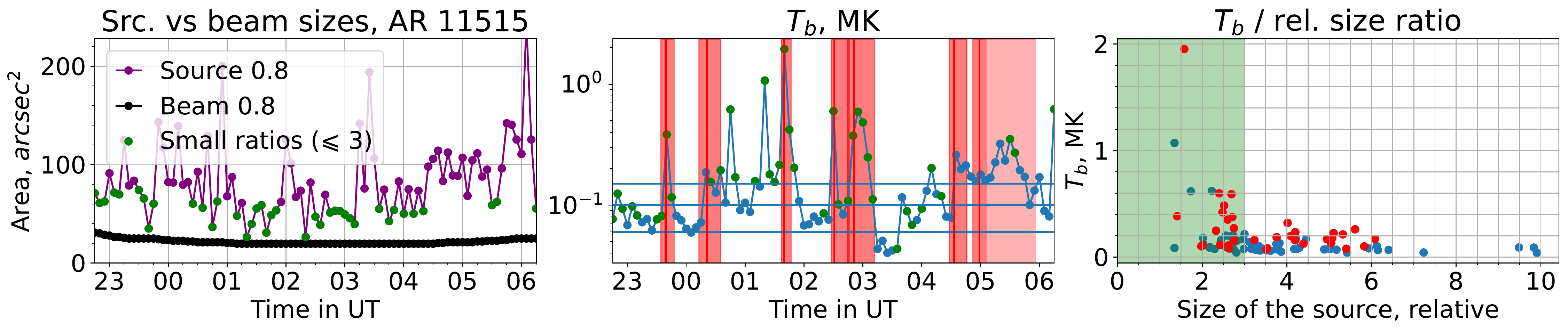}
\figsetgrpnote{NoRH radio source area, NoRH beam size area, and source peak brightness temperature relationship at 34 GHz. The beam area and the radio source area above 0.8 of the maximum values are used. Green colors indicate source areas less than 3 beam areas. Red colors feature solar flare episodes reported by HEK. Blue horizontal lines correspond to 0.06 MK, 0.1 MK, and 0.15 MK respectively.}
\figsetgrpend

\figsetgrpstart
\figsetgrpnum{11.47}
\figsetgrptitle{AR11515}
\figsetplot{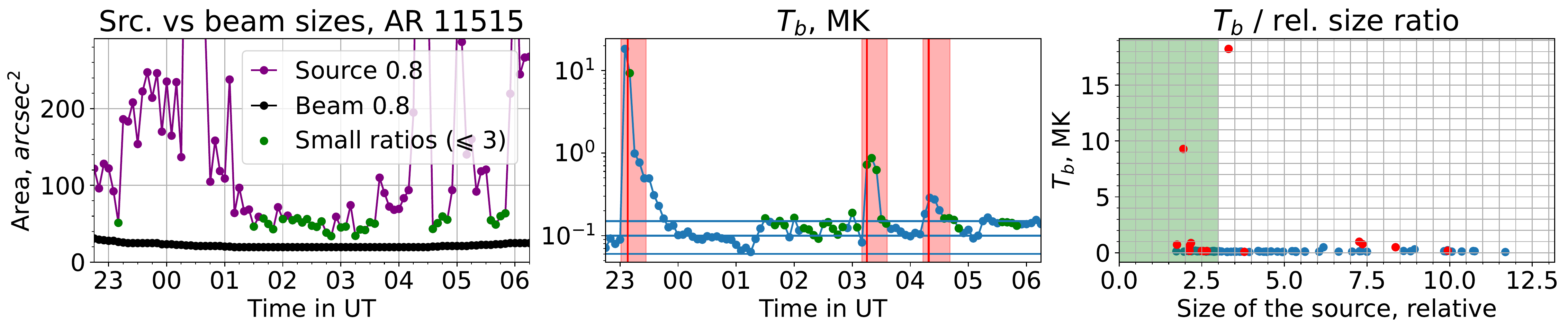}
\figsetgrpnote{NoRH radio source area, NoRH beam size area, and source peak brightness temperature relationship at 34 GHz. The beam area and the radio source area above 0.8 of the maximum values are used. Green colors indicate source areas less than 3 beam areas. Red colors feature solar flare episodes reported by HEK. Blue horizontal lines correspond to 0.06 MK, 0.1 MK, and 0.15 MK respectively.}
\figsetgrpend

\figsetgrpstart
\figsetgrpnum{11.48}
\figsetgrptitle{AR11515}
\figsetplot{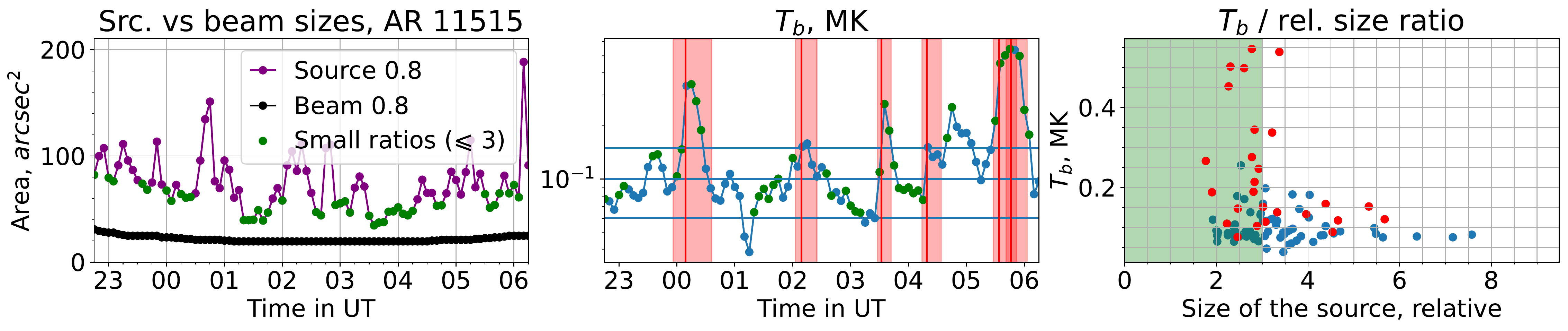}
\figsetgrpnote{NoRH radio source area, NoRH beam size area, and source peak brightness temperature relationship at 34 GHz. The beam area and the radio source area above 0.8 of the maximum values are used. Green colors indicate source areas less than 3 beam areas. Red colors feature solar flare episodes reported by HEK. Blue horizontal lines correspond to 0.06 MK, 0.1 MK, and 0.15 MK respectively.}
\figsetgrpend

\figsetgrpstart
\figsetgrpnum{11.49}
\figsetgrptitle{AR11882}
\figsetplot{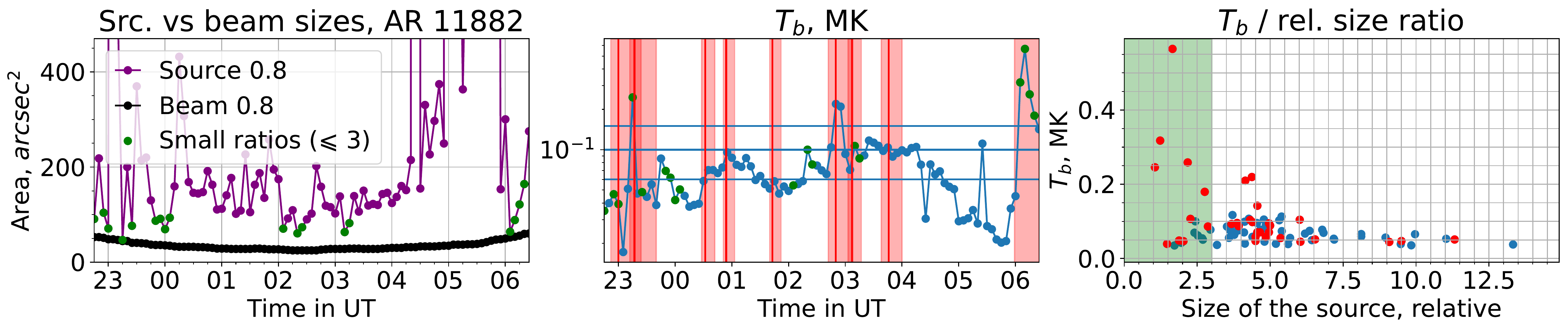}
\figsetgrpnote{NoRH radio source area, NoRH beam size area, and source peak brightness temperature relationship at 34 GHz. The beam area and the radio source area above 0.8 of the maximum values are used. Green colors indicate source areas less than 3 beam areas. Red colors feature solar flare episodes reported by HEK. Blue horizontal lines correspond to 0.06 MK, 0.1 MK, and 0.15 MK respectively.}
\figsetgrpend

\figsetgrpstart
\figsetgrpnum{11.50}
\figsetgrptitle{AR11967}
\figsetplot{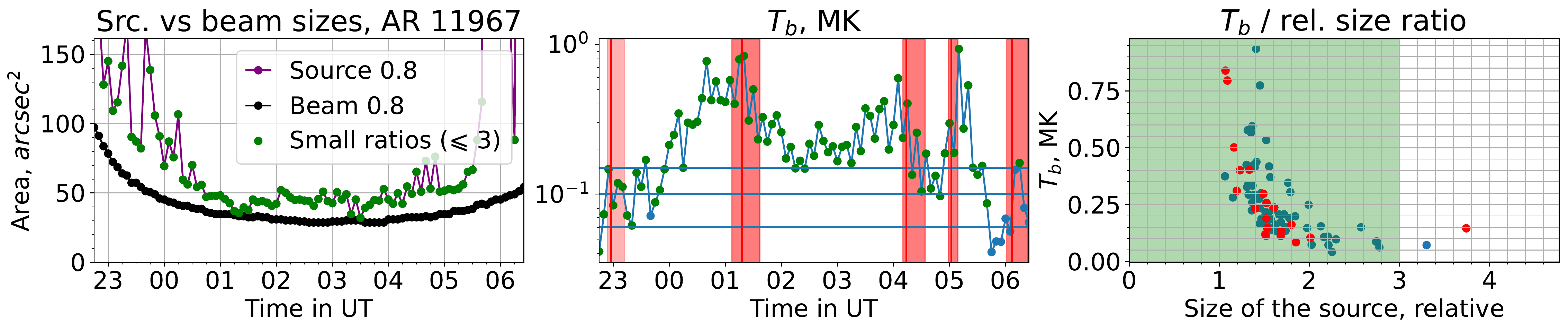}
\figsetgrpnote{NoRH radio source area, NoRH beam size area, and source peak brightness temperature relationship at 34 GHz. The beam area and the radio source area above 0.8 of the maximum values are used. Green colors indicate source areas less than 3 beam areas. Red colors feature solar flare episodes reported by HEK. Blue horizontal lines correspond to 0.06 MK, 0.1 MK, and 0.15 MK respectively.}
\figsetgrpend

\figsetgrpstart
\figsetgrpnum{11.51}
\figsetgrptitle{AR11974}
\figsetplot{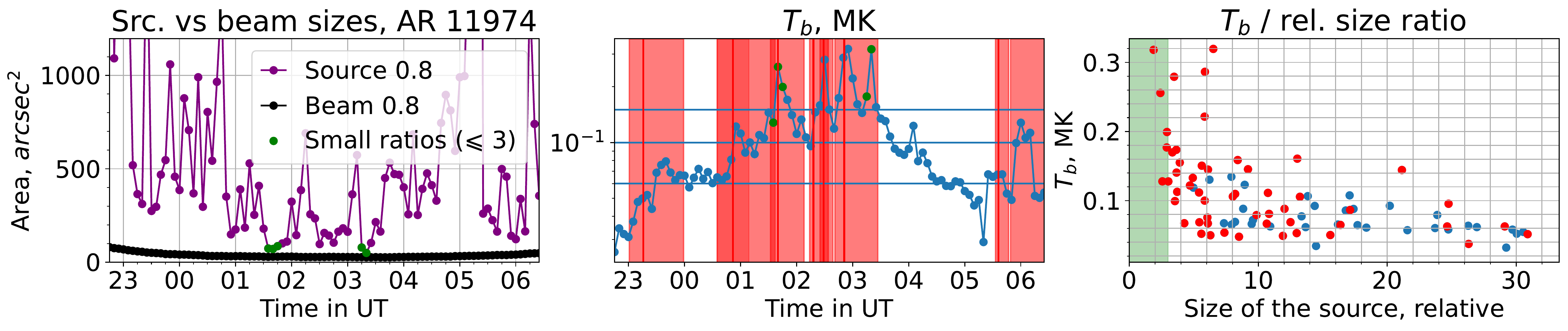}
\figsetgrpnote{NoRH radio source area, NoRH beam size area, and source peak brightness temperature relationship at 34 GHz. The beam area and the radio source area above 0.8 of the maximum values are used. Green colors indicate source areas less than 3 beam areas. Red colors feature solar flare episodes reported by HEK. Blue horizontal lines correspond to 0.06 MK, 0.1 MK, and 0.15 MK respectively.}
\figsetgrpend

\figsetgrpstart
\figsetgrpnum{11.52}
\figsetgrptitle{AR12192}
\figsetplot{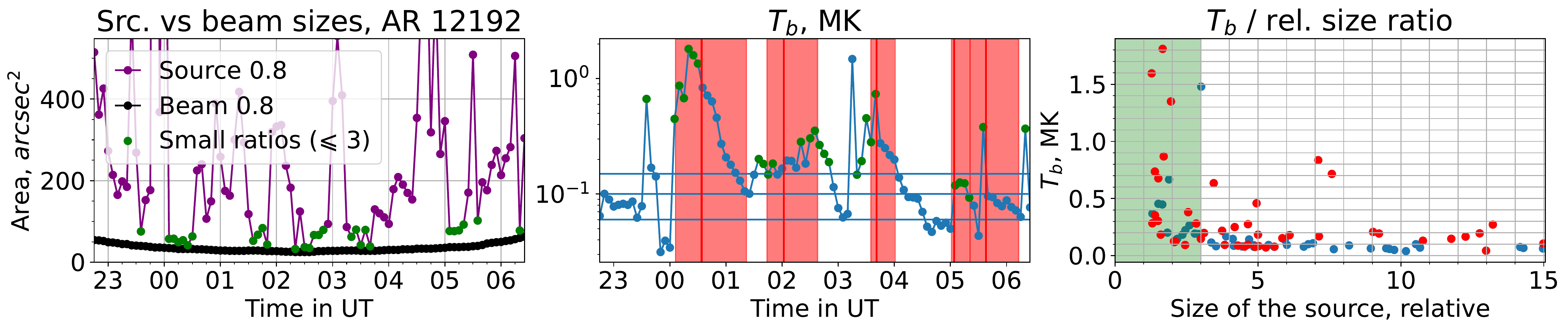}
\figsetgrpnote{NoRH radio source area, NoRH beam size area, and source peak brightness temperature relationship at 34 GHz. The beam area and the radio source area above 0.8 of the maximum values are used. Green colors indicate source areas less than 3 beam areas. Red colors feature solar flare episodes reported by HEK. Blue horizontal lines correspond to 0.06 MK, 0.1 MK, and 0.15 MK respectively.}
\figsetgrpend

\figsetgrpstart
\figsetgrpnum{11.53}
\figsetgrptitle{AR12297}
\figsetplot{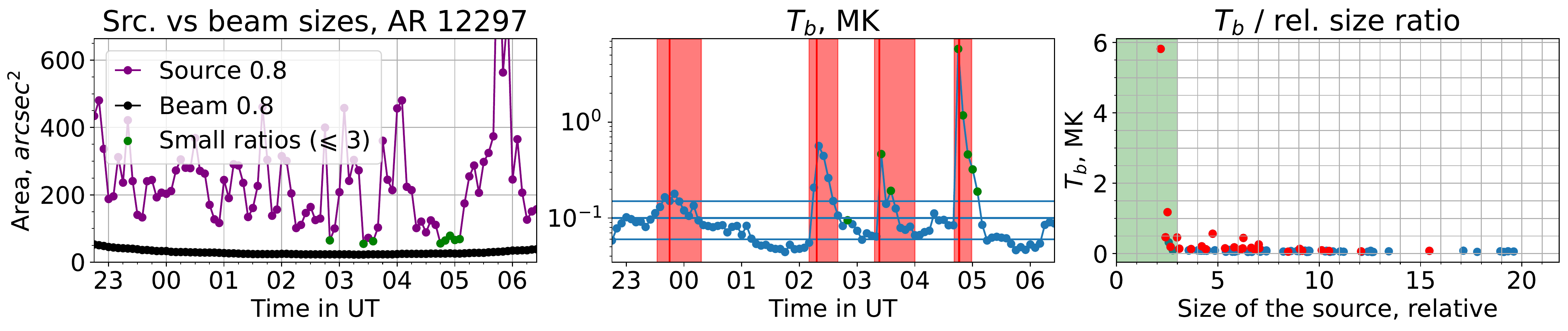}
\figsetgrpnote{NoRH radio source area, NoRH beam size area, and source peak brightness temperature relationship at 34 GHz. The beam area and the radio source area above 0.8 of the maximum values are used. Green colors indicate source areas less than 3 beam areas. Red colors feature solar flare episodes reported by HEK. Blue horizontal lines correspond to 0.06 MK, 0.1 MK, and 0.15 MK respectively.}
\figsetgrpend

\figsetgrpstart
\figsetgrpnum{11.54}
\figsetgrptitle{AR12644}
\figsetplot{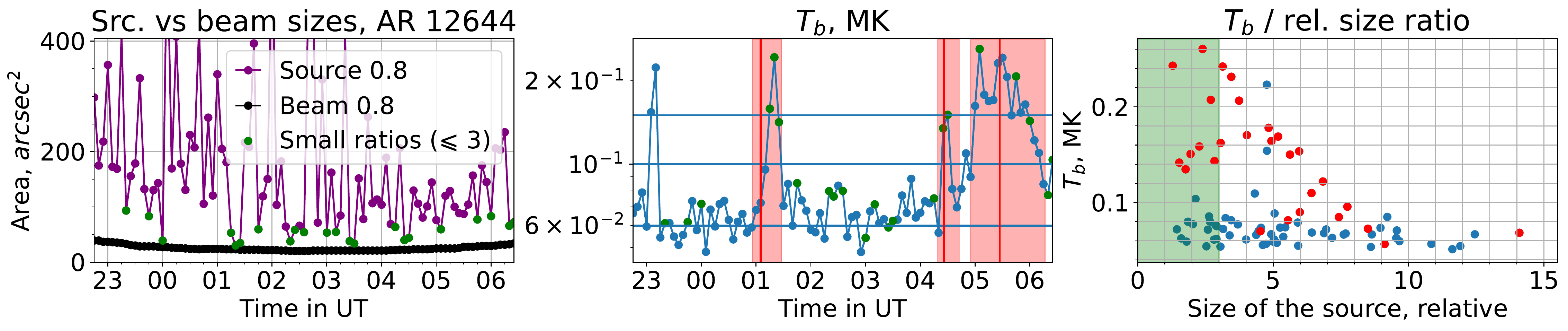}
\figsetgrpnote{NoRH radio source area, NoRH beam size area, and source peak brightness temperature relationship at 34 GHz. The beam area and the radio source area above 0.8 of the maximum values are used. Green colors indicate source areas less than 3 beam areas. Red colors feature solar flare episodes reported by HEK. Blue horizontal lines correspond to 0.06 MK, 0.1 MK, and 0.15 MK respectively.}
\figsetgrpend

\figsetgrpstart
\figsetgrpnum{11.55}
\figsetgrptitle{AR12673}
\figsetplot{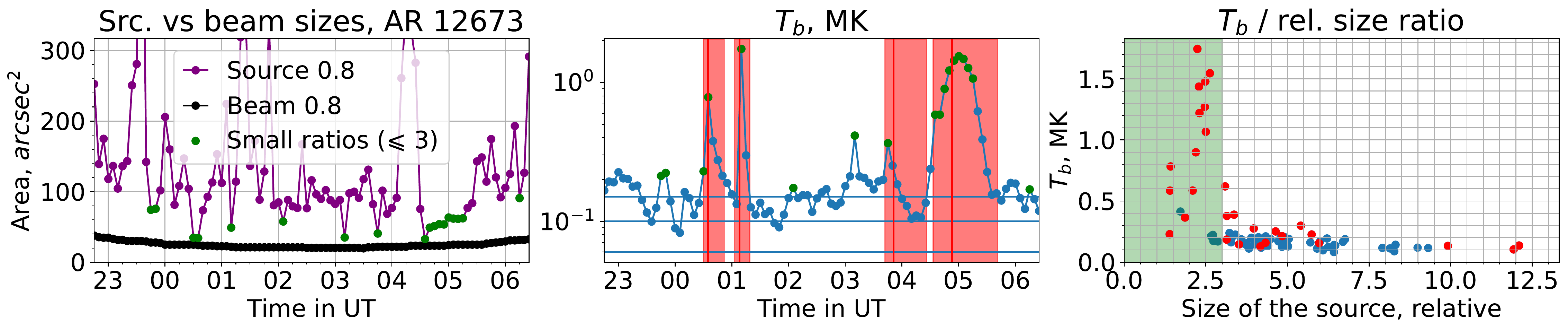}
\figsetgrpnote{NoRH radio source area, NoRH beam size area, and source peak brightness temperature relationship at 34 GHz. The beam area and the radio source area above 0.8 of the maximum values are used. Green colors indicate source areas less than 3 beam areas. Red colors feature solar flare episodes reported by HEK. Blue horizontal lines correspond to 0.06 MK, 0.1 MK, and 0.15 MK respectively.}
\figsetgrpend

\figsetgrpstart
\figsetgrpnum{11.56}
\figsetgrptitle{AR12673}
\figsetplot{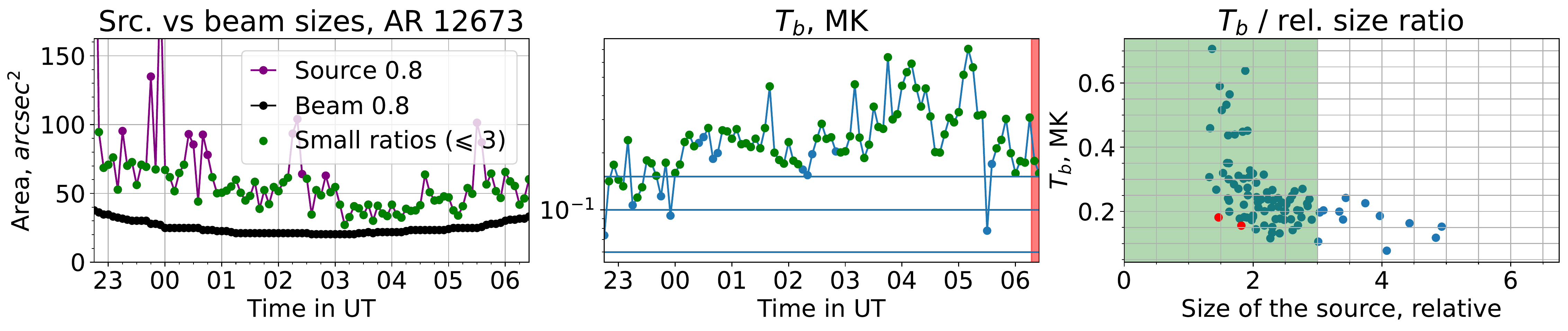}
\figsetgrpnote{NoRH radio source area, NoRH beam size area, and source peak brightness temperature relationship at 34 GHz. The beam area and the radio source area above 0.8 of the maximum values are used. Green colors indicate source areas less than 3 beam areas. Red colors feature solar flare episodes reported by HEK. Blue horizontal lines correspond to 0.06 MK, 0.1 MK, and 0.15 MK respectively.}
\figsetgrpend

\figsetgrpstart
\figsetgrpnum{11.57}
\figsetgrptitle{AR12673}
\figsetplot{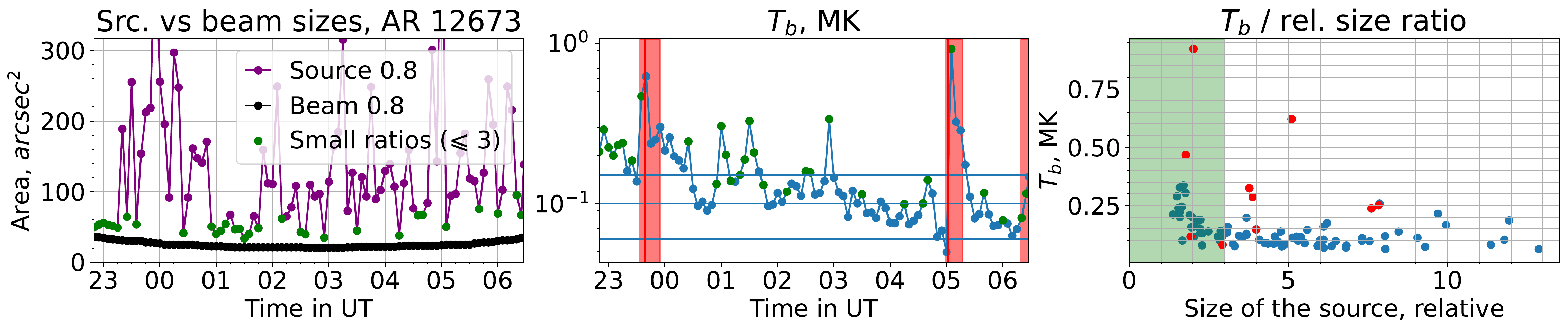}
\figsetgrpnote{NoRH radio source area, NoRH beam size area, and source peak brightness temperature relationship at 34 GHz. The beam area and the radio source area above 0.8 of the maximum values are used. Green colors indicate source areas less than 3 beam areas. Red colors feature solar flare episodes reported by HEK. Blue horizontal lines correspond to 0.06 MK, 0.1 MK, and 0.15 MK respectively.}
\figsetgrpend

\figsetend

\begin{figure*}[htbp] \centering
    \includegraphics[width=\linewidth]{"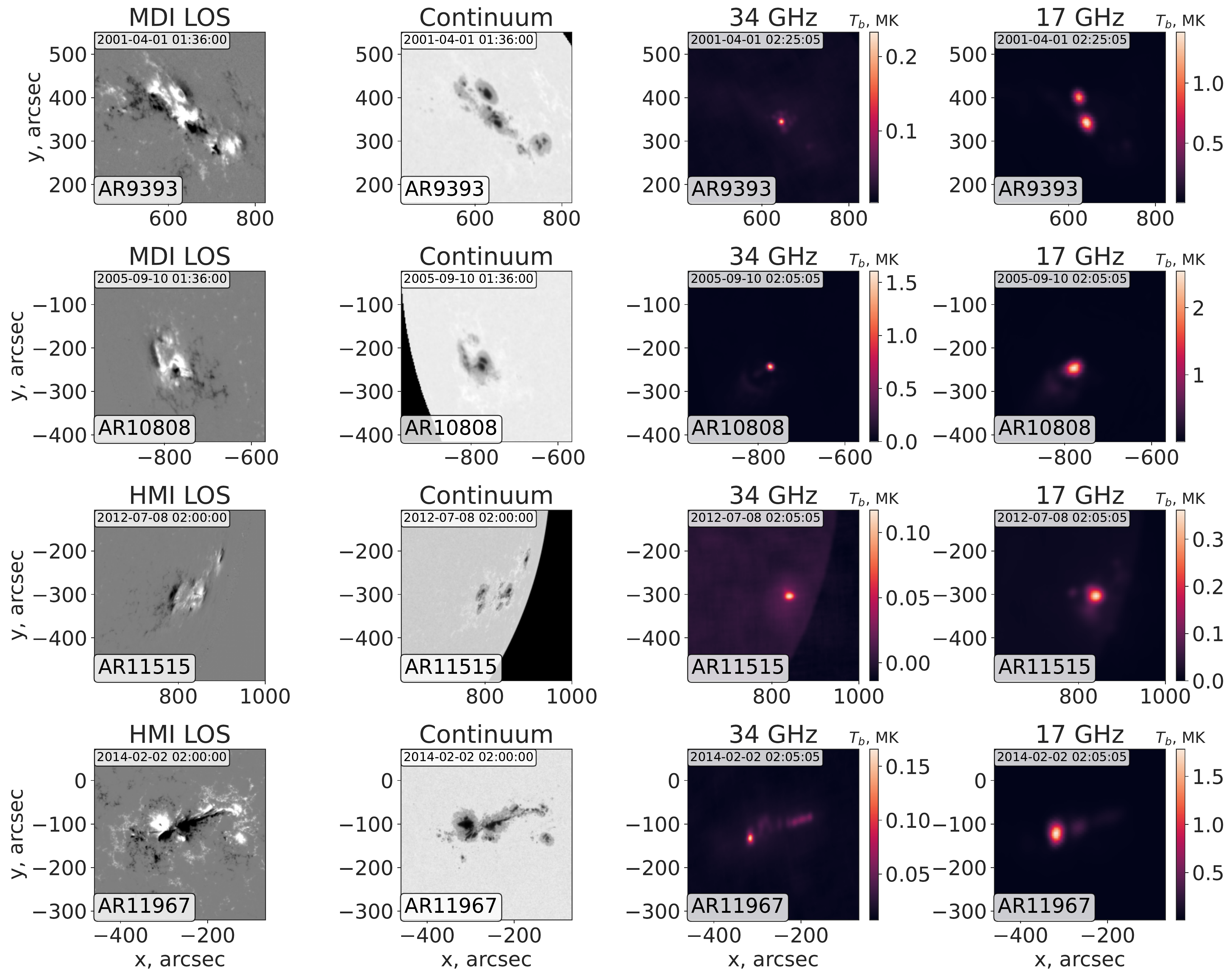"}
    \caption{Line of sight photospheric magnetograms, white-light images, and radio  images at 34 GHz and 17 GHz for optically thick sources from Table~\ref{results-table}. For AR\,9393 and AR\,10808 the magnetograms and continuum images were taken from SOHO/MDI, for AR\,11515 and AR\,11967 --- from SDO/HMI.}
    \label{fig:thick_detailed}
\end{figure*}

  Among selected events, we found only five cases (NOAA 9393, 10808, 11515, 11967 and 12673 in Table \ref{results-table}) with a different behavior, where the source area remains of the order of the beam size ($\leq 3$ beam areas) throughout the whole observational day or even several days. Thus in these five cases, we observe very compact unresolved radio sources with high brightness temperatures, which vary slowly due likely to slow evolution of the magnetic field, thermal plasma, or both. These cases are presented in Figures \ref{fig2017} and \ref{fig:thick_detailed}.

From this inspection of the source brightness and sizes, we found five cases, two and three in each solar cycle (NOAA ARs 9393, 10808, 11515, 11967 and 12673) (see Fig. \ref{spatialanalysis-timelines}), which fit the expectation for the optically thick GR-34 emission: they are bright and compact. Therefore, we classify these cases as ``confirmed'' optically thick GR-34 sources, although we cannot confidently exclude other cases from being optically thick (for example, if there are several displaced optically thick compact sources, which merge together in a large source due to insufficient instrumental spatial resolution). In only one case, AR\,10808, the measured brightness temperature is above 1\,MK indicative of a rather large, comparable with the beam size, optically thick GR-34 source.

Two optically thick cases in solar cycle 24, AR\,11967 and AR\,12673, for which photospheric vector magnetic data are available from \SDO/HMI and Hinode, both demonstrate the presence of photospheric magnetic field in excess of 5,000\,G. \citet{2018AGUFMSH41C3646O} reported the photospheric magnetic field up to 6,251\,G in AR\,11967 based on the Hinode data, while \citet{Wang_2018} reported values about 5,700\,G for AR\,12673 using GST data.

For the  two cases, registered in solar cycle 23, AR\,9393 and AR\,10808, we have not found any vector magnetic measurements, while MDI LOS measurements did not show exceptionally high values. AR\,11515 does not demonstrate presence of very strong field in SDO/HMI magnetograms, but, according to Hinode SOT/SP data, its peak magnetic field value is above 4000\,G, while \citet{2018AGUFMSH41C3646O} reported the value of 4,800\,G based on the Hinode data.

Relative to the total number of ARs with GR-34 emission (27), the probability of having the optically thick GR-34 source is $\frac{5}{27} \approx 18.5\%$.  The probability for a given active region to produce the optically thick GR-34 emission is, thus, $5 \div 4774 \times 100\% \approx 0.105\%$. Finally, we repeat that only one AR demonstrated brightness temperature above 1\,MK at 34\,GHz.

\subsection{Distribution of GR-34 sources over the visible solar disk}
\begin{figure*}[htbp] \centering
    \includegraphics[width=0.31\linewidth]{"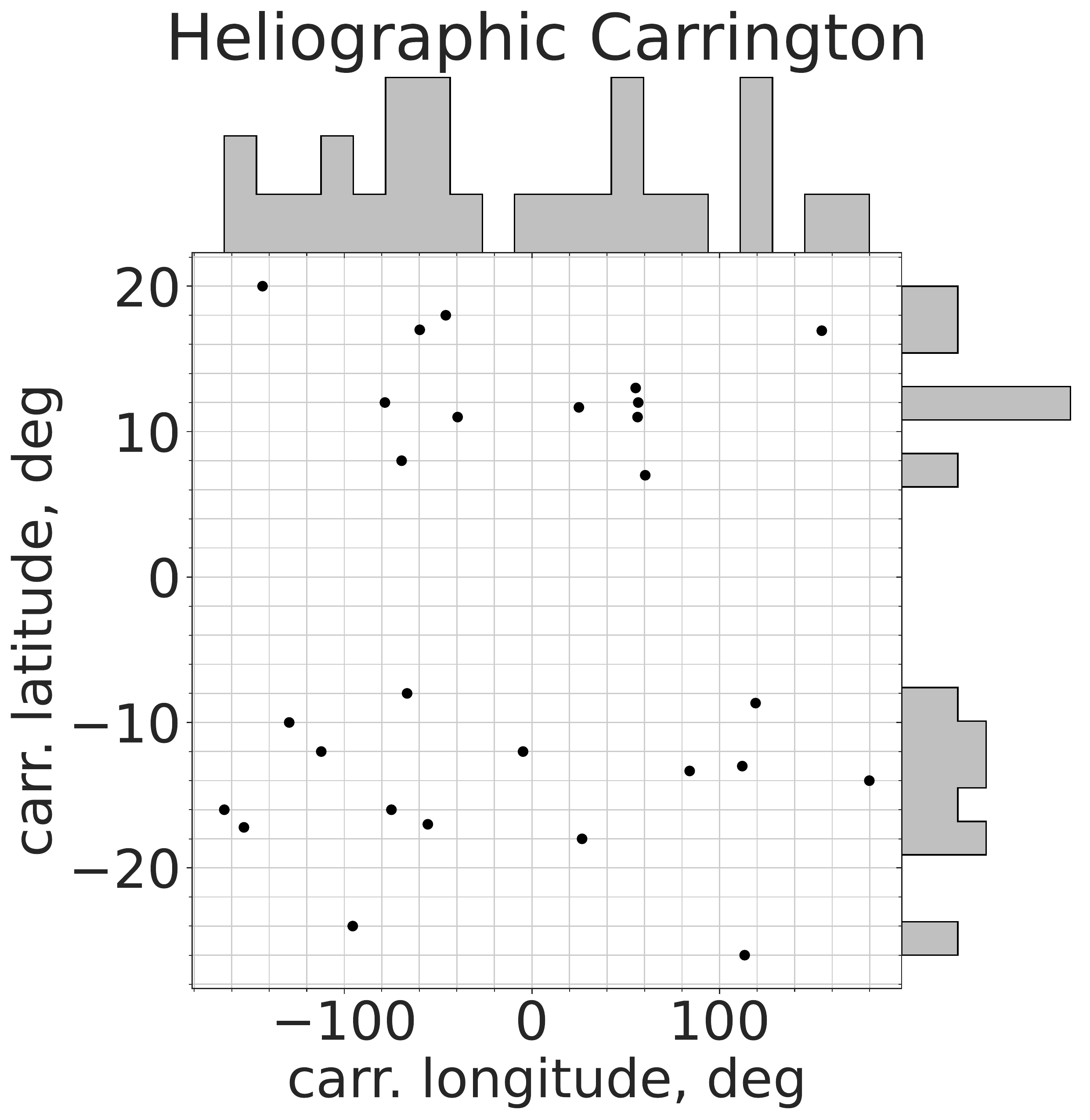"}
    \includegraphics[width=0.31\linewidth]{"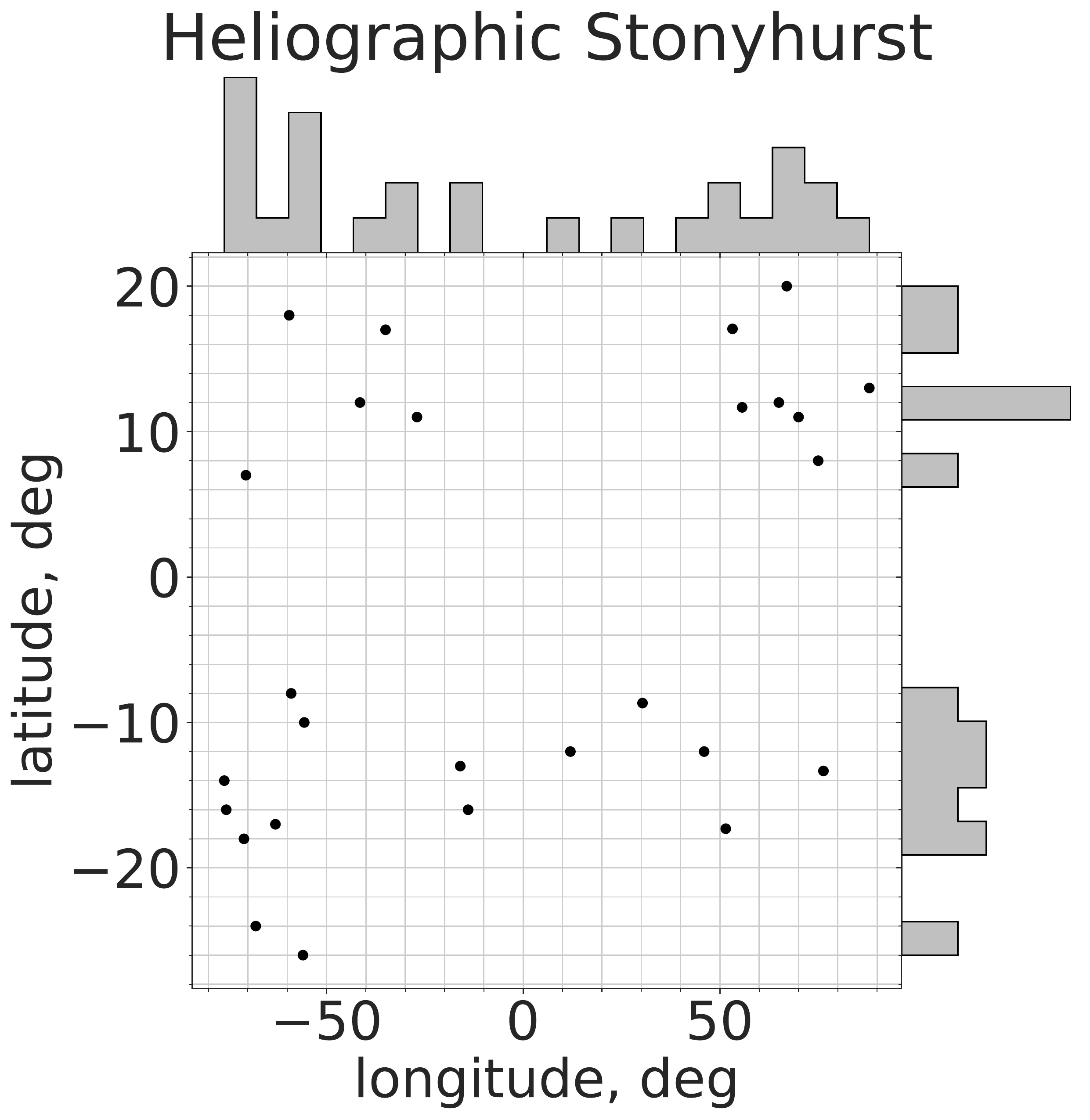"}
    \includegraphics[width=0.322\linewidth]{"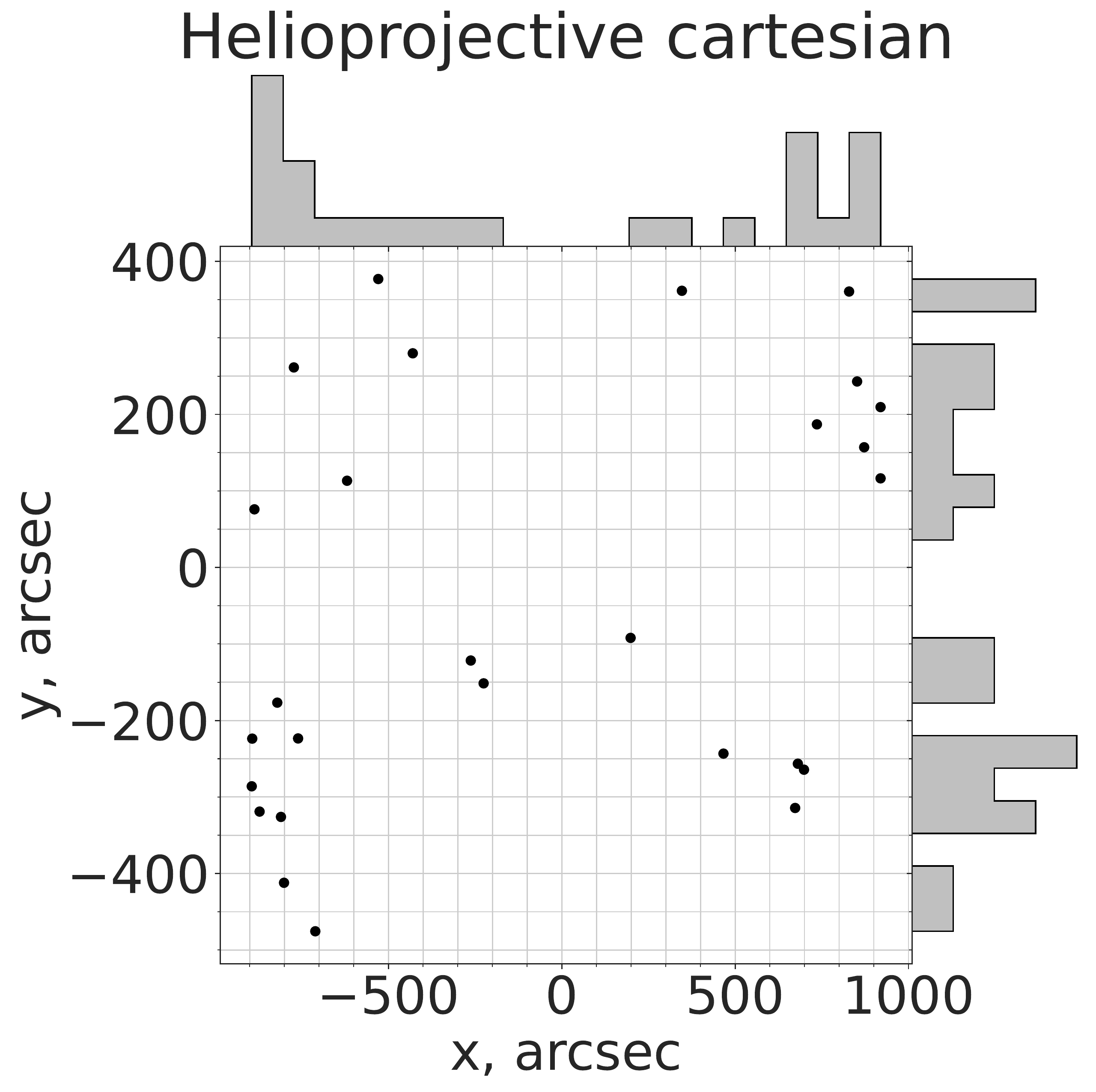"}
    \caption{Coordinate distribution of all GR-34 ARs. Several coordinate systems widely used in the solar physics are shown, including observer-oriented and absolute heliographic ones. Each point stands for a single active region, not the day of observation. If a microwave emission located in a particular active region appears more than once in the  table, average position through the all days listed in the table is taken.}
    \label{spatialdist}
\end{figure*}

 Most of the events are located between 10 and 20 degrees in both hemispheres which coincides with typical latitudes of solar active regions. The histogram of Carrington longitude (see left panel on Fig.\,\ref{spatialdist}) demonstrates no significant deviations from the uniform distribution.
 
\newpage

\startlongtable
\begin{deluxetable*}{lllllllllllll}
\tablecaption{Final GR-34 event list with properties and associations\label{results-table}}
\tablehead{
\colhead{N}&
\colhead{AR}&
\colhead{Date}&
\colhead{location}&
\colhead{$T_{34 med}$}&
\colhead{$T_{34 min}$}&
\colhead{$T_{17 med}$}&
\colhead{$T_{17 min}$}&
\colhead{$B_{max}$}&
\colhead{Thick}&
\colhead{SQ}&
\colhead{SEP}&
\colhead{Strongest}\\
\colhead{}&
\colhead{}&
\colhead{}&
\colhead{}&
\colhead{[MK]}&
\colhead{[MK]}&
\colhead{[MK]}&
\colhead{[MK]}&
\colhead{[G]}&
\colhead{GR}&
\colhead{}&
\colhead{}&
\colhead{flares}
}
\startdata
 1 &  8644 & 1999-07-24 & -801, -412 & 0.069 & 0.033 & 0.157 & 0.094 &                                                     940 (MDI) & - & N/A & - &                                                                                                                                                                   07-23 M1.2 \\
 2 &  9017 & 2000-05-24 & -892, -223 & 0.125 & 0.049 & 0.245 & 0.195 &                                                       no data & - & N/A & - &                                                                                                                                                                   05-25 C3.6 \\
 3 &  9077 & 2000-07-09 &  -808, 265 & 0.066 & 0.040 & 0.280 & 0.122 &                                                    1280 (MDI) & - & N/A & 1 &                                                                                                                                                                   07-12 X1.9 \\
 4 &  9077 & 2000-07-10 &  -737, 256 & 0.086 & 0.046 & 0.874 & 0.404 &                                                    2070 (MDI) & - & N/A & 1 &                                                                                                                                                                   07-12 X1.9 \\
 5 &  9066 & 2000-07-13 &   919, 209 & 0.078 & 0.037 & 0.251 & 0.122 &                                                    1380 (MDI) & - & N/A & - &                                                                                                                                                                            - \\
 6 &  9393 & 2001-03-28 &  -176, 386 & 0.080 & 0.046 & 0.634 & 0.346 &                                                    3380 (MDI) & - & N/A & 2 &                                                                                                                                                                   04-02 X20. \\
 7 &  9393 & 2001-03-30 &   285, 381 & 0.093 & 0.038 & 0.778 & 0.549 &                                                    3530 (MDI) & - & N/A & 2 &                                                                                                                                                                   04-02 X20. \\
 8 &  9393 & 2001-04-01 &   628, 356 & 0.154 & 0.031 & 1.394 & 0.984 &                                                    3430 (MDI) & + & N/A & 2 &                                                                                                                                                                   04-02 X20. \\
 9 &  9393 & 2001-04-02 &   766, 321 & 0.277 & 0.068 & 1.711 & 0.790 &                                                    2460 (MDI) & + & N/A & 2 &                                                                                                                                                                   04-02 X20. \\
10 &  9393 & 2001-04-03 &   868, 298 & 0.299 & 0.115 & 0.843 & 0.383 &                                                    2940 (MDI) & - & N/A & 2 &                                                                                                                                                                   04-02 X20. \\
11 &  9393 & 2001-04-04 &   904, 308 & 0.069 & 0.042 & 0.968 & 0.529 &                                                    2040 (MDI) & - & N/A & 2 &                                                                                                                                                                   04-02 X20. \\
12 &  9591 & 2001-08-23 & -810, -326 & 0.094 & 0.053 & 0.335 & 0.169 &                                                    1140 (MDI) & - & N/A & - &                                                                                                                                                                   08-25 X5.3 \\
13 &  9608 & 2001-09-07 & -711, -475 & 0.067 & 0.034 & 0.215 & 0.092 &                                                    1830 (MDI) & - & N/A & 1 &                                                                                                                                                                   09-17 M8.1 \\
14 &  9690 & 2001-11-06 & -871, -319 & 0.067 & 0.027 & 0.186 & 0.121 &                                                    3740 (MDI) & - & N/A & - &                                                                                                                                                                   11-08 M4.2 \\
15 &  9885 & 2002-03-31 &  -429, 279 & 0.086 & 0.050 & 0.162 & 0.093 &                                                    2880 (MDI) & - & N/A & - &                                                                                                                                                                   03-30 M3.4 \\
16 &  9893 & 2002-04-15 &   828, 360 & 0.075 & 0.029 & 0.166 & 0.068 &                                                    1390 (MDI) & - & N/A & - &                                                                                                                                                                   04-14 M3.7 \\
17 & 10095 & 2002-08-29 &   -916, 86 & 0.127 & 0.033 & 0.276 & 0.144 &                                                       no data & - & N/A & - &                                                                                                                                                                   08-30 X1.5 \\
18 & 10095 & 2002-08-30 &   -916, 86 & 0.075 & 0.026 & 0.310 & 0.135 &                                                    1100 (MDI) & - & N/A & - &                                                                                                                                                                   08-30 X1.5 \\
19 & 10095 & 2002-08-31 &   -856, 65 & 0.060 & 0.035 & 0.275 & 0.131 &                                                    1680 (MDI) & - & N/A & - &                                                                                                                                                                   08-30 X1.5 \\
20 & 10375 & 2003-06-10 &   491, 192 & 0.096 & 0.057 & 0.349 & 0.255 &                                                    2960 (MDI) & - & N/A & - &                                                                                                                                                                   06-09 X1.7 \\
21 & 10375 & 2003-06-11 &   667, 191 & 0.140 & 0.061 & 0.429 & 0.132 &                                                    2370 (MDI) & - & N/A & - &                                                                                                                                                                   06-09 X1.7 \\
22 & 10375 & 2003-06-12 &   817, 192 & 0.096 & 0.051 & 0.244 & 0.162 &                                                    1760 (MDI) & - & N/A & - &                                                                                                                                                                   06-09 X1.7 \\
23 & 10375 & 2003-06-13 &   888, 177 & 0.061 & 0.041 & 0.193 & 0.141 &                                                    1920 (MDI) & - & N/A & - &                                                                                                                                                                   06-09 X1.7 \\
24 & 10486 & 2003-10-22 & -916, -278 & 0.149 & 0.084 & 0.368 & 0.199 & 4200 (CrAO)\footnote{Reported by \citet{2006SoPh..239...41L}} & - & N/A & 2 &                                                                                                                                                                    11-04 X28 \\
25 & 10486 & 2003-10-23 & -916, -278 & 0.082 & 0.041 & 0.600 & 0.333 &                                                    1090 (MDI) & - & N/A & 2 &                                                                                                                                                                    11-04 X28 \\
26 & 10486 & 2003-10-24 & -872, -293 & 0.090 & 0.046 & 0.620 & 0.217 &                                                    1720 (MDI) & - & N/A & 2 &                                                                                                                                                                    11-04 X28 \\
27 & 10488 & 2003-11-03 &   889, 108 & 0.084 & 0.041 & 0.776 & 0.438 &                                                    1520 (MDI) & - & N/A & - &                                                                                                                                                                   11-03 X3.9 \\
28 & 10488 & 2003-11-04 &   949, 124 & 0.067 & 0.030 & 0.355 & 0.166 &                                                     960 (MDI) & - & N/A & - &                                                                                                                                                                   11-03 X3.9 \\
29 & 10656 & 2004-08-14 &  347, -311 & 0.096 & 0.037 & 0.395 & 0.083 &                                                    2420 (MDI) & - & N/A & - &                                                                                                                                                                   08-18 X1.8 \\
30 & 10656 & 2004-08-15 &  531, -300 & 0.072 & 0.046 & 0.238 & 0.133 &                                                    2080 (MDI) & - & N/A & - &                                                                                                                                                                   08-18 X1.8 \\
31 & 10656 & 2004-08-16 &  688, -284 & 0.061 & 0.027 & 0.267 & 0.119 &                                                    1760 (MDI) & - & N/A & - &                                                                                                                                                                   08-18 X1.8 \\
32 & 10656 & 2004-08-17 &  809, -264 & 0.070 & 0.049 & 0.197 & 0.149 &                                                    1410 (MDI) & - & N/A & - &                                                                                                                                                                   08-18 X1.8 \\
33 & 10656 & 2004-08-18 &  897, -252 & 0.091 & 0.054 & 0.283 & 0.192 &                                                     850 (MDI) & - & N/A & - &                                                                                                                                                                   08-18 X1.8 \\
34 & 10656 & 2004-08-19 &  923, -209 & 0.074 & 0.035 & 0.162 & 0.085 &                                                       no data & - & N/A & - &                                                                                                                                                                   08-18 X1.8 \\
35 & 10786 & 2005-07-13 &   872, 157 & 0.069 & 0.049 & 0.217 & 0.041 &                                                    1260 (MDI) & - & N/A & 1 &                                                                                                                                                                   07-14 X1.2 \\
36 & 10808 & 2005-09-08 & -925, -210 & 0.149 & 0.069 & 1.124 & 0.276 &                                                       no data & + & N/A & 1 &                                                                                                                                                                    09-07 X17 \\
37 & 10808 & 2005-09-09 & -925, -210 & 0.734 & 0.279 & 2.558 & 0.985 &                                                    1500 (MDI) & + & N/A & 1 &                                                                                                                                                                    09-07 X17 \\
38 & 10808 & 2005-09-10 & -763, -218 & 1.475 & 0.212 & 2.143 & 0.495 &                                                    2230 (MDI) & + & N/A & 1 &                                                                                                                                                                    09-07 X17 \\
39 & 10808 & 2005-09-11 & -656, -234 & 0.805 & 0.224 & 1.713 & 1.180 &                                                    2940 (MDI) & + & N/A & 1 &                                                                                                                                                                    09-07 X17 \\
40 & 10808 & 2005-09-12 & -472, -251 & 0.163 & 0.046 & 1.391 & 0.627 &                                                    3030 (MDI) & + & N/A & 1 &                                                                                                                                                                    09-07 X17 \\
41 & 11302 & 2011-09-25 &  -686, 120 & 0.069 & 0.035 & 0.740 & 0.323 &                                                    5000 (HMI) & - &   5 & 1 &                                                                                                                                                                   09-24 X1.9 \\
42 & 11302 & 2011-09-26 &  -552, 106 & 0.094 & 0.038 & 0.492 & 0.209 &                                                    4060 (HMI) & - &   5 & 1 &                                                                                                                                                                   09-24 X1.9 \\
43 & 11429 & 2012-03-06 &  -608, 369 & 0.117 & 0.055 & 0.365 & 0.209 &                                                    2730 (HMI) & - &   7 & 2 &                                                                                                                                                                   03-07 X5.4 \\
44 & 11429 & 2012-03-07 &  -450, 383 & 0.156 & 0.038 & 0.349 & 0.116 &                                                    3040 (HMI) & - &   7 & 2 &                                                                                                                                                                   03-07 X5.4 \\
45 & 11515 & 2012-07-05 &  401, -318 & 0.117 & 0.052 & 0.265 & 0.096 &                                                    3500 (HMI) & - &  13 & 1 &                                                                                                                                                                   07-06 X1.1 \\
46 & 11515 & 2012-07-06 &  523, -313 & 0.088 & 0.040 & 0.272 & 0.178 &                                                    3760 (HMI) & + &  13 & 1 &                                                                                                                                                                   07-06 X1.1 \\
47 & 11515 & 2012-07-07 &  689, -327 & 0.121 & 0.063 & 0.439 & 0.253 &                                                    3420 (HMI) & + &  13 & 1 &                                                                                                                                                                   07-06 X1.1 \\
48 & 11515 & 2012-07-08 &  805, -301 & 0.088 & 0.039 & 0.662 & 0.377 &                                                    3030 (HMI) & + &  13 & 1 &                                                                                                                                                                   07-06 X1.1 \\
49 & 11882 & 2013-10-26 & -820, -176 & 0.061 & 0.026 & 0.190 & 0.071 &                                                    4020 (HMI) & - &   - & - &                                                                                                                                                                   10-25 X2.1 \\
50 & 11967 & 2014-02-02 & -262, -121 & 0.207 & 0.041 & 1.481 & 0.448 &                                                    5000 (HMI) & + &   - & - & 01-30 M6.6\footnote{X1.2 flare has been registered on 2 February, 2014 in\\ AR 11944 which is the same AR as NOAA 11967, but\\ observed during the previous solar rotation.} \\
51 & 11974 & 2014-02-13 &   198, -92 & 0.065 & 0.026 & 0.130 & 0.051 &                                                    3290 (HMI) & - &   - & - &                                                                                                                                                                   02-12 M3.7 \\
52 & 12192 & 2014-10-27 &  680, -256 & 0.090 & 0.031 & 2.059 & 0.992 &                                                    3830 (HMI) & - &   6 & - &                                                                                                                                                                   10-24 X3.1 \\
53 & 12297 & 2015-03-12 & -225, -151 & 0.074 & 0.044 & 0.551 & 0.253 &                                                    3400 (HMI) & - &   6 & - &                                                                                                                                                                   03-11 X2.1 \\
54 & 12644 & 2017-04-03 &   851, 243 & 0.066 & 0.048 & 0.158 & 0.077 &                                                    2530 (HMI) & - &   - & - &                                                                                                                                                                   04-03 M5.8 \\
55 & 12673 & 2017-09-05 &  260, -246 & 0.154 & 0.083 & 0.751 & 0.286 &                                                    4790 (HMI) & - &   9 & 2 &                                                                                                                                                                   09-06 X9.3 \\
56 & 12673 & 2017-09-06 &  471, -251 & 0.226 & 0.074 & 1.072 & 0.428 &                                                    5000 (HMI) & + &   9 & 2 &                                                                                                                                                                   09-06 X9.3 \\
57 & 12673 & 2017-09-07 &  666, -232 & 0.117 & 0.062 & 0.896 & 0.421 &                                                    5000 (HMI) & + &   9 & 2 &                                                                                                                                                                   09-06 X9.3 
\enddata
\end{deluxetable*}

 Stonyhurst longitudes and Helioprojective cartesian solar-east (x) coordinates (shown in Fig. \ref{spatialdist}) both show  center-to-limb variations with more GR-34 sources located near the solar limb as compared with the solar disk center. This is consistent with cyclotron radiation opacity dependence on the line of sight, which is enhanced for the transverse directions.

\subsection{GR-34 sources and solar cycle}

\begin{figure*}[htbp] \centering
    \includegraphics[width=0.75\linewidth]{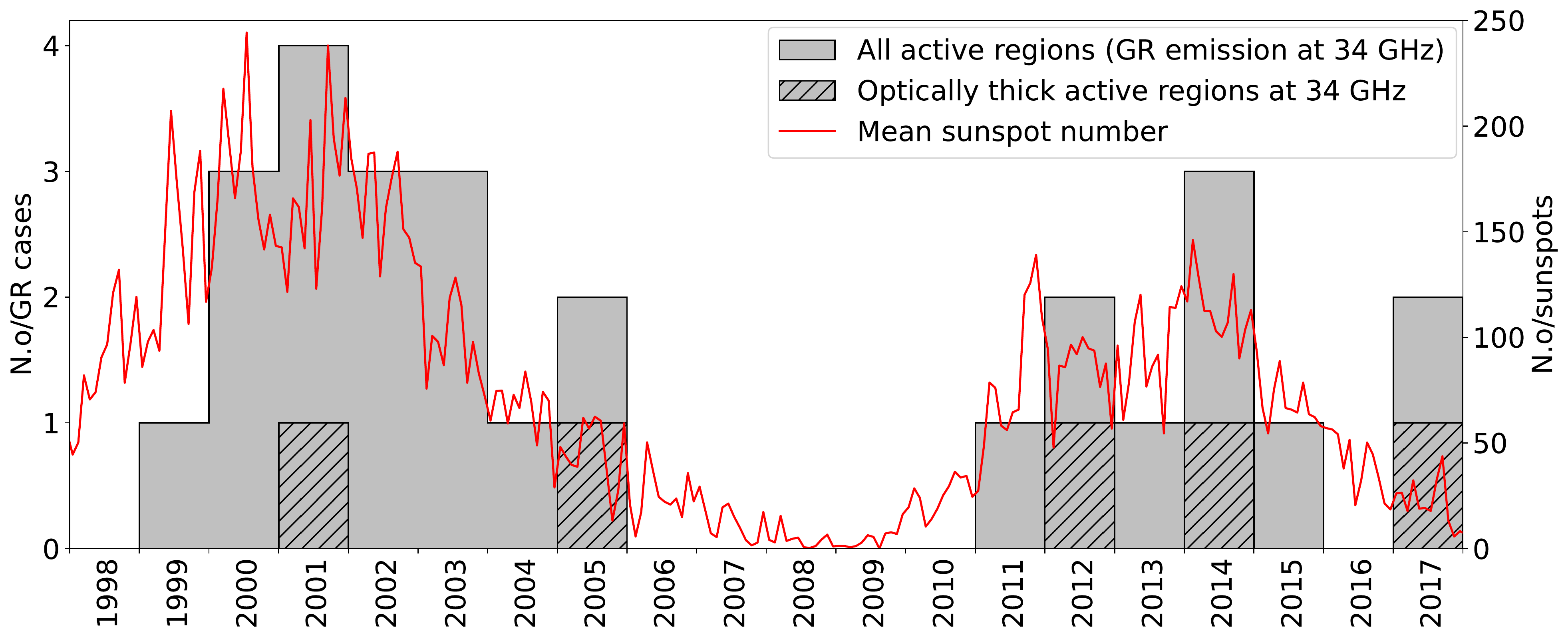}
    \caption{Discovered active regions with GR-34 emission in comparison to total mean sunspot number reported by \citep{belgiumsunspots}. The dates of GR-34 ARs were here taken as average dates of each individual GR-34 AR presented in Table \ref{results-table}.}
    \label{activity_check_fool}
\end{figure*}

Figure \ref{activity_check_fool} displays distribution of the GR-34 sources over years, thus showing dependence of the solar cycle phase. For the reference, the monthly averaged total number of sunspots is overplotted in the same figure. Although the total number of the GR-34 sources is not great, the Figure shows a clear correlation between the number of GR-34 cases and the solar cycle phase. In particular, solar cycle 24 is weaker than 23 in terms of the number of GR-34 sources, like in all other indices of the solar activity.

\subsection{Overall lifetime of GR-34 source}

\begin{figure}[htbp] \centering
    \includegraphics[width=0.54\linewidth]{"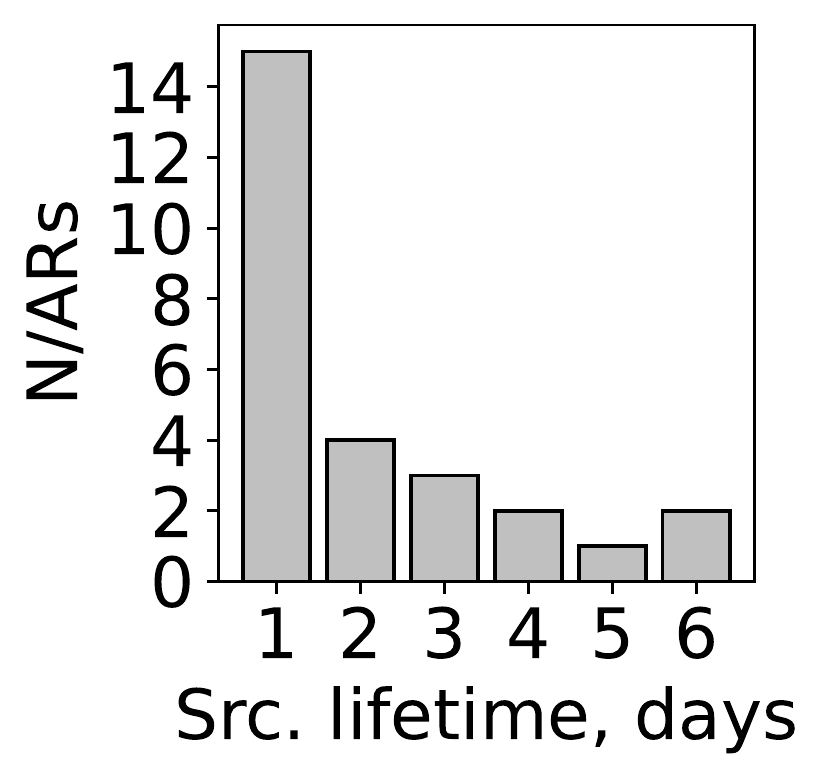"}
    \caption{Observational lifetime histogram of GR-34 sources according to the data reported in Table \ref{results-table}. Horizontal axis describes how many days the GR-34 emission was observed. Vertical axis stands for the number of active regions which show the corresponding lifetime.}
    \label{src_lifetime_days}
\end{figure}

Figure \ref{src_lifetime_days} shows how long the GR-34 emission can last in a given AR at 34\,GHz. We see that most ARs (15 of 27) show this emission during one day only, which implies a comparably fast evolution of the magnetic field in the AR on time scales shorter than one day. However, there are many cases (12 of 27), when it lasts many days. The longest duration is 6 days, which is detected in two cases. It would be interesting to investigate how this property corresponds to evolution of the photospheric magnetic field, flux emergence, cancelation etc.

\subsection{GR-34 lifetimes during observing day}

\begin{figure}[htbp] \centering
    \includegraphics[width=\linewidth]{"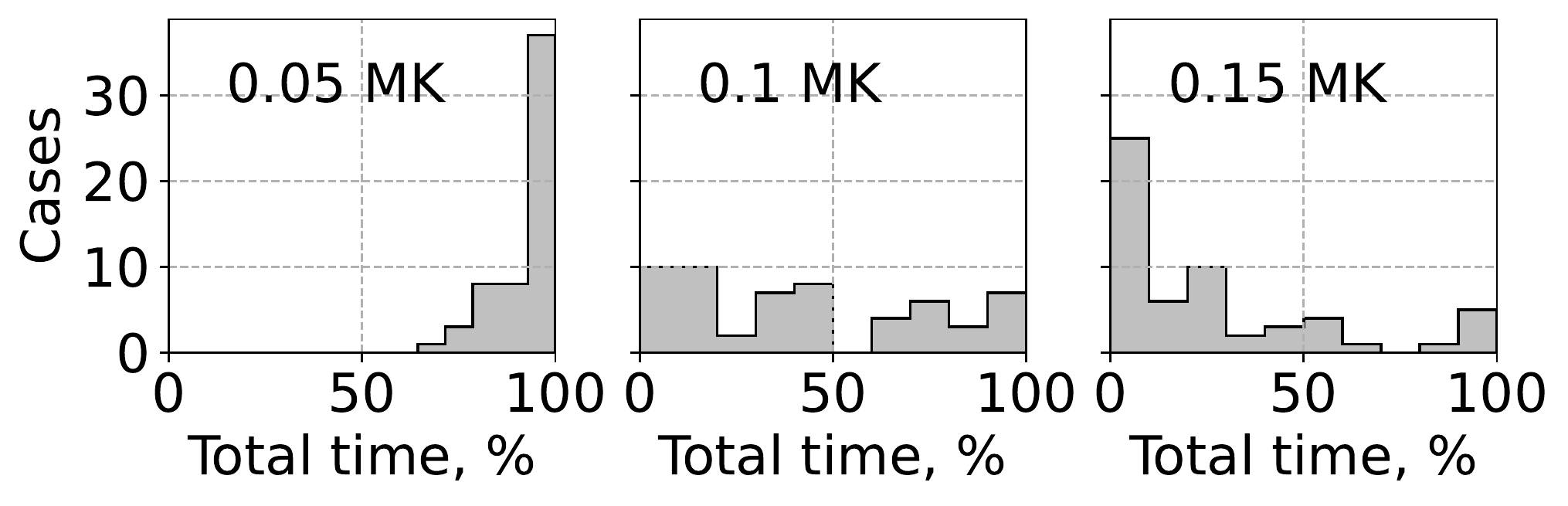"}
    \caption{Fraction of images per each day in which the peak brightness temperature exceeds certain brightness threshold indicated in the panels. Number of cases on vertical axis stands for number of days (events).}
    \label{lifetime_1}
\end{figure}

Let us consider the distribution of the events over the intervals of a time in which the radio source brightness remains above a given threshold relative to the duration of observational day. Most of the sources stay above the 100,000\,K threshold for a third to a half of day (not necessarily subsequently), see Fig. \ref{lifetime_1}.
  
\subsection{Brightness temperatures }
\begin{figure}[htbp] \centering
    \includegraphics[width=\linewidth]{"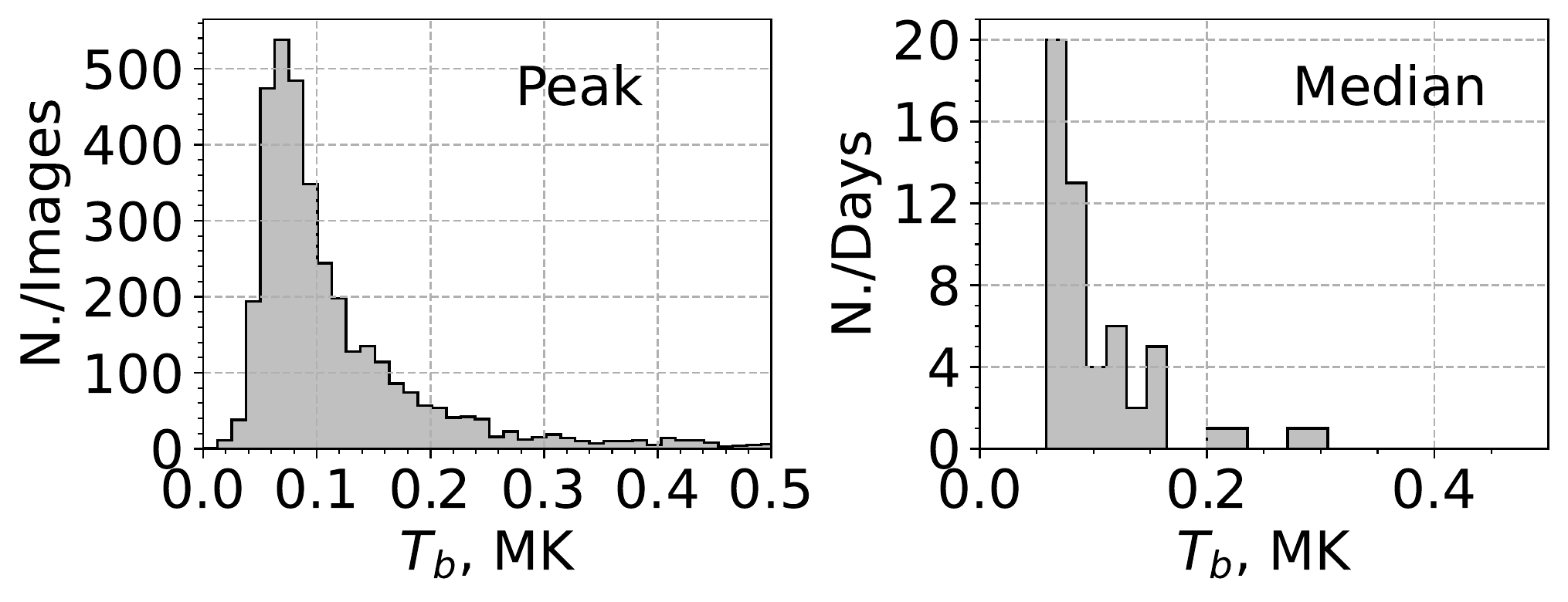"}
    \caption{Brightness temperature distributions across the cases presented in Table \ref{results-table}. Left panel shows the peak per-image $T_b$ distribution across all images of all cases, right panel represents the distribution of the daily median values of per-image peak brightness temperatures for each day (event). There are three events not shown here (corresponding to AR\,10808) with median daily $T_b$ greater than 0.5\,MK.}
    \label{peak_brightness_dist}
\end{figure}

\begin{figure}[htbp] \centering
    \includegraphics[width=\linewidth]{"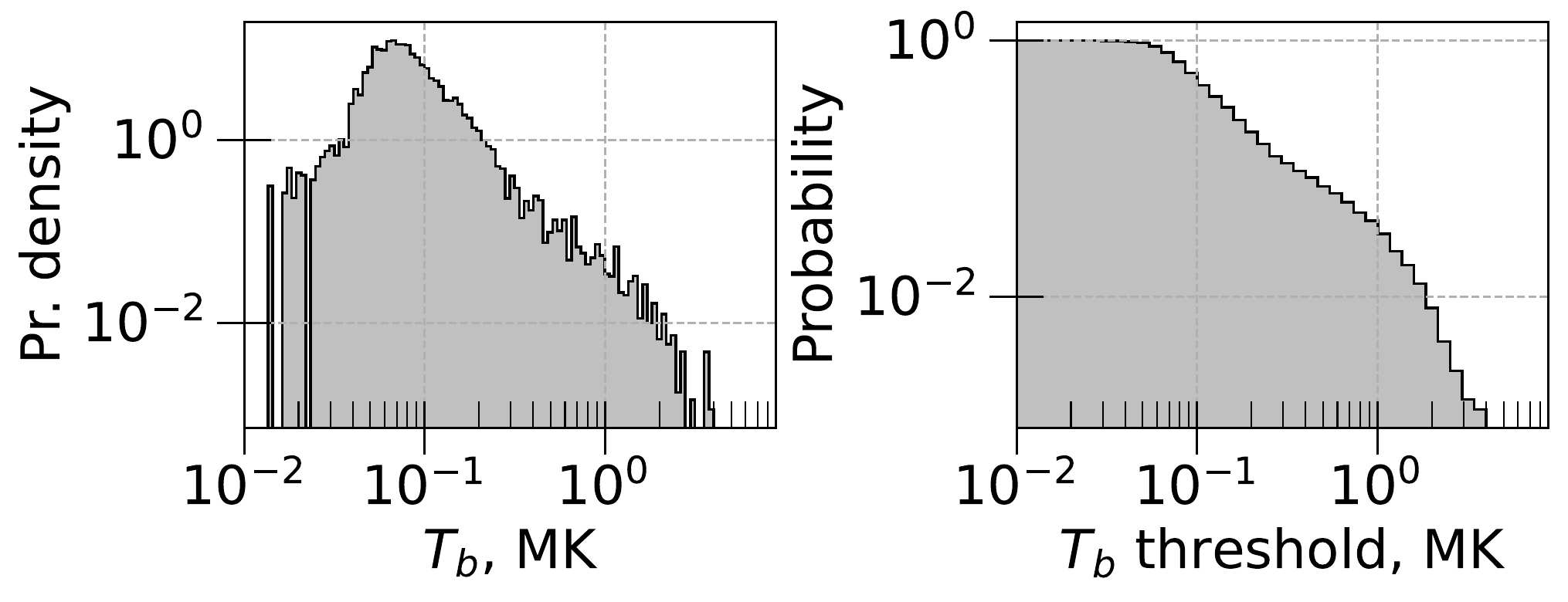"}
    \caption{Distribution of the the brightness temperature of all 5-min-cadence images across all candidates.
    Left panel represents the probability density of the peak brightness temperature. Right panel shows the fraction of images above the threshold.}
    \label{tempthreshold_all}
\end{figure}

Histogram of the peak brightness temperatures (left panel of Fig.~\ref{peak_brightness_dist}) shows that the most probable brightness temperatures for stable GR-34 sources remain in the interval between 50,000\,K and 300,000\,K with the mode value of approximately 74,000\,K. Figure \ref{tempthreshold_all} shows detailed brightness temperature distribution across all images.

\subsection{Solar activity manifestations}

GR-34 emission can be produced only by unusual ARs with exceptionally high magnetic field.
We inspect if those ARs are associated with other exceptional manifestations of solar activity. In last three columns of Table~\ref{results-table}, we list some of those manifestations including number of detected sunquakes, SEP events, or the strongest solar flare observed from the corresponding active region.

The sunquakes are observed in some flares including several famous events like 6 September 2017 X9.3 flare.
According to the list of sunquakes reported by \citet{2020ApJ...895...76S}, these events were detected in 6 of 10 ARs in our list observed in the SDO era since 2010. These 6 ARs hosted 46 sunquake events total. AR12673, the host of one of the optically thick GR-34 sources, produced 9 sunquakes, and AR11515 produced 13 sunquakes.

10 out of 27 active regions had at least 1 SEP event during their lifetime with total number of 14 SEP events. Four ARs (9393, 10808, 11515 and 12673), which hosted  optically thick GR-34 sources, produced 6 SEP events altogether. This means that 6 of 14 SEP events were produced by ARs with optically thick GR-34 sources (5 of 27 ARs).

17 of 27 (63\%) active regions in our list produced at least one X-class solar flare including 4 strongest flares in solar cycle 23 and 2 strongest flares in solar cycle 24. 4 of 5 optically thick GR-34 sources were observed in an AR that produced an X-class flare with an exception of AR 11967. The latter, however, was a host of an X1.2 flare during previous solar rotation when this AR had a number of 11944.

\section{Discussion and Conclusions}
\label{sec:discussion_and_conclusion}

This study reports a set of ARs that demonstrate bright GR emission at unusually high frequency, 34\,GHz, indicative of exceptionally high coronal magnetic field.
This coronal magnetic field must be around 4,000\,G if the GR emission is produced at the third gyroharmonic, or around  3,000\,G for the fourth one.

Despite we imply that GR-34 sources correspond to the corona, the measured brightness temperature of the detected bright radio sources is below the nominal coronal value of 1\,MK in most of the cases. Although this might be an effect of the  brightness dilution due to finite NoRH beam, the GR emission can partly (or in some cases
fully) originate from the cooler transition region.
Thus, this layer determines the lowest possible height where the magnetic field is estimated by our approach.

We investigated general statistical properties of these events such as the occurrence rate, dependence on the solar cycle phase, lifetime, sizes, brightness, and association with some manifestations of the solar activity. We found that the strong ($>4,000$\,G) coronal magnetic field happens more often than was previously appreciated. The occurrence of such cases is correlated with the phase of the solar cycle. Most of the identified ARs are the hosts of powerful X-class flares, sunquackes, or SEP events.  

The list of GR-34 ARs reported here offers several candidates for in depth case studies of the ARs with exceptionally strong magnetic field. In particular, three-dimensional models based on NLFFF reconstruction could be developed to investigate the magnetic and thermal structure of the ARs and distribution of electric currents (free energy) and twisted structures (flux ropes) there \citep[cf.][]{2019ApJ...880L..29A,2021ApJ...909...89F}. This will facilitate understanding the peculiar (neutral line) sources as they were proposed to be associated with such twisted structures \citep{2016SoPh..291..823K}. For example, the GR-34 source on 2017-Sep-06 was associated with the neutral line and a flux rope \citep{2019ApJ...880L..29A}; thus, some of the cases, reported here,  are likely too. In addition, this modeling will help determine the highest frequency, where the GR emission can be expected.

\cite{2006SoPh..239...41L} reported a historic record of strong (4 kG or more) photospheric magnetic field in ARs observed between 1917 and 2004. They found that 55 sunspot groups (about 0.2\% of ARs) have that strong magnetic field. Note that these fraction is lower than the fraction of ARs with 4\,kG coronal magnetic field identified in our study, which might indicate that \cite{2006SoPh..239...41L} overlooked a significant fraction of ARs with strong photospheric magnetic field due to incompleteness of historic vector magnetic measurements. They report only two cases for solar cycle 23, of which only one, 2003-Oct-22 with 4200\,G, is present in our list of the GR-34 sources.  

\cite{2013A&A...557A..24V} reported a very strong magnetic field of up to 7,000\,G in peripheral penumbral downflows in AR 10933 (January 05th 2007
from 1236–1310 UT) and  AR 10953
(April 30th 2007 from 1835–1939 UT). Neither of these ARs is present in our list, which implies absence of proportionally strong coronal magnetic field.

\cite{2018AGUFMSH41C3646O} reported 30 ARs with strong (4,000\,G or more) photospheric magnetic field based on Hinode observations. Our lists have 7 overlaps; it is interesting that most of top-10 ARs from the Okamoto list are present in our list, while only one AR from the remaining (bottom-20) part of the Okamoto list is present in our list.

This study was inspired by a discovery of record breaking ($\gtrsim4,000$\,G) coronal magnetic field in AR\,12673 on 2017-Sep-06 \citep{2019ApJ...880L..29A}. Here we found that high coronal magnetic field is not a unique case but appears regularly on the Sun, even though such ARs represent rather rare events: less than 1\% of all ARs show the GR-34 emission.
Nevertheless, many of these rare cases are associated with prominent manifestations of solar activity including powerful X- and M-class solar flares, sunquakes, or SEPs. This is consistent with several studies that reported unusually strong steady-state or flaring magnetic fields in the corona using various observational methods and techniques \citep{2006ApJ...641L..69B, 2019ApJ...874..126K,  2020FrASS...7...22K,  2020Sci...367..278F, 2020ApJ...900...17Y}; see Fig.\,\ref{f_all_coronal_B}. This also implies that future solar-dedicated radio instruments should cover high-frequency microwave range (perhaps, up to 40--50\,GHz or so) to monitor the whole range of the coronal magnetic field.

\acknowledgments
The authors thank Dr. Dale Gary for useful discussions. NoRH package routines from SolarSoft \citep{1998SoPh..182..497F} library of programming language IDL were used to synthesize NoRH imaging data from raw visibilities.  Python programming language and SunPy library \citep{2020ApJ...890...68S} have been used for the data analysis described above. 
GDF was supported in part
by NSF grants AGS-2121632, 
and AST-2206424  
and NASA grants 
80NSSC19K0068 to New Jersey Institute of Technology. VVF and SAA acknowledge support from the Russian Scientific Foundation under project 21-72-10139. This research benefited from support of the International Space Science Institute in ISSI Team 497.

\bibliographystyle{apj}
\bibliography{References,fleishman}

\end{document}